# Projective Loop Quantum Gravity
# II. Searching for Semi-Classical States

**Suzanne Lanéry[1,2] and Thomas Thiemann[1]**

[1] Institute for Quantum Gravity, Friedrich-Alexander University Erlangen-Nürnberg, Germany

[2] Mathematics and Theoretical Physics Laboratory, François-Rabelais University of Tours, France

**Abstract**

In [13] an extension of the Ashtekar-Lewandowski (AL) state space [2] of Loop Quantum Gravity was set up with the help a projective formalism introduced by Kijowski [12, 21, 15]. The motivation for this work was to achieve a more balanced treatment of the position and momentum variables (aka. holonomies and fluxes). Indeed, states in the AL Hilbert spaces describe discrete quantum excitations on top of a vacuum which is an eigenstate of the flux variables (a 'no-geometry' state): in such states, most holonomies are totally spread, making it difficult to approximate a smooth, classical 4-geometry.

However, going beyond the AL sector does not fully resolve this difficulty: one uncovers a deeper issue hindering the construction of states semi-classical with respect to a full set of observables. In the present article, we analyze this issue in the case of real-valued holonomies (we will briefly comment on the heuristic implications for other gauge groups, eg. $\mathcal{SU}(2)$ ). Specifically, we show that, in this case, there does *not* exist *any* state on the holonomy-flux algebra in which the variances of the holonomies and fluxes observables would *all* be *finite*, let alone *small*.

It is important to note that this obstruction cannot be bypassed by further enlarging the quantum state space, for it arises from the structure of the algebra itself: as there are too many (uncountably many) non-vanishing commutators between the holonomy and flux operators, the corresponding Heisenberg inequalities force the quantum uncertainties to blow up uncontrollably. A way out would be to suitably restrict the algebra of observables. In a companion paper [16] we take the first steps in this direction by developing a general framework to perform such a restriction without giving up the universality and diffeomorphism invariance of the theory.

# Contents

# 1 Introduction

This work is set up in the context of a *projective formalism* for Quantum Field Theory, first introduced by Jerzy Kijowski in the late '70s [12] and further developed by Andrzej Okołów more recently [20, 21, 22]. Instead of describing quantum states as density matrices over a single big Hilbert space, one constructs them as projective families of density matrices over small 'building block' Hilbert spaces [15, subsection 2.1]. The projections are given as *partial traces*, so that each small Hilbert space can be physically interpreted as selecting out specific degrees of freedom.

The description, within this projective framework, of a theory of Abelian connections had been developed by Okołów in [21, section 5] and [22], an important insight being to use 'building blocks' labeled by combinations of *edges and surfaces* (instead of edges only as in the case of the Ashtekar-Lewandowski (AL) Hilbert space). This construction was then generalized in [13] to an *arbitrary gauge group* $G$: in particular, $G$ is neither assumed to be Abelian nor compact, because the *factorization maps*, which are the central objects of the formalism, can be expressed *explicitly* in terms of the group operations [13, subsection 3.1]. The motivation to set up such a projective state space for the holonomy-flux algebra was to obtain better *semi-classical states*. As argued in the introduction of [13], the AL Hilbert space, being built out of a vacuum which is a *momentum eigenstate* (with vanishing fluxes), cannot accommodate states in which the quantum fluctuations would be *balanced* between position and momentum variables. However, now that this limitation could be overcome, we uncover a deeper issue, which has its root not in the restriction to a particular representation of the algebra of observables, but in the algebra *itself*.

As underlined in [13, subsection 3.2] the holonomy and flux variables do not come in independent, canonically conjugate pairs. In fact, any given flux observable has non trivial commutators with infinitely, *uncountably* many holonomies. As we will demonstrate below in the $G = \mathbb{R}$ case, the sheer amount of *Heisenberg uncertainty relations* arising from these non-vanishing commutators forces the quantum fluctuations to blow up uncontrollably as more and more degrees of freedom are taken into account: in the end, it is not even possible to keep the variance of all elementary variables *finite*, let alone small. Note that obstructions to the design of states with specific properties on the holonomy-flux algebra have been pointed out in earlier works [24]: in particular, the AL vacuum turns out to be the only diffeomorphism-invariant state [18], which notably forbids the design of a diffeomorphism-invariant coherent state.

These issues can be traced back to the fact that holonomy-flux algebra along analytical (or semi-analytical) edges and surfaces is generated by uncountably many elementary observables, which motivates the construction we will present in a companion article [16]: we will explore a strategy to drastically *reduce* the algebra of observables, without denaturing the *physical content* of the theory. Such a reduction then allows the *systematic* construction of projective states, in particular semi-classical ones. As a proof of principle, this strategy can easily be implemented in the case of a (slightly simplified) *one-dimensional* version of the holonomy-flux algebra, while the generalization to higher dimensions, especially the physically relevant $d = 3$ case, is currently a work in progress.

# 2 Obstructions to the construction of narrow states

We will, in subsection 2.1, work in the context of a general linear projective system $(\mathcal{L}, \mathcal{M}, \pi)$: namely, a projective system such that all small phase spaces $\mathcal{M}_\eta$ are *linear* and across which the distinction between configuration and momentum variables can be defined consistently (ie. all projections preserve this distinction). Note that this is precisely the setup of [21] and we recalled in [15, subsection 3.1] how to quantize such a classical projective limit into a projective system of quantum state space in the position representation.

In projective quantum state spaces of this kind, we can systematically study the construction of *narrow states*, namely states in which the measurement probabilities of *all* configuration and momentum variables have *finite variance*. In particular, we will spell out the conditions, arising from the Heisenberg uncertainty relations, for a certain assignment of variances to be realizable in a quantum state. Of special interest for the construction of semi-classical states are the *Gaussian states*, and, in fact, we will see that there exists, for any narrow state, a Gaussian state with the same variances.

In subsection 2.2, we will apply this study to the projective quantum state space set up in [13], in the special case $G = \mathbb{R}$ (where it constitutes a linear projective system in the sense above, as will be shown in prop. 2.13). We will then be able to prove (prop. 2.14) that there are not *any* narrow states in this state space: in other words, all states therein have infinite variance for at least some of the variables, and thus are not good candidates as semi-classical states. We will also formulate some weaker notions of semi-classicality that are excluded as well (props. 2.15 and 2.16).

## 2.1 Projective families of characteristic functions

The central tool of the present subsection will be Wigner characteristic functions (see [29, 11] and the review in appendix A): any quantum state can be represented by a function defined on the dual of the classical phase space, and this alternative representation is strictly equivalent to its representation as a density matrix $\rho$. This is the quantum version of characteristic distributions in classical statistical physics, in the sense that the moments of the probability distributions governing quantum measurements can be recovered from the derivatives of the characteristic function (prop. A.8). However, the positivity requirement for quantum characteristic functions (eq. (A.6.1), ensuring the positivity of $\rho$) differs from its classical counterpart (that would ensure positivity of the corresponding probability distribution), as it encodes the non-commutation of quantum observables (see the discussion preceding prop. A.6).

Unsurprisingly, the characteristic functions of the partial density matrices composing a projective state combine again into a projective family (the dual spaces $\mathcal{M}^*_\eta$ form naturally an inductive structure, with injections given by the pullback under the projections $\pi_{\eta'\to\eta}$, so functions on them arrange themselves in a projective structure, see also the discussion at the beginning of appendix A.2). This is the advantage of working with Wigner *characteristic* functions rather than Wigner quasi-probabilities, which would be in analogy to classical probability distributions (the computation of partial traces, or, at the classical level, of marginal probability distribution, is translated into a simple restriction when working with characteristic functions).

**Proposition 2.1** Let $(\mathcal{L}, \mathcal{M}, \pi)^{\downarrow}$ be a projective system of phase spaces such that:

1. $\forall \eta \in \mathcal{L}$, there exists two finite-dimensional real vector spaces $\mathcal{C}_\eta$, $\mathcal{P}_\eta$, an invertible linear map $\Xi_\eta : \mathcal{P}_\eta \to \mathcal{C}^*_\eta$ (where $\mathcal{C}^*_\eta$ denotes the dual of $\mathcal{C}_\eta$) and a symplectomorphism $L_\eta : \mathcal{M}_\eta \to \mathcal{C}_\eta \times \mathcal{P}_\eta$, with respect to the symplectic structure $\Omega_\eta$ defined on $\mathcal{C}_\eta \times \mathcal{P}_\eta$ by:
$$\forall (u, v),\, (u', v') \in \mathcal{C}_\eta \times \mathcal{P}_\eta\,,\ \Omega_\eta(u, v;\, u', v') := \Xi_\eta(v')(u) - \Xi_\eta(v)(u')\,;$$
2. $\forall \eta \preccurlyeq \eta' \in \mathcal{L}$, $\pi_{\eta' \to \eta} = L^{-1}_\eta \circ (Q_{\eta' \to \eta} \times P_{\eta' \to \eta}) \circ L_{\eta'}$ with $Q_{\eta' \to \eta}$, resp. $P_{\eta' \to \eta}$, a linear map $\mathcal{C}_{\eta'} \to \mathcal{C}_\eta$, resp. $\mathcal{P}_{\eta'} \to \mathcal{P}_\eta$.

We choose, for any $\eta \in \mathcal{L}$, a Lebesgue measure $\mu_\eta$ on $\mathcal{C}_\eta$. For any $\eta \preccurlyeq \eta' \in \mathcal{L}$, we define:

3. $\mathcal{C}_{\eta' \to \eta} := \operatorname{Ker} Q_{\eta' \to \eta}$;
4. $\begin{array}{rccl} \varphi_{\eta' \to \eta} : & \mathcal{C}_{\eta'} & \to & \mathcal{C}_{\eta' \to \eta} \times \mathcal{C}_\eta \\ & x' & \mapsto & R_{\eta' \to \eta}(x'),\ Q_{\eta' \to \eta}(x') \end{array}$, with $R_{\eta' \to \eta} : \mathcal{C}_{\eta'} \to \mathcal{C}_{\eta' \to \eta}$ the projection on $\mathcal{C}_{\eta' \to \eta}$ parallel to $\left(\Xi_\eta \circ P_{\eta' \to \eta} \circ \Xi^{-1}_{\eta'}\right)^* \langle \mathcal{C}_\eta \rangle$ (where $(\,\cdot\,)^*$ denotes the dual map).

Then, we can complete these elements into a factorizing system of measured manifolds [15, def. 3.1] $(\mathcal{L}, (\mathcal{C}, \mu), \varphi)^{\times}$, from which a projective system of quantum state spaces $(\mathcal{L}, \mathcal{H}, \Phi)^{\otimes}$ can be constructed as described in [15, prop. 3.3].

**Proof** Since a finite-dimensional vector space is in particular an additive simply-connected Lie group, this is a special case of [15, theorem 3.2]. The explicit expressions for $\mathcal{C}_{\eta' \to \eta}$ and $\varphi_{\eta' \to \eta}$ are obtained by adapting the ones from the proof of [15, theorem 3.2] to the slightly different notations we are using here. □

**Definition 2.2** We consider the same objects as in prop. 2.1. A projective family of characteristic functions is a family $\left(W_\eta\right)_{\eta \in \mathcal{L}}$ such that:

1. for any $\eta \in \mathcal{L}$, $W_\eta$ is a *continuous* function $\mathcal{C}^*_\eta \times \mathcal{P}^*_\eta \to \mathbb{C}$;
2. for any $\eta \in \mathcal{L}$, any $N \in \mathbb{N}$, any $(s_1, t_1), \ldots, (s_N, t_N) \in \mathcal{C}^*_\eta \times \mathcal{P}^*_\eta$ and any $z_1, \ldots, z_N \in \mathbb{C}$:
$$\sum_{i,j=1}^{N} \overline{z_j}\, z_i\, e^{i\,\xi_{ij}}\, W_\eta(s_i - s_j,\, t_i - t_j) \geqslant 0\,, \tag{2.2.1}$$
where $\xi_{ij} := \frac{1}{2}\left(t_i\left(\Xi^{-1}_\eta(s_j)\right) - t_j\left(\Xi^{-1}_\eta(s_i)\right)\right)$;
3. for any $\eta \preccurlyeq \eta' \in \mathcal{L}$, $W_\eta = W_{\eta'} \circ \left(Q^*_{\eta' \to \eta} \times P^*_{\eta' \to \eta}\right)$.

We denote by $\mathcal{W}^{\downarrow}_{(\mathcal{L}, (\mathcal{C}, \mathcal{P}), (Q, P))}$ the space of all projective families of characteristic functions.

**Proposition 2.3** We consider the same objects as in prop. 2.1. Let $\eta \in \mathcal{L}$. On $\mathcal{H}_\eta = L_2(\mathcal{C}_\eta, d\mu_\eta)$ we define, for any $(s, t) \in \mathcal{C}^*_\eta \times \mathcal{P}^*_\eta$, a unitary operator $\mathrm{T}_\eta(s, t)$ on $\mathcal{H}_\eta$ by:
$$\forall \psi \in \mathcal{H}_\eta\,,\ \forall x \in \mathcal{C}_\eta\,,\ [\mathrm{T}_\eta(s, t)\, \psi](x) = \exp\left(i\, s(x) + \tfrac{i}{2}\, t\left(\Xi^{-1}_\eta(s)\right)\right)\, \psi\left(x + \Xi^{*,-1}_\eta(t)\right),$$

as well as a densely defined, essentially self-adjoint operator $\mathrm{X}_\eta(s,t)$ with dense domain $\mathcal{D}_\eta := C_o^\infty(\mathcal{C}_\eta, \mathbb{C})$ (the space of smooth, compactly supported, complex-valued functions on $\mathcal{C}_\eta$) by:

$$\forall \psi \in \mathcal{D}_\eta\,,\ \forall x \in \mathcal{C}_\eta\,,\ [\mathrm{X}_\eta(s,t)\,\psi](x) = s(x)\,\psi(x) - i\left[T_x\psi\right]\left(\Xi_\eta^{*,-1}(t)\right)$$

(where $T_x\psi$ denotes the differential of $\psi$ at $x$ and $\Xi_\eta^{*,-1} = \left(\Xi_\eta^*\right)^{-1} = \left(\Xi_\eta^{-1}\right)^*$).

For any $\eta \preccurlyeq \eta' \in \mathcal{L}$ and any $(s,t) \in \mathcal{C}_\eta^* \times \mathcal{P}_{\eta'}^*$, we have:

$$\mathrm{T}_{\eta'}(s',t') = \Phi_{\eta'\to\eta}^{-1} \circ \left(\mathrm{id}_{\mathcal{H}_{\eta'\to\eta}} \otimes \mathrm{T}_\eta(s,t)\right) \circ \Phi_{\eta'\to\eta}\,,$$

$$\&\quad \mathrm{X}_{\eta'}(s',t') = \Phi_{\eta'\to\eta}^{-1} \circ \left(\mathrm{id}_{\mathcal{H}_{\eta'\to\eta}} \otimes \mathrm{X}_\eta(s,t)\right) \circ \Phi_{\eta'\to\eta}\,, \tag{2.3.1}$$

where $(s',t') := \left(Q_{\eta'\to\eta}^*(s),\ P_{\eta'\to\eta}^*(t)\right)$.

**Proof** Let $\eta \in \mathcal{L}$. For any $(s,t) \in \mathcal{C}_\eta^* \times \mathcal{P}_{\eta'}^*$, it has been proven in prop. A.5, that $\mathrm{T}_\eta(s,t)$, resp. $\mathrm{X}_\eta(s,t)$, is well-defined and is unitary, resp. essentially self-adjoint.

Let $\eta \preccurlyeq \eta' \in \mathcal{L}$ and $(s,t) \in \mathcal{C}_\eta^* \times \mathcal{P}_\eta^*$. Let $(s',t') := \left(Q_{\eta'\to\eta}^*(s),\ P_{\eta'\to\eta}^*(t)\right)$. For any $\psi_\eta \in \mathcal{H}_\eta$ and any $\psi_{\eta'\to\eta} \in \mathcal{H}_{\eta'\to\eta} = L_2(\mathcal{C}_{\eta'\to\eta}\,,\ d\mu_{\eta'\to\eta})$, we have, using the expression for $\Phi_{\eta'\to\eta}$ from [15, prop. 3.3]:

$$\begin{aligned}
\forall (y,x) \in \mathcal{C}_{\eta'\to\eta} \times \mathcal{C}_\eta,\quad & \left[\Phi_{\eta'\to\eta}\,\mathrm{T}_{\eta'}(s',t')\,\Phi_{\eta'\to\eta}^{-1}\left(\psi_{\eta'\to\eta} \otimes \psi_\eta\right)\right](y,x) = \\
&= \left[\mathrm{T}_{\eta'}(s',t')\,\Phi_{\eta'\to\eta}^{-1}\left(\psi_{\eta'\to\eta} \otimes \psi_\eta\right)\right] \circ \varphi_{\eta'\to\eta}^{-1}(y,x) \\
&= \exp\left(i\,s' \circ \varphi_{\eta'\to\eta}^{-1}(y,x) + \tfrac{i}{2}\,t'\left(\Xi_{\eta'}^{-1}(s')\right)\right)\ \Phi_{\eta'\to\eta}^{-1}(\psi_{\eta'\to\eta} \otimes \psi_\eta)\left(\varphi_{\eta'\to\eta}^{-1}(y,x) + \Xi_{\eta'}^{*,-1}(t')\right) \\
&= \exp\left(i\,s(x) + \tfrac{i}{2}\,t\left(\Xi_\eta^{-1}(s)\right)\right)\psi_{\eta'\to\eta}(y)\,\psi_\eta\left(x + \Xi_\eta^{*,-1}(t)\right) \\
&= \psi_{\eta'\to\eta}(y)\left[\mathrm{T}_\eta(s,t)\,\psi_\eta\right](x)\,,
\end{aligned}$$

where we have used:

$$\begin{aligned}
&\forall (y,x) \in \mathcal{C}_{\eta'\to\eta} \times \mathcal{C}_\eta, \\
&\qquad Q_{\eta'\to\eta} \circ \varphi_{\eta'\to\eta}^{-1}(y,x) = x\,, \\
&\qquad \&\quad \varphi_{\eta'\to\eta}\left(\varphi_{\eta'\to\eta}^{-1}(y,x) + \Xi_{\eta'}^{*,-1} \circ P_{\eta'\to\eta}^*(t)\right) = \left(y,\ x + Q_{\eta'\to\eta} \circ \Xi_{\eta'}^{*,-1} \circ P_{\eta'\to\eta}^*(t)\right)
\end{aligned}$$

(from the definition of $\varphi_{\eta'\to\eta}$ in prop. 2.1) as well as:

$$Q_{\eta'\to\eta} \circ \Xi_{\eta'}^{*,-1} \circ P_{\eta'\to\eta}^* = \Xi_\eta^{*,-1}\,,$$

as follows from the compatibility of $L_\eta^{-1} \circ \left(Q_{\eta'\to\eta} \times P_{\eta'\to\eta}\right) \circ L_{\eta'}$ with the symplectic structures $\Omega_{\eta'}$ and $\Omega_\eta$ (see [14, def. 2.1] and [15, eq. (3.2.1)]). Thus, we get:

$$\forall \psi_\eta \in \mathcal{H}_\eta\,,\ \forall \psi_{\eta'\to\eta} \in \mathcal{H}_{\eta'\to\eta}\,,\quad \Phi_{\eta'\to\eta}\,\mathrm{T}_{\eta'}(s',t')\,\Phi_{\eta'\to\eta}^{-1}\left(\psi_{\eta'\to\eta} \otimes \psi_\eta\right) = \psi_{\eta'\to\eta} \otimes \left[\mathrm{T}_\eta(s,t)\,\psi_\eta\right],$$

and therefore:

$$\mathrm{T}_{\eta'}(s',t') = \Phi_{\eta'\to\eta}^{-1} \circ \left(\mathrm{id}_{\mathcal{H}_{\eta'\to\eta}} \otimes \mathrm{T}_\eta(s,t)\right) \circ \Phi_{\eta'\to\eta}\,.$$

Similarly, we have, for any $\psi_\eta \in \mathcal{D}_\eta$ and any $\psi_{\eta'\to\eta} \in \mathcal{D}_{\eta'\to\eta} := C_o^\infty(\mathcal{C}_{\eta'\to\eta}, \mathbb{C})$, $\Phi_{\eta'\to\eta}^{-1}(\psi_{\eta'\to\eta} \otimes$

$\psi_\eta) \in \mathcal{D}_{\eta'}$ and:

$$\Phi_{\eta'\to\eta}\, \mathrm{X}_{\eta'}(s', t')\, \Phi^{-1}_{\eta'\to\eta}\, (\psi_{\eta'\to\eta} \otimes \psi_\eta) = \psi_{\eta'\to\eta} \otimes \left[\mathrm{X}_\eta(s, t)\, \psi_\eta\right],$$

using:

$$\forall (y, x) \in \mathcal{C}_{\eta'\to\eta} \times \mathcal{C}_\eta\,,$$
$$\left[T_{\varphi^{-1}_{\eta'\to\eta}(y,x)}\, \Phi^{-1}_{\eta'\to\eta} \left(\psi_{\eta'\to\eta} \otimes \psi_\eta\right)\right] = \psi_\eta(x)\, \left[T_y \psi_{\eta'\to\eta}\right] \circ R_{\eta'\to\eta} + \psi_{\eta'\to\eta}(y)\, \left[T_x\, \psi_\eta\right] \circ Q_{\eta'\to\eta}\,.$$

Moreover, $\mathcal{D}_{\eta'\to\eta} \otimes \mathcal{D}_\eta$ (understood as a tensor product of vector spaces, ie. *without* any completion) is a dense subspace of $\Phi_{\eta'\to\eta} \langle \mathcal{D}_{\eta'}\rangle$ and $\left(\Phi_{\eta'\to\eta}\, \mathrm{X}_{\eta'}(s', t')\, \Phi^{-1}_{\eta'\to\eta}\right)\Big|_{\mathcal{D}_{\eta'\to\eta}\otimes\mathcal{D}_\eta} = \mathrm{id}_{\mathcal{H}_{\eta'\to\eta}}\Big|_{\mathcal{D}_{\eta'\to\eta}} \otimes \mathrm{X}_\eta(s, t)$ is essentially self-adjoint (as a tensor product of essentially self-adjoint operators, see [23, theorem VIII.33]), hence the unique self-adjoint extensions of $\Phi_{\eta'\to\eta}\, \mathrm{X}_{\eta'}(s', t')\, \Phi^{-1}_{\eta'\to\eta}$ and $\mathrm{id}_{\mathcal{H}_{\eta'\to\eta}} \otimes \mathrm{X}_\eta(s, t)$ coincide. □

**Proposition 2.4** We consider the same objects as in def. 2.2 and prop. 2.3. For any $\eta \in \mathcal{L}$ and any $\rho_\eta \in \overline{\mathcal{S}}_\eta$ (with $\overline{\mathcal{S}}_\eta$ the space of non-negative, traceclass operators on $\mathcal{H}_\eta$), we define $W_{\rho_\eta}$ by:

$$\begin{array}{rccl} W_{\rho_\eta} & : \mathcal{C}^*_\eta \times \mathcal{P}^*_\eta & \to & \mathbb{C} \\ & (s, t) & \mapsto & \mathrm{Tr}_{\mathcal{H}_\eta}\, \rho_\eta\, \mathrm{T}_\eta(s, t) \end{array}.$$

The map $W : \left(\rho_\eta\right)_{\eta\in\mathcal{L}} \mapsto \left(W_{\rho_\eta}\right)_{\eta\in\mathcal{L}}$ is a bijection $\overline{\mathcal{S}}^{\otimes}_{(\mathcal{L},\mathcal{H},\Phi)} \to \mathcal{W}^{\downarrow}_{(\mathcal{L},\,(\mathcal{C},\mathcal{P}),\,(Q,P))}$ (where $\overline{\mathcal{S}}^{\otimes}_{(\mathcal{L},\mathcal{H},\Phi)}$ has been defined in [15, def. 2.2]).

**Proof** For any $\eta \in \mathcal{L}$, we denote by $\mathcal{W}_\eta$ the space of all continuous functions of positive type on $\mathcal{C}^*_\eta \times \mathcal{P}^*_\eta$ (prop. A.6). From props. A.7 and A.10, the map $W_\eta : \rho_\eta \mapsto W_{\rho_\eta}$ is a bijection $\overline{\mathcal{S}}_\eta \to \mathcal{W}_\eta$. Moreover, we have:

$$\overline{\mathcal{S}}^{\otimes}_{(\mathcal{L},\mathcal{H},\Phi)} = \left\{ \left(\rho_\eta\right)_{\eta\in\mathcal{L}} \,\middle|\, \forall \eta,\, \rho_\eta \in \overline{\mathcal{S}}_\eta \quad \& \quad \forall \eta \preccurlyeq \eta',\, \rho_\eta = \mathrm{Tr}_{\mathcal{H}_{\eta'\to\eta}} \left(\Phi_{\eta'\to\eta}\, \rho_{\eta'}\, \Phi^{-1}_{\eta'\to\eta}\right)\right\},$$

and:

$$\mathcal{W}^{\downarrow}_{(\mathcal{L},\,(\mathcal{C},\mathcal{P}),\,(Q,P))} = \left\{ \left(W_\eta\right)_{\eta\in\mathcal{L}} \,\middle|\, \forall \eta,\, W_\eta \in \mathcal{W}_\eta \quad \& \quad \forall \eta \preccurlyeq \eta',\, W_\eta = W_{\eta'} \circ \left(Q^*_{\eta'\to\eta} \times P^*_{\eta'\to\eta}\right)\right\}.$$

Now, from eq. (2.3.1), we have, for any $\eta \preccurlyeq \eta'$ and any $\rho_{\eta'} \in \overline{\mathcal{S}}_{\eta'}$ :

$$W_\eta \left(\mathrm{Tr}_{\mathcal{H}_{\eta'\to\eta}} \left(\Phi_{\eta'\to\eta}\, \rho_{\eta'}\, \Phi^{-1}_{\eta'\to\eta}\right)\right) = W_{\eta'}(\rho_{\eta'}) \circ \left(Q^*_{\eta'\to\eta} \times P^*_{\eta'\to\eta}\right).$$

Thus, $W$ is well-defined as a map $\overline{\mathcal{S}}^{\otimes}_{(\mathcal{L},\mathcal{H},\Phi)} \to \mathcal{W}^{\downarrow}_{(\mathcal{L},\,(\mathcal{C},\mathcal{P}),\,(Q,P))}$ and is bijective. □

Asking for a state to have finite variances in all position and configuration variables is expressed at the level of its Wigner characteristic function by asking the latter to be twice differentiable at 0, and the corresponding covariance matrix can be obtained from the Hessian of the characteristic function (prop. A.8). In particular, the positivity requirement for the characteristic function (eq. (A.6.1) ) gives rise to inequalities that have to be satisfied by the covariance matrix (def. 2.6.2): these are nothing but the Heisenberg uncertainty relations (as is manifest from the rewriting in prop. 2.7.2 or 2.7.3, since $V_\eta(s, s) = \Delta \mathrm{X}^2_\eta(s, 0)$ and $U_\eta(t, t) = \Delta \mathrm{X}^2_\eta(0, t)$ ). These inequalities will play a crucial role for the result of the next subsection.

Moreover, the projective structure binding the partial characteristic functions of a projective state together goes down to its covariance matrices on the various labels $\eta$, which arrange naturally into their own projective structure.

**Definition 2.5** Let $\rho = \left(\rho_\eta\right)_{\eta \in \mathcal{L}} \in \overline{\mathcal{S}}^{\otimes}_{(\mathcal{L},\mathcal{H},\Phi)}$ and let $\left(W_\eta\right)_{\eta \in \mathcal{L}} := W(\rho)$. We say that $\rho$ is a narrow state if there exist, for any $\eta \in \mathcal{L}$, a linear form $W^{(1)}_\eta$ and a symmetric bilinear form $W^{(2)}_\eta$ on $\mathcal{C}^*_\eta \times \mathcal{P}^*_\eta$ such that:

$$\forall (s,t) \in \mathcal{C}^*_\eta \times \mathcal{P}^*_\eta\,,\; W_\eta(\tau s,\, \tau t) = 1 + i\, \tau\, W^{(1)}_\eta(s,t) - \frac{\tau^2}{2}\, W^{(2)}_\eta(s,t;\, s,t) + o(\tau^2)\,. \tag{2.5.1}$$

From props. A.5 and A.8, we then have, for any $\eta \in \mathcal{L}$ :

1. $\mathrm{Tr}_{\mathcal{H}_\eta} \rho_\eta = 1$ (so that $\rho_\eta$ is a density matrix on $\mathcal{H}_\eta$);
2. $\forall (s,t) \in \mathcal{C}^*_\eta \times \mathcal{P}^*_\eta$, $\mathrm{Tr}_{\mathcal{H}_\eta} \rho_\eta\, \mathrm{X}_\eta(s,t) = W^{(1)}_\eta(s,t)\,$;
3. $\forall (s,t),\, (s',t') \in \mathcal{C}^*_\eta \times \mathcal{P}^*_\eta$, $\mathrm{Tr}_{\mathcal{H}_\eta} \dfrac{\mathrm{X}_\eta(s,t)\, \rho_\eta\, \mathrm{X}_\eta(s',t') + \mathrm{X}_\eta(s',t')\, \rho_\eta\, \mathrm{X}_\eta(s,t)}{2} = W^{(2)}_\eta(s,t;\, s',t')\,$.

We denote the space of all narrow states by $\hat{\mathcal{S}}^{\otimes}_{(\mathcal{L},\mathcal{H},\Phi)}$.

**Definition 2.6** We consider the same objects as in prop. 2.1. A projective family of variances is a family $\left(V_\eta\,,\, U_\eta\right)_{\eta \in \mathcal{L}}$ such that:

1. for any $\eta \in \mathcal{L}$, $V_\eta$, resp. $U_\eta$, is a strictly *positive* symmetric bilinear form on $\mathcal{C}^*_\eta$, resp. $\mathcal{P}^*_\eta$;
2. for any $\eta \in \mathcal{L}$ and any $(s,t) \in \mathcal{C}^*_\eta \times \mathcal{P}^*_\eta$, $V_\eta(s,\, s) + U_\eta(t,\, t) - t\big(\Xi^{-1}_\eta(s)\big) \geqslant 0$;
3. for any $\eta \preccurlyeq \eta' \in \mathcal{L}$, $V_\eta = V_{\eta'} \circ (Q^*_{\eta' \to \eta} \times Q^*_{\eta' \to \eta})$ & $U_\eta = U_{\eta'} \circ (P^*_{\eta' \to \eta} \times P^*_{\eta' \to \eta})\,$.

We denote by $\mathcal{V}^{\downarrow}_{(\mathcal{L},\,(\mathcal{C},\mathcal{P}),\,(Q,P))}$ the space of all projective families of variances.

**Proposition 2.7** Let $\eta \in \mathcal{L}$ and let $V_\eta$, resp. $U_\eta$, be a strictly positive symmetric bilinear form on $\mathcal{C}^*_\eta$, resp. $\mathcal{P}^*_\eta$. Let $V^{-1}_\eta$, resp. $U^{-1}_\eta$, be the strictly positive symmetric bilinear form on $\mathcal{C}_\eta$, resp. $\mathcal{P}_\eta$, characterized by:

$$\forall s, s' \in \mathcal{C}^*_\eta\,,\; V^{-1}_\eta\big(V_\eta(s,\,\cdot\,),\, V_\eta(s',\,\cdot\,)\big) = V_\eta(s,s')\,,$$

$$\text{resp. }\; \forall t, t' \in \mathcal{P}^*_\eta\,,\; U^{-1}_\eta\big(U_\eta(t,\,\cdot\,),\, U_\eta(t',\,\cdot\,)\big) = U_\eta(t,t')\,,$$

with the canonical identification $\mathcal{C}^{**}_\eta \approx \mathcal{C}_\eta$, resp. $\mathcal{P}^{**}_\eta \approx \mathcal{P}_\eta$.

Then, the three following conditions are equivalent:

1. $\forall (s,t) \in \mathcal{C}^*_\eta \times \mathcal{P}^*_\eta\,,\; V_\eta(s,\, s) + U_\eta(t,\, t) - t\big(\Xi^{-1}_\eta(s)\big) \geqslant 0$;
2. $\forall s \in \mathcal{C}^*_\eta\,,\; 2\, V_\eta(s,\, s) - \dfrac{1}{2}\, U^{-1}_\eta\big(\Xi^{-1}_\eta(s),\, \Xi^{-1}_\eta(s)\big) \geqslant 0$;
3. $\forall t \in \mathcal{P}^*_\eta\,,\; 2\, U_\eta(t,\, t) - \dfrac{1}{2}\, V^{-1}_\eta\big(\Xi^{*,-1}_\eta(t),\, \Xi^{*,-1}_\eta(t)\big) \geqslant 0$.

**Proof** Let $\widetilde{V}_\eta$ be the linear map $\mathcal{C}_\eta^* \to \mathcal{C}_\eta$ defined by:

$$\forall s \in \mathcal{C}_\eta^*,\ \widetilde{V}_\eta(s) = V_\eta(s,\,\cdot\,) \in \mathcal{C}_\eta^{**} \approx \mathcal{C}_\eta\,.$$

For any $s \neq 0$, we have $s\big(\widetilde{V}_\eta(s)\big) = V_\eta(s,\,s) > 0$ (for $V_\eta$ is strictly positive), hence $\widetilde{V}_\eta(s) \neq 0$. Thus, $\widetilde{V}_\eta$ is injective, and therefore bijective, since $\mathcal{C}_\eta^*$ and $\mathcal{C}_\eta$ are finite-dimensional vector spaces of the same dimension. $V_\eta^{-1}$ is then defined by:

$$\forall x, x' \in \mathcal{C}_\eta\,,\ V_\eta^{-1}(x,\,x') := V_\eta\big(\widetilde{V}_\eta^{-1}(x),\,\widetilde{V}_\eta^{-1}(x')\big)\,,$$

and is therefore a strictly positive symmetric bilinear form on $\mathcal{C}_\eta$. We have similarly $\widetilde{U}_\eta : \mathcal{P}_\eta^* \to \mathcal{P}_\eta$ and $U_\eta^{-1} : \mathcal{P}_\eta \times \mathcal{P}_\eta \to \mathbb{R}$.

*2.7.1 $\Rightarrow$ 2.7.2 & 2.7.1 $\Rightarrow$ 2.7.3.* We assume that 2.7.1 holds. Applying with $(s,t) = \big(s,\,\frac{1}{2}\,\widetilde{U}_\eta^{-1} \circ \Xi_\eta^{-1}(s)\big)$ for some $s \in \mathcal{C}_\eta^*$ yields:

$$V_\eta(s,\,s) + \frac{1}{4} U_\eta^{-1}\big(\Xi_\eta^{-1}(s),\,\Xi_\eta^{-1}(s)\big) - \frac{1}{2} U_\eta^{-1}\big(\Xi_\eta^{-1}(s),\,\Xi_\eta^{-1}(s)\big) \geqslant 0\,,$$

where we have used:

$$\forall p, p' \in \mathcal{P}_\eta\,,\ \big(\widetilde{U}_\eta^{-1}(p)\big)(p') = U_\eta\big(\widetilde{U}_\eta^{-1}(p),\,\widetilde{U}_\eta^{-1}(p')\big) = U_\eta^{-1}(p,\,p')\,.$$

Thus, we obtain the condition 2.7.2. Similarly, we can prove that 2.7.1 implies 2.7.3.

*2.7.2 $\Rightarrow$ 2.7.1 & 2.7.3 $\Rightarrow$ 2.7.1.* We assume that 2.7.2 holds. $V_\eta$ provides a scalar product on $\mathcal{C}_\eta^*$, so the strictly positive symmetric bilinear form $U_\eta^{-1}\big(\Xi_\eta^{-1}(\,\cdot\,),\,\Xi_\eta^{-1}(\,\cdot\,)\big)$ can be diagonalized in a $V_\eta$-orthonormal basis $(f_i)_{i\in\{1,\dots,n\}}$ (with $n := \dim \mathcal{C}_\eta$), ie. we have:

$$\forall i, j \in \{1, \dots, n\}\,,\ V_\eta(f_i\,,\,f_j) = \delta_{ij} \quad \& \quad U_\eta^{-1}\big(\Xi_\eta^{-1}(f_i),\,\Xi_\eta^{-1}(f_j)\big) = \lambda_{(i)}\,\delta_{ij}\,,$$

with $\forall i \in \{1, \dots, n\}\,,\ \lambda_{(i)} > 0$. Next, the condition 2.7.2 can be rewritten:

$$\forall i \in \{1, \dots, n\}\,,\ \lambda_{(i)} \leqslant 4\,.$$

Moreover, defining, for any $i \in \{1, \dots, n\}$, $g_i = \Xi_\eta^*(e_i)$ with $(e_i)_{i\in\{1,\dots,n\}}$ the basis in $\mathcal{C}_\eta$ dual to $(f_i)_{i\in\{1,\dots,n\}}$, we have:

$$\forall i, j \in \{1, \dots, n\}\,,\ \lambda_{(i)}\,\delta_{ij} = U_\eta^{-1}\big(\Xi_\eta^{-1}(f_i),\,\Xi_\eta^{-1}(f_j)\big) = U_\eta\big(\lambda_{(i)}\,g_i,\,\lambda_{(j)}\,g_j\big)$$

where we have used that, for any $i \in \{1, \dots, n\}$, $\widetilde{U}_\eta^{-1} \circ \Xi_\eta^{-1}(f_i) = \lambda_{(i)}\,g_i$. Hence, we get:

$$\forall i, j \in \{1, \dots, n\}\,,\ U_\eta\big(g_i,\,g_j\big) = \frac{\delta_{ij}}{\lambda_{(j)}}\,.$$

Let $s = s^i\,f_i \in \mathcal{C}_\eta^*$ and $t = t^j\,g_j \in \mathcal{P}_\eta^*$ (with implicit summation). We have:

$$V_\eta(s,\,s) + U_\eta(t,\,t) - t\big(\Xi_\eta^{-1}(s)\big) = s^i s^i + \frac{t^j t^j}{\lambda_{(j)}} - s^i\,t^i\,.$$

Now, for any $\sigma, \tau \in \mathbb{R}$, and any $\lambda \in \,]0,\,4]$, we have:

$$\sigma^2 + \frac{\tau^2}{\lambda} - \sigma\tau \geqslant 0\,,$$

so 2.7.1 is fulfilled. Similarly, we can prove that 2.7.3 implies 2.7.1. □

**Proposition 2.8** Let $\rho = \left(\rho_\eta\right)_{\eta\in\mathcal{L}} \in \hat{\mathcal{S}}^{\otimes}_{(\mathcal{L},\mathcal{H},\Phi)}$ and let $\left(W_\eta\right)_{\eta\in\mathcal{L}} := W(\rho)$. For any $\eta \in \mathcal{L}$, we define:

$$\begin{aligned} &\forall (s, s') \in \mathcal{C}^*_\eta\,,\ V_\eta(s,\ s') := W^{(2)}_\eta(s, 0;\ s', 0) - W^{(1)}_\eta(s, 0)\, W^{(1)}_\eta(s', 0)\,,\\ \&\quad &\forall (t, t') \in \mathcal{P}^*_\eta\,,\ U_\eta(t,\ t') := W^{(2)}_\eta(0, t;\ 0, t') - W^{(1)}_\eta(0, t)\, W^{(1)}_\eta(0, t')\,, \end{aligned}$$

with $W^{(1)}_\eta$ and $W^{(2)}_\eta$ as in def. 2.5.

Then, $\left(V_\eta\,,\ U_\eta\right)_{\eta\in\mathcal{L}} \in \mathcal{V}^{\downarrow}_{(\mathcal{L},\,(\mathcal{C},\mathcal{P}),\,(Q,P))}$. Accordingly, we define the map $V$ as:

$$\begin{array}{rcl} V\ :\ \hat{\mathcal{S}}^{\otimes}_{(\mathcal{L},\mathcal{H},\Phi)} & \to & \mathcal{V}^{\downarrow}_{(\mathcal{L},\,(\mathcal{C},\mathcal{P}),\,(Q,P))} \\ \rho & \mapsto & \left(V_\eta\,,\ U_\eta\right)_{\eta\in\mathcal{L}} \end{array}.$$

**Proof** Let $\eta \in \mathcal{L}$, $(s, t) \in \mathcal{C}^*_\eta \times \mathcal{P}^*_\eta$, and $\tau \in \mathbb{R}$. Applying eq. (2.2.1) for the points $(0, 0)$, $(\tau s, 0)$, $(0, \tau t)$ with respective coefficients $-(1 + i)$, $e^{-i\tau\, W^{(1)}_\eta(s,0)}$, $i\, e^{-i\tau\, W^{(1)}_\eta(0,t)}$ yields:

$$\begin{aligned} 4 + 2\,\mathrm{Re}\Big[&(i-1)\, e^{-i\tau\, W^{(1)}_\eta(s,0)}\, W_\eta(\tau s, 0) - (1+i)\, e^{-i\tau\, W^{(1)}_\eta(0,t)}\, W_\eta(0, \tau t) + \\ &+ i\, e^{-i\tau\, W^{(1)}_\eta(-s,t)}\, e^{i\frac{\tau^2}{2}\, t\circ\Xi^{-1}_\eta(s)}\, W_\eta(-\tau s, \tau t)\Big] \geqslant 0\,, \end{aligned}$$

where we have used that $W_\eta(0, 0) = 1$ and that, for any $(s', t') \in \mathcal{C}^*_\eta \times \mathcal{P}^*_\eta$, $W_\eta(-s', -t') = \overline{W(s', t')}$ (prop. A.6; note that this in particular accounts for the reality of $W^{(1)}_\eta$ and $W^{(2)}_\eta$). Now, for any $(s', t') \in \mathcal{C}^*_\eta \times \mathcal{P}^*_\eta$, we have the expansion:

$$e^{-i\tau\, W^{(1)}_\eta(s',t')}\, W_\eta(\tau s', \tau t') = 1 - \frac{\tau^2}{2} W^{(2)}_\eta(s', t';\ s', t') + \frac{\tau^2}{2}\left[W^{(1)}_\eta(s', t')\right]^2 + o(\tau^2)\,.$$

Inserting in the previous inequality, we get:

$$4 + 2\left[-2 + \frac{\tau^2}{2} V_\eta(s,\ s) + \frac{\tau^2}{2} U_\eta(t,\ t) - \frac{\tau^2}{2}\, t\circ\Xi^{-1}_\eta(s) + o(\tau^2)\right] \geqslant 0\,.$$

Hence, we have, for any $(s, t) \in \mathcal{C}^*_\eta \times \mathcal{P}^*_\eta$:

$$V_\eta(s,\ s) + U_\eta(t,\ t) - t\circ\Xi^{-1}_\eta(s) \geqslant 0\,. \tag{2.8.1}$$

Let $s \in \mathcal{C}^*_\eta$ with $s \neq 0$. We have $\Xi^{-1}_\eta(s) \neq 0$, so there exists $t_s \in \mathcal{P}^*_\eta$ such that $t_s \circ \Xi^{-1}_\eta(s) = 1$. Applying eq. (2.8.1) with $(0, t_s)$ yields $u := U_\eta(t_s,\ t_s) \geqslant 0$, allowing us to define $t := \frac{1}{1+u}\, t_s$. Applying again eq. (2.8.1), now with $(s, t)$, we get:

$$V_\eta(s,\ s) - \frac{1}{(1+u)^2} \geqslant 0\,,$$

so $V_\eta(s,\ s) > 0$. Similarly, we have, for any $t \in \mathcal{P}^*_\eta$, $t \neq 0 \Rightarrow U_\eta(t,\ t) > 0$. Thus, def. 2.6.1 and 2.6.2 are fulfilled (actually, we have shown that 2.6.1 is implied by 2.6.2).

Let $\eta \preccurlyeq \eta' \in \mathcal{L}$. From $W_\eta = W_{\eta'} \circ (Q^*_{\eta'\to\eta} \times P^*_{\eta'\to\eta})$ together with eq. (2.5.1) implies, for any

$(s,t) \in \mathcal{C}^*_\eta \times \mathcal{P}^*_\eta$:

$$W^{(1)}_\eta(s,t) = W^{(1)}_{\eta'}(s',t') \quad \& \quad W^{(2)}_\eta(s,t;\, s,t) = W^{(2)}_{\eta'}(s',t';\, s',t'),$$

with $(s',\, t') := \left(Q^*_{\eta'\to\eta}(s),\, P^*_{\eta'\to\eta}(t)\right)$. Hence, def. 2.6.3 holds. □

The Heisenberg uncertainty relations that a projective family of covariance matrices need to satisfy if there exists a narrow quantum state with these covariances turn out to also be a *sufficient* condition for the existence of such a state. Indeed, we can construct a Gaussian operator combining the covariance matrix for the configuration variables with the one for the momentum variables, and the uncertainty relations are precisely what is required for this operator to be a *positive semi-definite* operator of unit trace, ie. a density matrix. Moreover, the projective conditions between the covariance matrices ensure that these Gaussian operators assemble into a projective quantum state. Although this is remarkably easy to check at the level of the characteristic functions (where the projections are simply restrictions so that the compatibility of the Gaussian states on various labels can be directly read out from the expression of their characteristic functions), it is instructive to understand how taking the partial trace of Gaussian states works, in particular how the form of factorization $\mathcal{C}_{\eta'} \approx \mathcal{C}_{\eta'\to\eta} \times \mathcal{C}_\eta$ conspires with the expression for the Gaussian states so that the correct projections get respectively applied on the covariance matrices for the positions and momenta (in accordance with the projection $\mathcal{M}_{\eta'} \to \mathcal{M}_\eta$).

Note that this is the point where it is critical to be working on *linear* configuration spaces. While the Wigner transform machinery could be adapted, for example, to the case of a compact group $G$ [1], the nice projection property of Gaussian distributions is what makes the construction of projective coherent states in the linear case much easier than in the $L_2\left(G^N\right)$ case: even in the easiest – $G = \mathcal{U}(1)$ – case, the equivalent of Gaussian states, namely Hall states [8], do not have such a nice behavior under partial trace.

**Proposition 2.9** Let $\left(V_\eta,\, U_\eta\right)_{\eta\in\mathcal{L}} \in \mathcal{V}^{\downarrow}_{(\mathcal{L},\,(\mathcal{C},\mathcal{P}),\,(Q,P))}$. Then, there exists $\rho \in \hat{\mathcal{S}}^{\otimes}_{(\mathcal{L},\mathcal{H},\Phi)}$ such that $V(\rho) = \left(V_\eta,\, U_\eta\right)_{\eta\in\mathcal{L}}$, with $V$ the map introduced in prop. 2.8. In other words, the map $V$ is surjective.

**Proof** Let $\eta \in \mathcal{L}$. Let $(e_i)_{i\in\{1,\dots,n\}}$, $(f_i)_{i\in\{1,\dots,n\}}$ and $\left(\lambda_{(i)}\right)_{i\in\{1,\dots,n\}}$ be as in the proof of prop. 2.7. For any $k_1, \dots, k_n \in \mathbb{N}$, we define $\psi_{k_1,\dots,k_n}$ as:

$$\forall x = x^i\, e_i \in \mathcal{C}_\eta,\ \psi_{k_1,\dots,k_n}(x) := \frac{1}{\sqrt{\alpha_\eta}} \prod_{i=1}^{n} \frac{1}{\pi^{1/4}\, \lambda_{(i)}^{1/8}\, \sqrt{k_i!\, 2^{k_i}}}\, e^{-\frac{(x^i)^2}{2\sqrt{\lambda_{(i)}}}}\, H_{k_i}\left(\frac{x^i}{\lambda_{(i)}^{1/4}}\right),$$

where for any $k \in \mathbb{N}$, $H_k$ is the $k$-th Hermite polynomial:

$$\forall t \in \mathbb{R},\ H_k(t) := (-1)^k\, e^{t^2}\, \frac{d^k}{dt^k} e^{-t^2},$$

and $\alpha_\eta$ is such that $d\mu_\eta(x^i\, e_i) = \alpha_\eta\, dx^1 \dots dx^n$. Then, for any $k_1, \dots, k_n \in \mathbb{N}$, $\psi_{k_1,\dots,k_n} \in \mathcal{H}_\eta = L_2(\mathcal{C}_\eta,\, d\mu_\eta)$ with $\|\psi_{k_1,\dots,k_n}\|_{\mathcal{H}_\eta} = 1$. Next, we define:

$$\rho_\eta := \sum_{k_1,\dots,k_n=0}^{\infty} \left[\prod_{i=1}^{n} \frac{2\sqrt{\lambda_{(i)}}}{2+\sqrt{\lambda_{(i)}}} \left(\frac{2-\sqrt{\lambda_{(i)}}}{2+\sqrt{\lambda_{(i)}}}\right)^{k_i}\right] |\psi_{k_1,\dots,k_n}\rangle \langle \psi_{k_1,\dots,k_n}| .$$

Using that $\forall i \in \{1, \dots, n\}$, $\lambda_{(i)} \in ]0, 4]$ (from the proof of prop. 2.7), $\rho_\eta$ is a non-negative operator on $\mathcal{H}_\eta$ and we have:

$$\|\rho_\eta\|_1 \leqslant \sum_{k_1,\dots,k_n=0}^{\infty} \left|\prod_{i=1}^{n} \frac{2\sqrt{\lambda_{(i)}}}{2+\sqrt{\lambda_{(i)}}} \left(\frac{2-\sqrt{\lambda_{(i)}}}{2+\sqrt{\lambda_{(i)}}}\right)^{k_i}\right| \|\psi_{k_1,\dots,k_n}\|^2_{\mathcal{H}_\eta} = 1 ,$$

where $\|\cdot\|_1$ denotes the trace norm [23, theorem VI.20]. Hence, $\rho$ is a (self-adjoint), positive semi-definite, traceclass operator on $\mathcal{H}_\eta$. Using Mehler formula [19], its kernel is given by:

$$\forall x = x^i\, e_i\, ,\ y = y^i\, e_i \in \mathcal{C}_\eta\, ,$$

$$\begin{aligned}\rho_\eta(x; y) &:= \sum_{k_1,\dots,k_n=0}^{\infty} \left[\prod_{i=1}^{n} \frac{2\sqrt{\lambda_{(i)}}}{2+\sqrt{\lambda_{(i)}}} \left(\frac{2-\sqrt{\lambda_{(i)}}}{2+\sqrt{\lambda_{(i)}}}\right)^{k_i}\right] \psi_{k_1,\dots,k_n}(x)\, \overline{\psi_{k_1,\dots,k_n}(y)} \\ &= \frac{1}{\alpha_\eta\, (2\pi)^{n/2}} \exp\left[-\sum_{i=1}^{n} \frac{1}{2} \left(\frac{x^i + y^i}{2}\right)^2 + \frac{1}{2\,\lambda_{(i)}} (x^i - y^i)^2\right] \\ &= \frac{1}{\alpha_\eta\, (2\pi)^{n/2}} \exp\left[-\frac{1}{2}\, V_\eta^{-1}\left(\frac{x+y}{2}\, ,\, \frac{x+y}{2}\right) - \frac{1}{2}\, U_\eta\left(\Xi^*_\eta(x-y)\, ,\, \Xi^*_\eta(x-y)\right)\right].\end{aligned}$$

Thus, for any $(s, t) \in \mathcal{C}^*_\eta \times \mathcal{P}^*_\eta$, we get, using the expression for $\mathrm{T}_\eta(s, t)$ from prop. 2.3:

$$\begin{aligned}W_{\rho_\eta}(s, t) &:= \mathrm{Tr}_{\mathcal{H}_\eta}\, \rho_\eta\, \mathrm{T}_\eta(s, t) \\ &= \frac{1}{\alpha_\eta\, (2\pi)^{n/2}} \int d\mu_\eta(x)\, \exp\Bigg[i\, s(x) + \frac{i}{2}\, t\left(\Xi^{-1}_\eta(s)\right) + \\ &\qquad -\frac{1}{2}\, V_\eta^{-1}\left(x + \frac{\Xi^{*,-1}_\eta(t)}{2}\, ,\, x + \frac{\Xi^{*,-1}_\eta(t)}{2}\right) - \frac{1}{2}\, U_\eta(t,\, t)\Bigg] \\ &= \exp\left[-\frac{1}{2}\, V_\eta(s,\, s) - \frac{1}{2}\, U_\eta(t,\, t)\right].\end{aligned}$$

From 2.6.3, we have, for any $\eta \preccurlyeq \eta' \in \mathcal{L}$, $W_{\rho_\eta} = W_{\rho_{\eta'}} \circ \left(Q^*_{\eta'\to\eta} \times P^*_{\eta'\to\eta}\right)$, hence, from the proof of prop. 2.4, $\rho_\eta = \mathrm{Tr}_{\mathcal{H}_{\eta'\to\eta}}\left(\Phi_{\eta'\to\eta}\, \rho_{\eta'}\, \Phi^{-1}_{\eta'\to\eta}\right)$. So $\rho := \left(\rho_\eta\right)_{\eta\in\mathcal{L}} \in \overline{\mathcal{S}}^{\otimes}_{(\mathcal{L},\mathcal{H},\Phi)}$, and we have $\left(W_{\rho_\eta}\right)_{\eta\in\mathcal{L}} = W(\rho)$. Moreover, for any $\eta \in \mathcal{L}$, we get the expansion:

$$\forall (s, t) \in \mathcal{C}^*_\eta \times \mathcal{P}^*_\eta,\ W_{\rho_\eta}(\tau s,\, \tau t) = 1 - \frac{\tau^2}{2}\, V_\eta(s,\, s) - \frac{\tau^2}{2}\, U_\eta(t,\, t) + o(\tau^2) .$$

Therefore, $\rho \in \hat{\mathcal{S}}^{\otimes}_{(\mathcal{L},\mathcal{H},\Phi)}$ with:

$$\forall \eta \in \mathcal{L},\ W^{(1)}_\eta(s, t) = 0 \quad \& \quad W^{(2)}_\eta(s, t;\, s, t) = V_\eta(s,\, s) + U_\eta(t,\, t) .$$

In particular, this implies $V(\rho) = \left(V_\eta\, ,\, U_\eta\right)_{\eta\in\mathcal{L}}$.

*Note.* One can also check directly that, for any $\eta \preccurlyeq \eta' \in \mathcal{L}$, $\rho_\eta = \mathrm{Tr}_{\mathcal{H}_{\eta'\to\eta}}\left(\Phi_{\eta'\to\eta}\,\rho_{\eta'}\,\Phi^{-1}_{\eta'\to\eta}\right)$. Indeed, denoting by $\rho^{(\eta')}_\eta$ the integral kernel of $\mathrm{Tr}_{\mathcal{H}_{\eta'\to\eta}}\left(\Phi_{\eta'\to\eta}\,\rho_{\eta'}\,\Phi^{-1}_{\eta'\to\eta}\right)$, we have:

$$\forall x, y \in \mathcal{C}_\eta\,,\; \rho^{(\eta')}_\eta(x;\, y) = \int d\mu_{\eta'\to\eta}(z)\,\rho_{\eta'}\left(\varphi^{-1}_{\eta'\to\eta}(z, x);\; \varphi^{-1}_{\eta'\to\eta}(z, y)\right).$$

Now, for any $x, y \in \mathcal{C}_\eta$ and any $z \in \mathcal{C}_{\eta'\to\eta}$, the definition of $\varphi_{\eta'\to\eta}$ together with $Q_{\eta'\to\eta} \circ \Xi^{*,-1}_{\eta'} \circ P^*_{\eta'\to\eta} = \Xi^{*,-1}_\eta$ (from the proof of prop. 2.3) yields:

$$\varphi^{-1}_{\eta'\to\eta}(z, x) = K_{\eta'\to\eta}(z) + J_{\eta'\to\eta}(x)\,,$$

with $J_{\eta'\to\eta} := \Xi^{*,-1}_{\eta'} \circ P^*_{\eta'\to\eta} \circ \Xi^*_\eta$ and $K_{\eta'\to\eta}$ the canonical injection of $\mathcal{C}_{\eta'\to\eta} = \mathrm{Ker}\, Q_{\eta'\to\eta}$ in $\mathcal{C}_{\eta'}$. So we get:

$$\begin{aligned}
U_{\eta'}&\left(\Xi^*_{\eta'}\left[\varphi^{-1}_{\eta'\to\eta}(z, x) - \varphi^{-1}_{\eta'\to\eta}(z, y)\right]\,,\, \Xi^*_{\eta'}\left[\varphi^{-1}_{\eta'\to\eta}(z, x) - \varphi^{-1}_{\eta'\to\eta}(z, y)\right]\right) = \\
&= U_{\eta'}\left(P^*_{\eta'\to\eta} \circ \Xi^*_\eta(x - y)\,,\, P^*_{\eta'\to\eta} \circ \Xi^*_\eta(x - y)\right) \\
&= U_\eta\left(\Xi^*_\eta(x - y)\,,\, \Xi^*_\eta(x - y)\right),
\end{aligned}$$

as well as:

$$\begin{aligned}
V^{-1}_{\eta'}&\left(\frac{\varphi^{-1}_{\eta'\to\eta}(z, x) + \varphi^{-1}_{\eta'\to\eta}(z, y)}{2}\,,\, \frac{\varphi^{-1}_{\eta'\to\eta}(z, x) + \varphi^{-1}_{\eta'\to\eta}(z, y)}{2}\right) = \\
&= \left[\widetilde{Z}^{(\eta')}_\eta(z)\right](z) + 2\left[\widetilde{X}^{(\eta')}_\eta\left(\frac{x+y}{2}\right)\right](z) + \left[\widetilde{Y}^{(\eta')}_\eta\left(\frac{x+y}{2}\right)\right]\left(\frac{x+y}{2}\right),
\end{aligned}$$

with:

$$\widetilde{X}^{(\eta')}_\eta := K^*_{\eta'\to\eta} \circ \widetilde{V}^{-1}_{\eta'} \circ J_{\eta'\to\eta}\,,\quad \widetilde{Y}^{(\eta')}_\eta := J^*_{\eta'\to\eta} \circ \widetilde{V}^{-1}_{\eta'} \circ J_{\eta'\to\eta} \quad \& \quad \widetilde{Z}^{(\eta')}_\eta := K^*_{\eta'\to\eta} \circ \widetilde{V}^{-1}_{\eta'} \circ K_{\eta'\to\eta}\,,$$

where $\widetilde{V}_{\eta'}$ is defined as in the proof of prop. 2.7. For any $z \in \mathcal{C}_{\eta'\to\eta} \setminus \{0\}$, def. 2.6.1 implies:

$$\left[\widetilde{Z}^{(\eta')}_\eta(z)\right](z) = V_{\eta'}\left(\widetilde{V}^{-1}_{\eta'} \circ K_{\eta'\to\eta}(z),\, \widetilde{V}^{-1}_{\eta'} \circ K_{\eta'\to\eta}(z)\right) > 0,$$

hence $\widetilde{Z}^{(\eta')}_\eta : \mathcal{C}_{\eta'\to\eta} \to \mathcal{C}^*_{\eta'\to\eta}$ is invertible, and the above equation becomes:

$$\begin{aligned}
V^{-1}_{\eta'}&\left(\frac{\varphi^{-1}_{\eta'\to\eta}(z, x) + \varphi^{-1}_{\eta'\to\eta}(z, y)}{2}\,,\, \frac{\varphi^{-1}_{\eta'\to\eta}(z, x) + \varphi^{-1}_{\eta'\to\eta}(z, y)}{2}\right) = \\
&= \left[\widetilde{Z}^{(\eta')}_\eta\left(z + (\widetilde{Z}^{(\eta')}_\eta)^{-1} \circ \widetilde{X}^{(\eta')}_\eta\left(\frac{x+y}{2}\right)\right)\right]\left(z + (\widetilde{Z}^{(\eta')}_\eta)^{-1} \circ \widetilde{X}^{(\eta')}_\eta\left(\frac{x+y}{2}\right)\right) + \\
&\qquad + \left[\left(\widetilde{Y}^{(\eta')}_\eta - (\widetilde{X}^{(\eta')}_\eta)^* \circ (\widetilde{Z}^{(\eta')}_\eta)^{-1} \circ \widetilde{X}^{(\eta')}_\eta\right)\left(\frac{x+y}{2}\right)\right]\left(\frac{x+y}{2}\right).
\end{aligned}$$

On the other hand, using:

$$K_{\eta'\to\eta} \circ R_{\eta'\to\eta} + J_{\eta'\to\eta} \circ Q_{\eta'\to\eta} = \mathrm{id}_{\mathcal{C}_{\eta'}}\,,\quad Q_{\eta'\to\eta} \circ K_{\eta'\to\eta} = 0 \quad \& \quad Q_{\eta'\to\eta} \circ J_{\eta'\to\eta} = \mathrm{id}_{\mathcal{C}_\eta}$$

(as follows from the expressions for $\varphi_{\eta'\to\eta}$ and $\varphi^{-1}_{\eta'\to\eta}$), together with $\widetilde{V}_\eta = Q_{\eta'\to\eta} \circ \widetilde{V}_{\eta'} \circ Q^*_{\eta'\to\eta}$, we can check that:

$$\widetilde{V}_\eta \left( \widetilde{Y}^{(\eta')}_\eta - (\widetilde{X}^{(\eta')}_\eta)^* \circ (\widetilde{Z}^{(\eta')}_\eta)^{-1} \circ \widetilde{X}^{(\eta')}_\eta \right) = \mathrm{id}_{\mathcal{C}_\eta} \tag{2.9.1}$$

(noting that $\widetilde{V}_\eta$ is the lower right block of $\left(R_{\eta'\to\eta}\,,\, Q_{\eta'\to\eta}\right) \circ \widetilde{V}_{\eta'} \circ \left(R^*_{\eta'\to\eta}\,,\, Q^*_{\eta'\to\eta}\right)$, which is the inverse of $\left(K^*_{\eta'\to\eta}\,,\, J^*_{\eta'\to\eta}\right) \circ \widetilde{V}^{-1}_{\eta'} \circ \left(K_{\eta'\to\eta}\,,\, J_{\eta'\to\eta}\right)$, eq. (2.9.1) can be visualized by performing a blockwise matrix inversion of the latter). Putting everything together and carrying out the Gaussian integration over $z$, we get:

$$\forall x, y \in \mathcal{C}_\eta\,,\; \rho^{(\eta')}_\eta(x;\, y) \propto \exp\left[ -\frac{1}{2}\, V^{-1}_\eta \left( \frac{x+y}{2}, \frac{x+y}{2} \right) - \frac{1}{2}\, U_\eta \left( \Xi^*_\eta(x-y), \Xi^*_\eta(x-y) \right) \right],$$

so $\mathrm{Tr}_{\mathcal{H}_{\eta'\to\eta}} \left( \Phi_{\eta'\to\eta}\, \rho_{\eta'}\, \Phi^{-1}_{\eta'\to\eta} \right) \propto \rho_\eta$, the proportionality factor being determined to be 1 from:

$$\mathrm{Tr}_{\mathcal{H}_\eta} \mathrm{Tr}_{\mathcal{H}_{\eta'\to\eta}} \left( \Phi_{\eta'\to\eta}\, \rho_{\eta'}\, \Phi^{-1}_{\eta'\to\eta} \right) = \mathrm{Tr}_{\mathcal{H}_{\eta'}}\, \rho_{\eta'} = 1 = \mathrm{Tr}_{\mathcal{H}_\eta}\, \rho_\eta\,.$$

□

## 2.2 A no-go result in the $G = \mathbb{R}$ case

To prove the advertised no-go result in the case of the holonomy-flux algebra with $G = \mathbb{R}$ (using the projective system set up in [13]), it will be enough to concentrate on a certain (uncountable) subset of observables out of this algebra. So, we will choose an edge, that we will identify with the line segment $[0, 1]$, and a continuous stack of surfaces intersecting this edge, one for each point in $]0, 1]$ (see fig. 2.1). We will keep, for each surface, only its face looking at 0. Moreover, for all holonomies included in the selected edge to have finite variance, it would be sufficient that all holonomies *ending at* 0 would have finite variance, since the holonomy between $e$ and $e'$ can be expressed as a composition of the one between $e$ and 0 and the one between $e'$ and 0 (hence would have finite variance if those two had).

Thus, we will attach, to each point $e$ in $]0, 1]$, the holonomy starting at $e$ (and ending at 0), as well as the flux that acts at the onset of this holonomy. It is manifest from the symplectic structure in prop. 2.10 that these pairs of variables attached to the various points in $]0, 1]$ are *not independent* canonically conjugate pairs: instead, the flux at some point $e$ acts on *all* holonomies that start *at or above* $e$. As announced in section 1, these uncountably many non-zero commutators will play a decisive role in the proof.

Now, suppose that it would be possible to construct a quantum state in which all those holonomies and fluxes would have *finite* variances. Then, the points in $]0, 1]$ could be organized into countably many (overlapping) classes: a point would belong to the class indexed by some $A \in \mathbb{N}$ if the variance of both the holonomy and flux attached to this point would be less than $A$. The assumption of all those variances being finite would ensure that each point belongs at least to one of those classes, so, as there are *uncountably many* points in $]0, 1]$, there should exist some $A$ whose class $\mathcal{K}$ would be uncountable, hence *infinite*.

Finally given $n$ points in this infinite set $\mathcal{K}$, we will, in the proof of lemma 2.12, combine the variances of the holonomies and fluxes attached to these points into a quadratic expression, that should, if all these variances were bounded by a constant $A$, be bounded *independently* of $n$. However, we will show that, by virtue of the Heisenberg uncertainty relations, this quadratic expression can be bounded *below* by a *diverging* expression of $n$.

In the present subsection, $\Sigma$ will denote a finite-dimensional, analytic manifold (without boundary) [17] and $d$ its dimension ($d \geqslant 2$).

**Proposition 2.10** We define the label set $\mathcal{L}^{(\text{aux})}$ as the set of all *finite* subsets of points in $]0, 1]$:

$$\mathcal{L}^{(\text{aux})} := \{\kappa \subset ]0, 1] \mid \#\kappa < \infty\}.$$

We equip $\mathcal{L}^{(\text{aux})}$ with the partial order $\subset$ (set inclusion). For any $\kappa = (e_1, \ldots, e_n) \in \mathcal{L}^{(\text{aux})}$, with $e_1 < \ldots < e_n$, we define:

$$\mathcal{C}^{(\text{aux})}_\kappa := \{h : \kappa \to \mathbb{R}\} \quad \& \quad \mathcal{P}^{(\text{aux})}_\kappa := \{P : \kappa \to \mathbb{R}\},$$

as well as the linear map $\Xi^{(\text{aux})}_\kappa : \mathcal{P}^{(\text{aux})}_\kappa \to \mathcal{C}^{(\text{aux}),*}_\kappa$ given by:

$$\forall P \in \mathcal{P}^{(\text{aux})}_\kappa,\ \forall h \in \mathcal{C}^{(\text{aux})}_\kappa,\ \ \Xi^{(\text{aux})}_\kappa(P)(h) := \sum_{k=1}^{n} P(e_k)\, h(e_k) - \sum_{k=1}^{n-1} P(e_{k+1})\, h(e_k).$$

$\Xi^{(\text{aux})}_\kappa$ is invertible, with $\Xi^{(\text{aux}),-1}_\kappa$ such that:

$$\forall s \in \mathcal{C}^{(\text{aux}),*}_\kappa,\ \forall e \in \kappa,\ \left[\Xi^{(\text{aux}),-1}_\kappa(s)\right](e) = s\left(e' \mapsto \begin{cases} 1 & \text{if } e \leqslant e' \\ 0 & \text{else} \end{cases}\right).$$

We equip $\mathcal{M}^{(\text{aux})}_\kappa := \mathcal{C}^{(\text{aux})}_\kappa \times \mathcal{P}^{(\text{aux})}_\kappa$ with the symplectic structure $\Omega^{(\text{aux})}_\kappa$ defined from $\Xi^{(\text{aux})}_\kappa$ as in prop. 2.1.

For any $\kappa \subset \kappa' \in \mathcal{L}^{(\text{aux})}$, we define:

$$\begin{array}{rcccl} Q^{(\text{aux})}_{\kappa'\to\kappa} & : & \mathcal{C}^{(\text{aux})}_{\kappa'} & \to & \mathcal{C}^{(\text{aux})}_\kappa \\ & & h & \mapsto & h|_\kappa \end{array} \quad \& \quad \begin{array}{rcccl} P^{(\text{aux})}_{\kappa'\to\kappa} & : & \mathcal{P}^{(\text{aux})}_{\kappa'} & \to & \mathcal{P}^{(\text{aux})}_\kappa \\ & & P & \mapsto & P|_\kappa \end{array},$$

and $\pi^{(\text{aux})}_{\kappa'\to\kappa} = Q^{(\text{aux})}_{\kappa'\to\kappa} \times P^{(\text{aux})}_{\kappa'\to\kappa}$. Then, $\left(\mathcal{L}^{(\text{aux})}, \mathcal{M}^{(\text{aux})}, \pi^{(\text{aux})}\right)^\downarrow$ is a projective system of phase spaces and fulfills prop. 2.1.1 and 2.1.2. We denote by $\left(\mathcal{L}^{(\text{aux})}, \mathcal{H}^{(\text{aux})}, \Phi^{(\text{aux})}\right)^\otimes$ the corresponding projective system of quantum state spaces.

**Proof** $\mathcal{L}^{(\text{aux})}$ is a directed set, since any two finite subsets of $]0, 1]$ are included in their union and this union is finite. Let $\kappa = (e_1, \ldots, e_n) \in \mathcal{L}^{(\text{aux})}$ with $e_1 < \ldots < e_n$. $\Xi^{(\text{aux})}_\kappa$ being invertible can be checked from the expression given for $\Xi^{(\text{aux}),-1}_\kappa$ and this ensures that $\Omega^{(\text{aux})}_\kappa$ is indeed a symplectic form (aka. an anti-symmetric, non-degenerate form).

For any $\kappa \subset \kappa' \in \mathcal{L}^{(\text{aux})}$, $\pi^{(\text{aux})}_{\kappa'\to\kappa}$ is a surjective map $\mathcal{M}_{\kappa'} \to \mathcal{M}_\kappa$ and is compatible with the symplectic structures, for we have:

$$\Xi^{(\text{aux}),-1}_\kappa = P^{(\text{aux})}_{\kappa'\to\kappa} \circ \Xi^{(\text{aux}),-1}_{\kappa'} \circ Q^{(\text{aux}),*}_{\kappa'\to\kappa}.$$

Moreover, for any $\kappa \subset \kappa' \subset \kappa'' \in \mathcal{L}^{(\text{aux})}$, $\pi^{(\text{aux})}_{\kappa''\to\kappa} = \pi^{(\text{aux})}_{\kappa'\to\kappa} \circ \pi^{(\text{aux})}_{\kappa''\to\kappa'}$. Therefore, $\left(\mathcal{L}^{(\text{aux})}, \mathcal{M}^{(\text{aux})}, \pi^{(\text{aux})}\right)$ is a projective system of phase spaces. Prop. 2.1.1 and 2.1.2 are fulfilled by construction with

$\forall \kappa \in \mathcal{L}^{(\mathrm{aux})},\ L_\kappa = \mathrm{id}_{\mathcal{M}_\kappa}$. □

**Theorem 2.11** With the notations of def. 2.6 and prop. 2.10, $\mathcal{V}^{\downarrow}_{(\mathrm{aux})} := \mathcal{V}^{\downarrow}_{\left(\mathcal{L}^{(\mathrm{aux})},\, (\mathcal{C}^{(\mathrm{aux})}, \mathcal{P}^{(\mathrm{aux})}),\, (Q^{(\mathrm{aux})}, P^{(\mathrm{aux})})\right)} = \varnothing$. Hence, prop. 2.8 implies that $\hat{\mathcal{S}}^{\otimes}_{(\mathrm{aux})} := \hat{\mathcal{S}}^{\otimes}_{(\mathcal{L}^{(\mathrm{aux})}, \mathcal{H}^{(\mathrm{aux})}, \Phi^{(\mathrm{aux})})} = \varnothing$.

**Lemma 2.12** Let $\mathcal{K}$ be a countably *infinite* subset of $]0, 1]$ and let $A > 0$. We define:

$$\mathcal{V}^{\downarrow}_{(\mathcal{K},A)} := \left\{ (V_\kappa,\, U_\kappa)_\kappa \in \mathcal{V}^{\downarrow}_{(\mathrm{aux})} \,\middle|\, \forall \kappa \subset \mathcal{K},\, \forall e \in \kappa,\ V_\kappa(\mathrm{h}^{[e]}_\kappa,\, \mathrm{h}^{[e]}_\kappa) \leqslant A \quad \& \quad U_\kappa(\mathrm{P}^{[e]}_\kappa,\, \mathrm{P}^{[e]}_\kappa) \leqslant A \right\},$$

with $\forall \kappa \in \mathcal{L}^{(\mathrm{aux})}, \forall e \in \kappa,\ \ \mathrm{h}^{[e]}_\kappa := \big(h \mapsto h(e)\big) \quad \& \quad \mathrm{P}^{[e]}_\kappa := \big(P \mapsto P(e)\big)$. Then, $\mathcal{V}^{\downarrow}_{(\mathcal{K},A)} = \varnothing$.

**Proof** Proceeding by contradiction, we suppose that there exists $(V_\kappa,\, U_\kappa)_\kappa \in \mathcal{V}^{\downarrow}_{(\mathcal{K},A)}$. Let $n \in \mathbb{N}$. Since $\mathcal{K}$ is infinite, there exist $\kappa = (e_1,\, \dots,\, e_n) \subset \mathcal{K}$ with $e_1 < \dots < e_n$. Then, we have:

$$\frac{1}{n} \sum_{k=1}^{n} V_\kappa \left(\mathrm{h}^{[e_k]}_\kappa,\, \mathrm{h}^{[e_k]}_\kappa\right) + U_\kappa \left(\mathrm{P}^{[e_k]}_\kappa,\, \mathrm{P}^{[e_k]}_\kappa\right) \leqslant 2A.$$

We define the $n$ by $n$ matrix $M^{(n)}$ by:

$$\forall k, l \in \{1, \dots, n\},\ M^{(n)}_{kl} := \mathrm{P}^{[e_k]}_\kappa \left(\Xi^{(\mathrm{aux}),-1}_\kappa\big(\mathrm{h}^{[e_l]}_\kappa\big)\right) = \begin{cases} 1 & \text{if } k \leqslant l \\ 0 & \text{else} \end{cases}.$$

Performing a singular value decomposition of $M^{(n)}$ there exist two $n$ by $n$ orthogonal matrices $O^{(n)}$ and $O'^{(n)}$, and a diagonal matrix $\Delta^{(n)}$ with non-negative real entries, such that:

$$M^{(n)} = O'^{(n)\top}\, \Delta^{(n)}\, O^{(n)},$$

where $(\cdot)^\top$ denotes the transpose matrix. Defining, for any $k \in \{1, \dots, n\}$, $s_k \in \mathcal{C}^{(\mathrm{aux}),*}_\kappa$ and $t_k \in \mathcal{P}^{(\mathrm{aux}),*}_\kappa$ by:

$$s_k := \sum_{l=1}^{n} O^{(n)}_{kl}\, \mathrm{h}^{[e_l]}_\kappa \quad \& \quad t_k := \sum_{l=1}^{n} O'^{(n)}_{kl}\, \mathrm{P}^{[e_l]}_\kappa,$$

we get:

$$\frac{1}{n} \sum_{k=1}^{n} V_\kappa\,(s_k\,,\, s_k) + U_\kappa\,(t_k\,,\, t_k) = \frac{1}{n} \sum_{l=1}^{n} V_\kappa \left(\mathrm{h}^{[e_l]}_\kappa,\, \mathrm{h}^{[e_l]}_\kappa\right) + U_\kappa \left(\mathrm{P}^{[e_l]}_\kappa,\, \mathrm{P}^{[e_l]}_\kappa\right) \leqslant 2A.$$

On the other hand, from def. 2.6.2, we have:

$$\frac{1}{n} \sum_{k=1}^{n} V_\kappa\,(s_k\,,\, s_k) + U_\kappa\,(t_k\,,\, t_k) \geqslant \frac{1}{n} \sum_{k=1}^{n} t_k \left(\Xi^{(\mathrm{aux}),-1}_\kappa(s_k)\right)$$

$$= \frac{1}{n} \mathrm{Tr}\left[O'^{(n)}\, M^{(n)}\, O^{(n)\top}\right] = \frac{1}{n} \mathrm{Tr}\, \Delta^{(n)} = \frac{1}{n} \mathrm{Tr} \sqrt{M^{(n)\top}\, M^{(n)}}.$$

Now, $N^{(n)} := \dfrac{1}{n^2} M^{(n)\top} M^{(n)}$ is given by:

$$\forall k, l \in \{1, \dots, n\},\ N^{(n)}_{kl} = \frac{1}{n^2} \sum_{m=1}^{n} M^{(n)}_{mk} M^{(n)}_{ml} = \frac{1}{n^2} \sum_{m=1}^{\min(k,l)} 1 = \frac{1}{n} \min\left(\frac{k}{n},\, \frac{l}{n}\right),$$

and the previous inequalities requires $\mathrm{Tr}\sqrt{N^{(n)}} \leqslant 2A$.

On the complex Hilbert space $\mathfrak{J} := L_2\left([0,1], d\mu_{[0,1]}\right)$ (with $d\mu_{[0,1]}$ the usual, normalized measure on $[0,1]$), we define, for any $n \in \mathbb{N}$, a bounded operator $N^{[n]}$ by:

$$\forall \zeta \in \mathfrak{J},\ \forall x \in [0,1],\ \left[N^{[n]}\zeta\right](x) := \sum_{l=1}^{n} \min\left([x]_n, \frac{l}{n}\right) \int_{\frac{l-1}{n}}^{\frac{l}{n}} dy\ \zeta(y)$$
$$= \int_0^1 dy\ \min\left([x]_n, [y]_n\right)\ \zeta(y),$$

with $[x]_n := \frac{1}{n}\lceil n\,x \rceil$ and $\lceil \cdot \rceil$ the ceiling function. For any $m \in \{1, \ldots, n\}$, we define $u_m^{[n]} \in \mathfrak{J}$ by:

$$\forall x \in [0,1],\ u_m^{[n]}(x) := \sqrt{n}\, O^{(n)}_{m\,\lceil nx \rceil}.$$

We have $\left\langle u_m^{[n]} \,\middle|\, u_{m'}^{[n]} \right\rangle_{\mathfrak{J}} = \delta_{mm'}$ and, using $N^{(n)} = O^{(n)\top} \left(\frac{\Delta^{(n)}}{n}\right)^2 O^{(n)}$:

$$N^{[n]} = \sum_{m=1}^{n} \left(\frac{\Delta^{(n)}_{mm}}{n}\right)^2 \left| u_m^{[n]} \right\rangle \left\langle u_m^{[n]} \right|.$$

Now, we define a bounded operator $N^{[\infty]}$ on $\mathfrak{J}$ by:

$$\forall \zeta \in \mathfrak{J},\ \forall x \in [0,1],\ \left[N^{[\infty]}\zeta\right](x) = \int_0^1 dy\ \min(x, y)\ \zeta(y)$$
$$= \int_0^x dy\ y\,\zeta(y) + x \int_x^1 dy\ \zeta(y),$$

and, for any $m \in \mathbb{N} \setminus \{0\}$, we define $u_m^{[\infty]} \in \mathfrak{J}$ by:

$$\forall x \in [0,1],\ u_m^{[\infty]}(x) := \sqrt{2}\,\sin\left[\left(m - \frac{1}{2}\right)\pi\,x\right].$$

We have $\left\langle u_m^{[\infty]} \,\middle|\, u_{m'}^{[\infty]} \right\rangle_{\mathfrak{J}} = \delta_{mm'}$ and:

$$N^{[\infty]} u_m^{[\infty]} = \frac{4}{\pi^2\,(2m-1)^2}\, u_m^{[\infty]}.$$

Since $\sum_{m=1}^{\infty} \frac{2}{\pi\,(2m-1)} = \infty$, there exists $M \in \mathbb{N} \setminus \{0\}$ such that:

$$\sum_{m=1}^{M} \frac{2}{\pi\,(2m-1)} > 2A + 1.$$

Let $n \in \mathbb{N}$ such that $n > 8\,\pi^2 M^3$ and complete $\left(u_m^{[n]}\right)_{m\in\{1,\ldots,n\}}$ into an orthonormal basis $\left(u_m^{[n]}\right)_{m\in\mathbb{N}\setminus\{0\}}$ of $\mathfrak{J}$. For any $m \in \{1, \ldots, M\}$, we have:

$$\left\| N^{[\infty]}\, u_m^{[\infty]} - N^{[n]}\, u_m^{[\infty]} \right\| \leqslant \left\| N^{[\infty]} - N^{[n]} \right\| \left\| u_m^{[\infty]} \right\| \leqslant \frac{1}{n}.$$

But we also have:

$$\left\| N^{[\infty]} u_m^{[\infty]} - N^{[n]} u_m^{[\infty]} \right\|^2 = \frac{1}{n^4} \sum_{m'=1}^{\infty} \left| \left(\Delta_{mm}^{(\infty)}\right)^2 - \left(\Delta_{m'm'}^{(n)}\right)^2 \right|^2 \left| \left\langle u_{m'}^{[n]} \,\middle|\, u_m^{[\infty]} \right\rangle_{\mathfrak{J}} \right|^2$$

where $\Delta_{mm}^{(\infty)} := \dfrac{2\,n}{\pi\,(2m-1)}$ and, for any $m' > n$, $\Delta_{m'm'}^{(n)} := 0$. Thus, we get:

$$\frac{1}{n^4} \left[ \inf_{m' \in \mathbb{N} \setminus \{0\}} \left| \left(\Delta_{mm}^{(\infty)}\right)^2 - \left(\Delta_{m'm'}^{(n)}\right)^2 \right| \right]^2 \sum_{m'=1}^{\infty} \left| \left\langle u_{m'}^{[n]} \,\middle|\, u_m^{[\infty]} \right\rangle_{\mathfrak{J}} \right|^2 \leqslant \frac{1}{n^2}\,,$$

and therefore:

$$\inf_{m' \in \mathbb{N} \setminus \{0\}} \left| \left(\Delta_{mm}^{(\infty)}\right)^2 - \left(\Delta_{m'm'}^{(n)}\right)^2 \right| \leqslant n\,.$$

Hence, for any $m \in \{1, \ldots, M\}$, there exists $\widetilde{m} \in \mathbb{N} \setminus \{0\}$, such that:

$$\left| \frac{4}{\pi^2\,(2m-1)^2} - \left( \frac{\Delta_{\widetilde{m}\widetilde{m}}^{(n)}}{n} \right)^2 \right| \leqslant \frac{2}{n}\,.$$

Using $\Delta_{\widetilde{m}\widetilde{m}}^{(n)} \geqslant 0$ and $n > 8\,\pi^2\,M^3$, this implies:

$$\left| \frac{2}{\pi\,(2m-1)} - \frac{\Delta_{\widetilde{m}\widetilde{m}}^{(n)}}{n} \right| = \left| \frac{4}{\pi^2\,(2m-1)^2} - \left( \frac{\Delta_{\widetilde{m}\widetilde{m}}^{(n)}}{n} \right)^2 \right| \left| \frac{2}{\pi\,(2m-1)} + \frac{\Delta_{\widetilde{m}\widetilde{m}}^{(n)}}{n} \right|^{-1} \leqslant \frac{1}{4\,\pi\,M^2}\,.$$

On the other hand, for any $m' > n$, we have:

$$\left| \frac{2}{\pi\,(2m-1)} - \frac{\Delta_{m'm'}^{(n)}}{n} \right| \geqslant \frac{1}{\pi\,M} \geqslant \frac{1}{\pi\,M^2}\,,$$

so $\widetilde{m} \leqslant n$. Next, for any $m \neq m' \in \{1, \ldots, M\}$, the inequality:

$$\left| \frac{2}{\pi\,(2m-1)} - \frac{2}{\pi\,(2m'-1)} \right| \geqslant \frac{1}{\pi\,M^2}$$

holds, so $\widetilde{m} \neq \widetilde{m}'$. Therefore, we obtain:

$$\sum_{\widetilde{m}=1}^{n} \frac{\Delta_{\widetilde{m}\widetilde{m}}^{(n)}}{n} \geqslant \sum_{m=1}^{M} \frac{\Delta_{\widetilde{m}\widetilde{m}}^{(n)}}{n} \geqslant \sum_{m=1}^{M} \frac{2}{\pi\,(2m-1)} - \frac{1}{4\,\pi\,M} > 2A\,,$$

in contradiction with $\dfrac{1}{n} \operatorname{Tr} \Delta^{(n)} \leqslant 2A$, whence the initial assumption on the existence of $(V_\kappa,\, U_\kappa)_\kappa \in \mathcal{V}^{\downarrow}_{(\mathcal{K},A)}$ is proven wrong. □

**Proof of theorem 2.11** Again, we proceed by contradiction and we suppose that there exists $(V_\kappa\,,\, U_\kappa)_\kappa \in \mathcal{V}^{\downarrow}_{(\text{aux})}$. For any $e \in \,]0, 1]$, we define $V^{[e]} := V_\kappa\left(\mathsf{h}_\kappa^{[e]},\, \mathsf{h}_\kappa^{[e]}\right)$ and $U^{[e]} := U_\kappa\left(\mathsf{P}_\kappa^{[e]},\, \mathsf{P}_\kappa^{[e]}\right)$ for some $\kappa \in \mathcal{L}^{(\text{aux})}$ such that $e \in \kappa$. If $\kappa' \in \mathcal{L}^{(\text{aux})}$ is such that $e \in \kappa'$, then $\kappa, \kappa' \subset \kappa \cup \kappa' \in \mathcal{L}^{(\text{aux})}$ and:

$$V_{\kappa'}\left(\mathrm{h}^{[e]}_{\kappa'}, \mathrm{h}^{[e]}_{\kappa'}\right) = V_{\kappa\cup\kappa'}\left(Q^{(\mathrm{aux}),*}_{\kappa\cup\kappa'\to\kappa'}\left(\mathrm{h}^{[e]}_{\kappa'}\right), Q^{(\mathrm{aux}),*}_{\kappa\cup\kappa'\to\kappa'}\left(\mathrm{h}^{[e]}_{\kappa'}\right)\right) = V_{\kappa\cup\kappa'}\left(\mathrm{h}^{[e]}_{\kappa\cup\kappa'}, \mathrm{h}^{[e]}_{\kappa\cup\kappa'}\right)$$
$$= V_{\kappa}\left(\mathrm{h}^{[e]}_{\kappa}, \mathrm{h}^{[e]}_{\kappa}\right) = V^{[e]}.$$

Similarly, $U_{\kappa'}\left(\mathrm{P}^{[e]}_{\kappa'}, \mathrm{P}^{[e]}_{\kappa'}\right) = U^{[e]}$. Now, we have:

$$]0, 1] = \bigcup_{A=1}^{\infty} \left\{e \in ]0, 1] \,\middle|\, V^{[e]} \leqslant A \quad \& \quad U^{[e]} \leqslant A\right\},$$

and since $]0, 1]$ is *uncountably* infinite, there exists $A \in \mathbb{N} \setminus \{0\}$ such that:

$$\left\{e \in ]0, 1] \,\middle|\, V^{[e]} \leqslant A \quad \& \quad U^{[e]} \leqslant A\right\}$$

is infinite, hence contains some countably infinite subset $\mathcal{K}$. Then, we have $(V_\kappa, U_\kappa)_\kappa \in \mathcal{V}^{\downarrow}_{(\mathcal{K},A)}$, but this is impossible since $\mathcal{V}^{\downarrow}_{(\mathcal{K},A)} = \varnothing$ from the previous lemma. □

**Proposition 2.13** Let $\mathcal{L}_{\mathrm{HF}}$ be the directed label set defined in [13, def. 2.14]. For any $\eta \in \mathcal{L}_{\mathrm{HF}}$, let $\mathcal{C}^{\mathbb{R}}_{\eta} := \{h : \gamma(\eta) \to \mathbb{R}\}$, $\mathcal{P}^{\mathbb{R}}_{\eta} := \{P : \mathcal{F}(\eta) \to \mathbb{R}\}$ (with $\gamma(\eta)$ and $\mathcal{F}(\eta)$ as in [13, def. 2.14]), $\mathcal{M}^{\mathbb{R}}_{\eta} := T^*(\mathcal{C}^{\mathbb{R}}_{\eta})$ and, for any $\eta \preccurlyeq \eta' \in \mathcal{L}_{\mathrm{HF}}$, let $\pi^{\mathbb{R}}_{\eta'\to\eta} : \mathcal{M}^{\mathbb{R}}_{\eta'} \to \mathcal{M}^{\mathbb{R}}_{\eta}$ be defined as in [13, prop. 3.8] in the special case $G = (\mathbb{R}, +)$. Then, $\left(\mathcal{L}_{\mathrm{HF}}, \mathcal{M}^{\mathbb{R}}, \pi^{\mathbb{R}}_{\eta'\to\eta}\right)^{\downarrow}$ is a projective system of phase spaces fulfilling the hypotheses of prop. 2.1, with:

1. $\forall \eta \in \mathcal{L}_{\mathrm{HF}}, \forall P \in \mathcal{P}^{\mathbb{R}}_{\eta}, \Xi^{\mathbb{R}}_{\eta}(P) : h \mapsto \sum_{e\in\gamma(\eta)} P \circ \chi_\eta(e)\ h(e)$ (with $\chi_\eta$ as in [13, def. 2.14]);
2. $\forall \eta \in \mathcal{L}_{\mathrm{HF}}, \forall (h, p) \in \mathcal{M}^{\mathbb{R}}_{\eta}, L^{\mathbb{R}}_{\eta}(h, p) := \left(h, F \mapsto p\left(\delta_{\chi_\eta^{-1}(F)}\right)\right)$ (with, for any $F \in \mathcal{F}(\eta)$ and any $e \in \gamma(\eta)$, $\delta_{\chi_\eta^{-1}(F)}(e) = 1$ if $\chi_\eta(e) = F$, and 0 otherwise);
3. $\forall \eta \preccurlyeq \eta' \in \mathcal{L}_{\mathrm{HF}}, \forall h_{\eta'} \in \mathcal{C}^{\mathbb{R}}_{\eta'}, Q^{\mathbb{R}}_{\eta'\to\eta}(h_{\eta'}) : e \mapsto \sum_{k=1}^{n_{\eta'\to\eta,e}} \epsilon_{\eta'\to\eta,e}(k)\ h_{\eta'} \circ a_{\eta'\to\eta,e}(k)$ (with the notations of [13, prop. 3.5]);
4. $\forall \eta \preccurlyeq \eta' \in \mathcal{L}_{\mathrm{HF}}, \forall P_{\eta'} \in \mathcal{P}^{\mathbb{R}}_{\eta'}, P^{\mathbb{R}}_{\eta'\to\eta}(P_{\eta'}) : F \mapsto \sum_{F'\in H^{(1,3)}_{\eta'\to\eta,F}} P_{\eta'}(F')$ (with $H^{(1,3)}_{\eta'\to\eta,F}$ also from [13, prop. 3.5]).

The projective system of quantum state spaces $\left(\mathcal{L}_{\mathrm{HF}}, \mathcal{H}^{\mathbb{R}}, \Phi^{\mathbb{R}}\right)^{\otimes}$ provided by [13, prop. 3.10] (in the special case $G = (\mathbb{R}, +)$) can be identified with the one provided by prop. 2.1. Moreover, for any $\eta \in \mathcal{L}_{\mathrm{HF}}$ and any $(s, t) \in \mathcal{C}^{\mathbb{R},*}_{\eta} \times \mathcal{P}^{\mathbb{R},*}_{\eta}$,

$$\mathrm{X}^{\mathbb{R}}_{\eta}(s, t) = \sum_{e\in\gamma(\eta)} s(\delta_e)\ \widehat{\mathrm{h}^{(e,\mathrm{id}_{\mathbb{R}})}_{\eta}} - \sum_{F\in\mathcal{F}(\eta)} t(\delta_F)\ \widehat{\mathrm{P}^{(\overline{F},1)}_{\eta}},$$

using the densely defined, essentially self-adjoint operators introduced in [13, eqs. (3.11.1) and (3.11.2)] (the minus sign is the result of conflicting conventions in [15, prop. A.14] and prop. A.5).

**Proof** *Assertions 2.1.1 and 2.1.2.* Let $\eta \in \mathcal{L}_{\mathrm{HF}}$. $\Xi^{\mathbb{R}}_{\eta}$ is an invertible linear map $\mathcal{P}_\eta \to \mathcal{C}^*_\eta$ with:

$$\forall p \in \mathcal{C}^*_\eta,\ \Xi^{\mathbb{R},-1}_\eta(p) : F \mapsto p\left(\delta_{\chi_\eta^{-1}(F)}\right).$$

Next, $L^{\mathbb{R}}_\eta$ is an invertible linear map $\mathcal{M}^{\mathbb{R}}_\eta \to \mathcal{C}^{\mathbb{R}}_\eta \times \mathcal{P}^{\mathbb{R}}_\eta$, with:

$$\forall (h,\, P) \in \mathcal{C}^{\mathbb{R}}_\eta \times \mathcal{P}^{\mathbb{R}}_\eta,\ L^{\mathbb{R},-1}_\eta(h,\, P) = \Big(h,\, h' \mapsto \sum_{e \in \gamma(\eta)} P \circ \chi_\eta(e)\ h'(e)\Big).$$

Let $\Omega^{\mathbb{R}}_\eta$ be the symplectic structure on $\mathcal{C}^{\mathbb{R}}_\eta \times \mathcal{P}^{\mathbb{R}}_\eta$ defined by:

$$\forall h, h' \in \mathcal{C}_\eta\,,\ \forall P, P' \in \mathcal{P}_\eta\,,\ \ \Omega^{\mathbb{R}}_\eta(h, P;\, h', P') := \Xi^{\mathbb{R}}_\eta(P')(h) - \Xi^{\mathbb{R}}_\eta(P)(h').$$

We have:

$$\forall (h, p),\, (h', p') \in \mathcal{M}^{\mathbb{R}}_\eta,\ \big(L^{\mathbb{R},*}_\eta \Omega^{\mathbb{R}}_\eta\big)(h, p;\, h', p') = p'(h) - p(h').$$

Hence, $L^{\mathbb{R}}_\eta$ is a symplectomorphism (with respect to the canonical symplectic structure on $\mathcal{M}^{\mathbb{R}}_\eta = T^*(\mathcal{C}^{\mathbb{R}}_\eta)$, see [14, eq. (2.16.1)]). Moreover, it coincides with the map defined in [13, prop. 3.8] in the case $G = (\mathbb{R},\, +)$.

Let $\eta \preccurlyeq \eta' \in \mathcal{L}_{\text{HF}}$. Using [13, prop. 3.8] we have, for any $h_{\eta'}\,,\ P_{\eta'} \in \mathcal{C}^{\mathbb{R}}_{\eta'} \times \mathcal{P}^{\mathbb{R}}_{\eta'}$:

$$L^{\mathbb{R}}_\eta \circ \pi^{\mathbb{R}}_{\eta' \to \eta} \circ L^{\mathbb{R},-1}_{\eta'}(h_{\eta'}\,,\ P_{\eta'}) = \left(Q^{\mathbb{R}}_{\eta' \to \eta}(h_{\eta'}),\ P^{\mathbb{R}}_{\eta' \to \eta}(P_{\eta'})\right).$$

*Projective of quantum state spaces.* For any $\eta \preccurlyeq \eta' \in \mathcal{L}_{\text{HF}}$, let $\mathcal{C}^{\mathbb{R}}_{\eta' \to \eta}$ and $\varphi^{\mathbb{R}}_{\eta' \to \eta}$, resp. $\mathcal{C}_{\eta' \to \eta}$ and $\varphi_{\eta' \to \eta}$, be defined as in [13, prop. 3.6] (with $G = (\mathbb{R}, +)$), resp. as in prop. 2.1. It follows from the uniqueness part of [14, prop. 2.10] and from the connectedness of $\mathcal{M}_{\eta'}$ that the tangential lifts of the maps $\varphi^{\mathbb{R}}_{\eta' \to \eta}$ and $\varphi_{\eta' \to \eta}$ must coincide modulo a suitable symplectomorphic identification of the cotangent bundles $T^*(\mathcal{C}^{\mathbb{R}}_{\eta' \to \eta})$ and $T^*(\mathcal{C}_{\eta' \to \eta})$. Hence, the maps $\varphi^{\mathbb{R}}_{\eta' \to \eta}$ and $\varphi_{\eta' \to \eta}$ themselves coincide modulo a diffeomorphic identification of the manifolds $\mathcal{C}^{\mathbb{R}}_{\eta' \to \eta}$ and $\mathcal{C}_{\eta' \to \eta}$. Therefore, the associated projective systems of quantum state spaces coincide modulo a unitary identification of the Hilbert spaces $\mathcal{H}^{\mathbb{R}}_{\eta' \to \eta} := L_2(\mathcal{C}^{\mathbb{R}}_{\eta' \to \eta})$ and $\mathcal{H}_{\eta' \to \eta} := L_2(\mathcal{C}_{\eta' \to \eta})$.

*Observables.* Let $(s, t) \in \mathcal{C}^{\mathbb{R},*}_\eta \times \mathcal{P}^{\mathbb{R},*}_\eta$ and $\psi \in \mathcal{D}^{\mathbb{R}}_\eta := C^\infty_o(\mathcal{C}^{\mathbb{R}}_\eta,\, \mathbb{C})$. For any $h \in \mathcal{C}^{\mathbb{R}}_\eta$, we have:

$$\begin{aligned}\left[\mathrm{X}^{\mathbb{R}}_\eta(s, t)\, \psi\right](h) &:= s(h)\, \psi(h) - i\, [T_h \psi]\left(\Xi^{\mathbb{R},*,-1}_\eta(t)\right) \\ &= \sum_{e \in \gamma(\eta)} s(\delta_e)\, h(e)\, \psi(h) - \sum_{F \in \mathcal{F}(\eta)} i\, t(\delta_F)\, [T_h \psi](\delta_{\chi_\eta^{-1}(F)}),\end{aligned}$$

where we have used that, for any $F \in \mathcal{F}(\eta)$, $\Xi^{\mathbb{R},*,-1}_\eta(t)(\chi_\eta^{-1}(F)) = t(\delta_F)$. Now, specializing the definitions from [13, prop. 3.11] to the case $G = (\mathbb{R}, +)$, this can be rewritten as:

$$\left[\mathrm{X}^{\mathbb{R}}_\eta(s, t)\, \psi\right](h) = \sum_{e \in \gamma(\eta)} s(\delta_e) \left[\widehat{\mathrm{h}^{(e, \mathrm{id}_{\mathbb{R}})}_\eta}\, \psi\right](h) - \sum_{F \in \mathcal{F}(\eta)} t(\delta_F) \left[\widehat{\mathrm{P}^{(\overline{F}, 1)}_\eta}\, \psi\right](h).$$

□

**Proposition 2.14** With the notations of def. 2.5 and prop. 2.13, $\hat{\mathcal{S}}^{\otimes}_{(\mathcal{L}_{\text{HF}}, \mathcal{H}^{\mathbb{R}}, \Phi^{\mathbb{R}})} = \varnothing$.

**Proof** *Directed pre-order on $\mathcal{L}_{\text{HF}} \sqcup \mathcal{L}^{(\text{aux})}$.* Let $\Psi : U \to V$ be an analytical coordinate patch on $\Sigma$, with $U$ an open neighborhood of 0 in $\mathbb{R}^d$ (recall that $d := \dim(\Sigma) \geqslant 1$). Let $\epsilon$ such that $B_\epsilon^{(d)} \subset U$. For any $\tau \neq \tau' \in [0,2]$ we define:

$$\begin{array}{rcl} \breve{e}_{\tau,\tau'} \;:\; U_{\tau,\tau'} \subset \mathbb{R}^d \approx \mathbb{R} \times \mathbb{R}^{d-1} & \to & V \\ (x,y) & \mapsto & \Psi\left(\frac{\epsilon}{2}\big(\tau + x\,(\tau'-\tau)\big), \frac{\epsilon}{2}\, y\right) \end{array},$$

where $U_{\tau,\tau'} = \left\{(x,y) \in \mathbb{R}^d \approx \mathbb{R} \times \mathbb{R}^{d-1} \,\middle|\, \big(\tau + x\,(\tau'-\tau),\, y\big) \in \frac{2}{\epsilon}\, U\right\}$ is an open neighborhood of $[0,1] \times \{0\}^{d-1}$ in $\mathbb{R}^d$. We denote by $e_{\tau,\tau'}$ be the corresponding edge. Next, for any $\tau \in [0,1]$ we define $\breve{S}_\tau \equiv \breve{e}_{\tau,2}$ and we note that $U_{\tau,2}$ is an open neighborhood of $\{0\} \times B^{(d-1)}$ in $\mathbb{R}^d$. We denote by $S_\tau$ the corresponding surface.

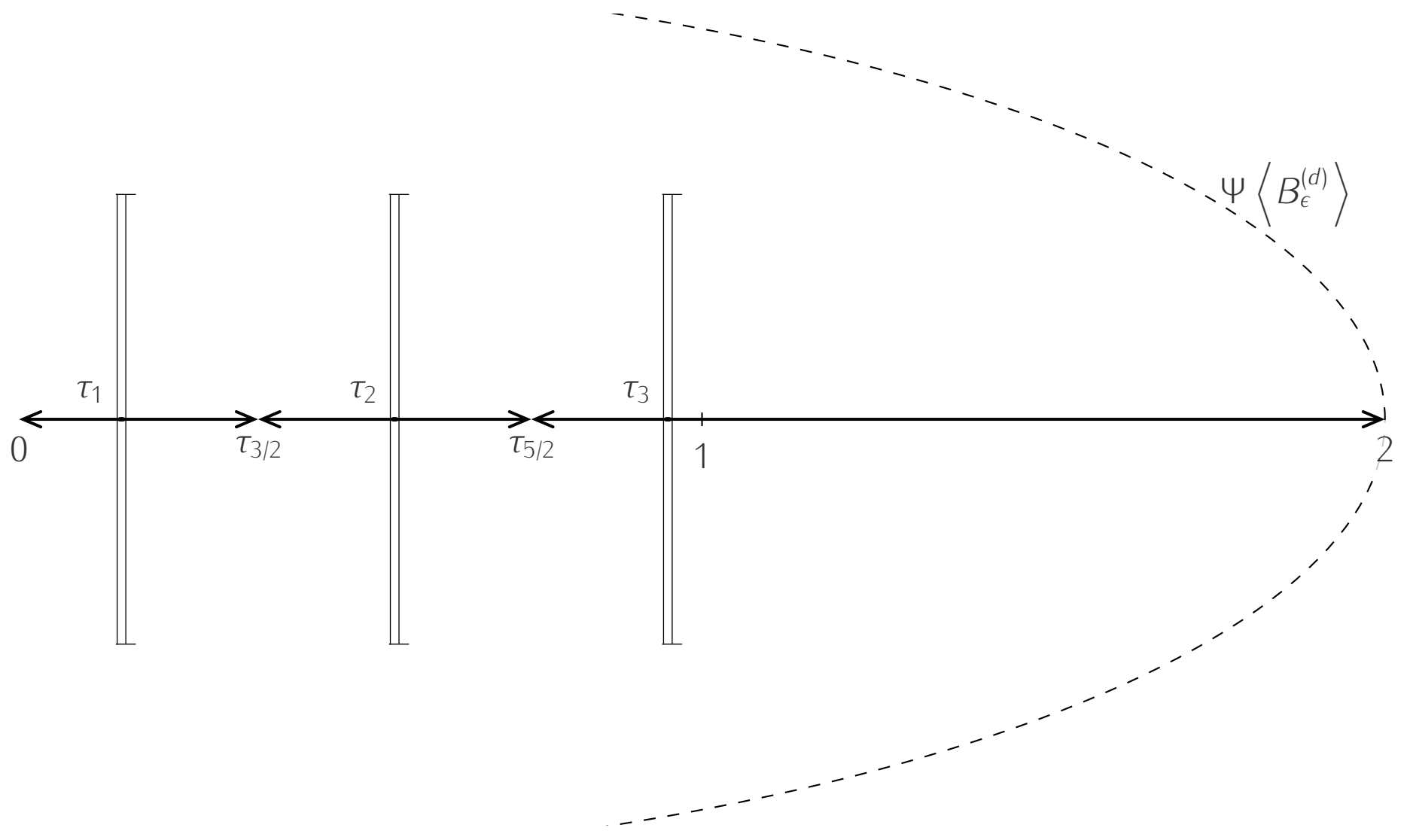

Figure 2.1 – Constructing $\eta_\kappa$

Let $\kappa = (\tau_1, \ldots, \tau_n) \in \mathcal{L}^{(\text{aux})}$, with $0 < \tau_1 < \ldots < \tau_n \leqslant 1$. We define (fig. 2.1):

$$\gamma_\kappa := \left\{e_{\tau_i\,,\,\tau_{i-1/2}} \,\middle|\, i \in \{1,\ldots,n\}\right\} \cup \left\{e_{\tau_i\,,\,\tau_{i+1/2}} \,\middle|\, i \in \{1,\ldots,n\}\right\} \in \mathcal{L}_{\text{graphs}}\,,$$

where $\tau_{1/2} := 0$, $\tau_{n+1/2} := 2$ and, for any $i \in \{1,\ldots,n-1\}$, $\tau_{i+1/2} := \frac{1}{2}(\tau_i + \tau_{i+1})$. We also define:

$$\lambda_\kappa := \Big[\,\{S_\tau \mid \tau \in \kappa\}\,\Big]_\sim \in \mathcal{L}_{\text{profls}}\,,$$

where $\sim$ denotes the equivalence relation from [13, def. 2.12]. Since, for any $\tau \neq \tau'$, $r(S_\tau) \cap r(S_{\tau'}) = \varnothing$, we have:

$$\mathcal{F}(\lambda_\kappa) = \left\{F_\diamond^\tau(\kappa) \,\middle|\, \tau \in \kappa,\, \diamond \in \{\uparrow, \downarrow\}\right\},$$

where, for any $\tau \in \kappa$ and any $\diamond \in \{\uparrow, \downarrow\}$:

$$F_\diamond^\tau(\kappa) := \{e \in \mathcal{L}_{\text{edges}} \mid e \diamond S_\tau \;\;\&\;\; \forall \tau' \in \kappa \setminus \{\tau\}\,,\; e \curvearrowright S_{\tau'}\}\,.$$

Thus, $\eta_\kappa := (\gamma_\kappa\,,\, \lambda_\kappa) \in \mathcal{L}_{\text{HF}}$ with:

$$\forall i \in \{1,\ldots,n\}\;,\; \chi_{\eta_\kappa}(e_{\tau_i\,,\,\tau_{i-1/2}}) = F_\downarrow^{\tau_i}(\kappa) \quad \& \quad \chi_{\eta_\kappa}(e_{\tau_i\,,\,\tau_{i+1/2}}) = F_\uparrow^{\tau_i}(\kappa)\,.$$

Moreover, for any $i \in \{1,\ldots,n\}$, we have:

$$e_{\tau_i\,,0} = e_{\tau_1\,,\,\tau_{1/2}} \circ e^{-1}_{\tau_1\,,\,\tau_{3/2}} \circ \ldots \circ e_{\tau_i\,,\,\tau_{i-1/2}}\,,$$

and $\overline{F^{\tau_i}_{\downarrow}(\kappa)} = \overline{F^{\tau_i}} \in \mathcal{L}_{\text{faces}}$, with $F^{\tau_i} := \left\{ e \in \mathcal{L}_{\text{edges}} \,\middle|\, e \downarrow S_{\tau_i} \right\}$,

where, for $F \subset \mathcal{L}_{\text{edges}}$, $F^{\perp}$ and $\overline{F} := F^{\perp} \circ F$ are defined as in [13, prop. 3.3], and the equality $\overline{F^{\tau_i}_{\downarrow}(\kappa)} = \overline{F^{\tau_i}}$ can be proved using prop. [13, 2.11.4].

Now, let $\mathcal{L} := \mathcal{L}_{\text{HF}} \sqcup \mathcal{L}^{(\text{aux})}$ and extend the pre-orders on $\mathcal{L}_{\text{HF}}$ and $\mathcal{L}^{(\text{aux})}$ with:

$$\forall \eta = (\gamma, \lambda) \in \mathcal{L}_{\text{HF}},\ \forall \kappa \in \mathcal{L}^{(\text{aux})},$$
$$\kappa \preccurlyeq \eta \Leftrightarrow \left(\forall \tau \in \kappa,\ \gamma \in \mathcal{L}_{\text{graphs}/e_{\tau,0}} \ \ \&\ \ \lambda \in \mathcal{L}_{\text{profls}/\overline{F^{\tau}}}\right),$$

where $\mathcal{L}_{\text{graphs}/e}$ for $e \in \mathcal{L}_{\text{edges}}$, resp. $\mathcal{L}_{\text{profls}/\overline{F}}$ for $\overline{F} \in \mathcal{L}_{\text{faces}}$, has been defined in [13, prop. 3.2], resp. in [13, prop. 3.3]. To check that this defines a pre-order $\mathcal{L}$, we note that:

$$\forall e \in \mathcal{L}_{\text{edges}},\ \forall \gamma \preccurlyeq \gamma' \in \mathcal{L}_{\text{graphs}}\,,\ \ \gamma \in \mathcal{L}_{\text{graphs}/e} \Rightarrow \gamma' \in \mathcal{L}_{\text{graphs}/e}\,,$$
$$\forall \overline{F} \in \mathcal{L}_{\text{faces}},\ \forall \lambda \preccurlyeq \lambda' \in \mathcal{L}_{\text{profls}}\,,\ \ \lambda \in \mathcal{L}_{\text{profls}/\overline{F}} \Rightarrow \lambda' \in \mathcal{L}_{\text{profls}/\overline{F}}\,.$$

Since, for any $\kappa \in \mathcal{L}^{(\text{aux})}$, $\kappa \preccurlyeq \eta_\kappa$ with $\eta_\kappa$ the label constructed above, $\mathcal{L}_{\text{HF}}$ is cofinal in $\mathcal{L}$ and, in particular, $\mathcal{L}$ is directed.

*Projective system on $\mathcal{L}$.* For any $\kappa \in \mathcal{L}^{(\text{aux})}$ and any $\eta = (\gamma, \lambda) \in \mathcal{L}_{\text{HF}}$ such that $\kappa \preccurlyeq \eta$, we define:

$$\begin{array}{rcrcl} Q_{\eta\to\kappa} & : & \mathcal{C}^{\mathbb{R}}_{\eta} & \to & \mathcal{C}^{(\text{aux})}_{\kappa} \\ & & h & \mapsto & \left[\tau \mapsto \sum_{k=1}^{n_{\gamma\to e_{\tau,0}}} \epsilon_{\gamma\to e_{\tau,0}}(k)\ h \circ a_{\gamma\to e_{\tau,0}}(k)\right] \end{array},$$

as well as:

$$\begin{array}{rcrcl} P_{\eta\to\kappa} & : & \mathcal{P}^{\mathbb{R}}_{\eta} & \to & \mathcal{P}^{(\text{aux})}_{\kappa} \\ & & P & \mapsto & \left[\tau \mapsto \sum_{F' \in H_{\lambda\to\overline{F^{\tau}}}} P(F')\right] \end{array},$$

where $n_{\gamma\to e_{\tau,0}}$, $\epsilon_{\gamma\to e_{\tau,0}}$ and $a_{\gamma\to e_{\tau,0}}$ have been defined in [13, prop. 3.2], and $H_{\lambda\to\overline{F^{\tau}}}$ in [13, prop. 3.3].

Defining, for any $\kappa \in \mathcal{L}^{(\text{aux})}$ and any $\eta \in \mathcal{L}_{\text{HF}}$, $\pi_{\eta\to\kappa} : \mathcal{M}^{\mathbb{R}}_{\eta} \to \mathcal{M}^{(\text{aux})}_{\kappa}$ as:

$$\pi_{\eta\to\kappa} = \left(Q_{\eta\to\kappa} \times P_{\eta\to\kappa}\right) \circ L^{\mathbb{R}}_{\eta}\,,$$

and specializing the definitions from [13, prop. 3.9] to the case $G = (\mathbb{R}, +)$, we have:

$$\forall (h,\,p) \in \mathcal{M}^{\mathbb{R}}_{\eta}\,,\ \pi_{\eta\to\kappa}(h,\,p) = \left(\tau \mapsto \mathrm{h}^{(e_{\tau,0}\,,\,\mathrm{id}_{\mathbb{R}})}_{\eta}(h,\,p),\ \tau \mapsto \mathrm{P}^{(\overline{F^{\tau}},1)}_{\eta}(h,\,p)\right).$$

For any $\kappa \preccurlyeq \kappa' \in \mathcal{L}^{(\text{aux})}$ and any $\eta \in \mathcal{L}_{\text{HF}}$ with $\kappa' \preccurlyeq \eta$, we thus have, using the definition of $\pi^{(\text{aux})}_{\kappa'\to\kappa}$ from prop. 2.10:

$$\pi_{\eta\to\kappa} = \pi^{(\text{aux})}_{\kappa'\to\kappa} \circ \pi_{\eta\to\kappa'}\,.$$

Moreover, for any $\kappa \in \mathcal{L}^{(\text{aux})}$ and any $\eta \preccurlyeq \eta' \in \mathcal{L}_{\text{HF}}$, with $\kappa \preccurlyeq \eta$, we have shown in [13, prop. 3.9] that:

$$\forall \tau \in \kappa,\ \mathrm{h}^{(e_{\tau,0}\,,\,\mathrm{id}_{\mathbb{R}})}_{\eta} \circ \pi^{\mathbb{R}}_{\eta'\to\eta} = \mathrm{h}^{(e_{\tau,0}\,,\,\mathrm{id}_{\mathbb{R}})}_{\eta'} \quad \& \quad \mathrm{P}^{(\overline{F^{\tau}},1)}_{\eta} \circ \pi^{\mathbb{R}}_{\eta'\to\eta} = \mathrm{P}^{(\overline{F^{\tau}},1)}_{\eta'}\,,$$

therefore, we also have:

$$\pi_{\eta'\to\kappa} = \pi_{\eta\to\kappa} \circ \pi^{\mathbb{R}}_{\eta'\to\eta} \,.$$

Let $\kappa = (\tau_1\,, \ldots\,, \tau_n) \in \mathcal{L}^{(\mathrm{aux})}$ with $0 < \tau_1 < \ldots < \tau_n \leqslant 1$ and let $\eta_\kappa = (\gamma_\kappa\,, \lambda_\kappa)$ be constructed as above. We have:

$$\forall h \in \mathcal{C}^{\mathbb{R}}_{\eta_\kappa}\,, \forall \tau_j \in \kappa,\ \ \big[Q_{\eta_\kappa\to\kappa}(h)\big](\tau_j) = \sum_{i=1}^{j} h(e_{\tau_i,\tau_{i-1/2}}) - \sum_{i=1}^{j-1} h(e_{\tau_i,\tau_{i+1/2}})\,,$$

as well as:

$$\forall P \in \mathcal{P}^{\mathbb{R}}_{\eta_\kappa}\,, \forall \tau_i \in \kappa,\ \ \big[P_{\eta_\kappa\to\kappa}(P)\big](\tau_i) = P\big(F^{\tau_i}_{\downarrow}(\kappa)\big)\,.$$

Inserting the expression for $\Xi^{\mathbb{R},-1}_{\eta_\kappa}$ from the proof of prop. 2.13, we get, for any $s \in \mathcal{C}^{(\mathrm{aux}),*}_{\kappa}$:

$$P_{\eta_\kappa\to\kappa} \circ \Xi^{\mathbb{R},-1}_{\eta_\kappa} \circ Q^*_{\eta_\kappa\to\kappa}(s) : \tau_i \mapsto s \circ Q_{\eta_\kappa\to\kappa}\left(\delta_{\chi^{-1}_{\eta_\kappa}\left(F^{\tau_i}_{\downarrow}(\kappa)\right)}\right) = s\left(\tau_j \mapsto \sum_{k=1}^{j} \delta_{ik}\right).$$

Thus, $P_{\eta_\kappa\to\kappa} \circ \Xi^{\mathbb{R},-1}_{\eta_\kappa} \circ Q^*_{\eta_\kappa\to\kappa} = \Xi^{(\mathrm{aux}),-1}_{\kappa}$, in other words $\pi_{\eta_\kappa\to\kappa}$ is a projection compatible with the symplectic structures. Then, for any $\eta \in \mathcal{L}_{\mathrm{HF}}$ such that $\eta \succcurlyeq \kappa$, there exists $\eta' \succcurlyeq \eta, \eta_\kappa$ such that $\pi_{\eta\to\kappa} \circ \pi^{\mathbb{R}}_{\eta'\to\eta} = \pi_{\eta'\to\kappa} = \pi_{\eta_\kappa\to\kappa} \circ \pi^{\mathbb{R}}_{\eta'\to\eta_\kappa}$, so $\pi_{\eta\to\kappa}$ is a projection compatible with the symplectic structures.

Thus, we can combine $\big(\mathcal{L}^{(\mathrm{aux})}, \mathcal{M}^{(\mathrm{aux})}, \pi^{(\mathrm{aux})}\big)^{\downarrow}$ and $\big(\mathcal{L}_{\mathrm{HF}}, \mathcal{M}^{\mathbb{R}}, \pi^{\mathbb{R}}\big)^{\downarrow}$ into a projective system of phase spaces on $\mathcal{L}$, fulfilling prop. 2.1.1 and 2.1.2. We denote by $(\mathcal{L}, \mathcal{H}, \Phi)^{\otimes}$ the corresponding projective system of quantum state spaces.

*Mapping narrow states on $\mathcal{L}_{\mathrm{HF}}$ to narrow states on $\mathcal{L}^{(\mathrm{aux})}$.* Applying [15, prop. 2.6] twice, to go from $\mathcal{L}_{\mathrm{HF}}$ to $\mathcal{L}$ (using that $\mathcal{L}_{\mathrm{HF}}$ is cofinal in $\mathcal{L}$) and from $\mathcal{L}$ to $\mathcal{L}^{(\mathrm{aux})}$, there exist a map $\sigma : \overline{\mathcal{S}}^{\otimes}_{(\mathcal{L}_{\mathrm{HF}}, \mathcal{H}^{\mathbb{R}}, \Phi^{\mathbb{R}})} \to \overline{\mathcal{S}}^{\otimes}_{(\mathcal{L}^{(\mathrm{aux})}, \mathcal{H}^{(\mathrm{aux})}, \Phi^{(\mathrm{aux})})}$ and a map $\alpha : \overline{\mathcal{A}}^{\otimes}_{(\mathcal{L}^{(\mathrm{aux})}, \mathcal{H}^{(\mathrm{aux})}, \Phi^{(\mathrm{aux})})} \to \overline{\mathcal{A}}^{\otimes}_{(\mathcal{L}_{\mathrm{HF}}, \mathcal{H}^{\mathbb{R}}, \Phi^{\mathbb{R}})}$ such that:

$$\forall \rho \in \overline{\mathcal{S}}^{\otimes}_{(\mathcal{L}_{\mathrm{HF}}, \mathcal{H}^{\mathbb{R}}, \Phi^{\mathbb{R}})}\,, \forall A \in \overline{\mathcal{A}}^{\otimes}_{(\mathcal{L}^{(\mathrm{aux})}, \mathcal{H}^{(\mathrm{aux})}, \Phi^{(\mathrm{aux})})}\,,\ \ \mathrm{Tr}\big(\rho\, \alpha(A)\big) = \mathrm{Tr}\big(\sigma(\rho)\, A\big)\,.$$

Moreover, for any $\kappa \in \mathcal{L}^{(\mathrm{aux})}$ and any $(s, t) \in \mathcal{C}^{(\mathrm{aux}),*}_{\kappa} \times \mathcal{P}^{(\mathrm{aux}),*}_{\kappa}$, we have from eq. (2.3.1):

$$\begin{aligned}\alpha\left(\big[\mathrm{T}^{(\mathrm{aux})}_{\kappa}(s, t)\big]_{\sim,\mathcal{L}^{(\mathrm{aux})}}\right) &= \Big[\Phi^{-1}_{\eta_\kappa\to\kappa} \circ \big(\mathrm{id}_{\mathcal{H}_{\eta_\kappa\to\kappa}} \otimes \mathrm{T}^{(\mathrm{aux})}_{\kappa}(s, t)\big) \circ \Phi_{\eta_\kappa\to\kappa}\Big]_{\sim,\mathcal{L}_{\mathrm{HF}}} \\ &= \Big[\mathrm{T}^{\mathbb{R}}_{\eta_\kappa}\left(s \circ Q_{\eta_\kappa\to\kappa}\,, t \circ P_{\eta_\kappa\to\kappa}\right)\Big]_{\sim,\mathcal{L}_{\mathrm{HF}}}\,.\end{aligned}$$

Thus, for any $\rho \in \overline{\mathcal{S}}^{\otimes}_{(\mathcal{L}_{\mathrm{HF}}, \mathcal{H}^{\mathbb{R}}, \Phi^{\mathbb{R}})}$ and any $\kappa \in \mathcal{L}^{(\mathrm{aux})}$, we get:

$$\forall (s, t) \in \mathcal{C}^{(\mathrm{aux}),*}_{\kappa} \times \mathcal{P}^{(\mathrm{aux}),*}_{\kappa}\,,\ \ W_{\sigma(\rho)_\kappa}(s, t) = W_{\rho_{\eta_\kappa}}\left(s \circ Q_{\eta_\kappa\to\kappa}\,, t \circ P_{\eta_\kappa\to\kappa}\right),$$

hence $\sigma\left\langle \hat{\mathcal{S}}^{\otimes}_{(\mathcal{L}_{\mathrm{HF}},\mathcal{H}^{\mathbb{R}},\Phi^{\mathbb{R}})}\right\rangle \subset \hat{\mathcal{S}}^{\otimes}_{(\mathcal{L}^{(\mathrm{aux})},\mathcal{H}^{(\mathrm{aux})},\Phi^{(\mathrm{aux})})} = \varnothing$. □

If we cannot have Gaussian states, the next best thing would be to have some quantum analogue of the classical probability distributions whose characteristic function takes the form:

$$\big\langle e^{-is(X)}\big\rangle_X = \exp\Big[ia(s) - \big(b(s)\big)^{\alpha/2}\Big],$$

with $\alpha \in ]0, 2]$, $a$ a linear form and $b$ a (symmetric) non-negative bilinear form on the dual space. These distributions have the nice property that their marginal probabilities (the classical equivalent of the partial trace of a quantum state) are again of the same form, with the same $\alpha$ (this is manifest by recalling that the characteristic function for a marginal probability is the restriction of the full characteristic function to the corresponding vector subspace of the dual space, in complete analogy to def. 2.2.3). For $\alpha = 2$ the distribution is Gaussian, while for $\alpha < 2$ it is heavy-tailed (aka. has infinite variance), and these distributions generalize the Gaussian distribution in the sense that they appear as attractors in heavy-tailed generalizations of the central limit theorem [6].

However, as prop. 2.15 below shows, the previous arguments excludes in the same stroke a large class of states.

**Proposition 2.15** We consider the same objects as in prop. 2.4 and we suppose that there exists a state $\rho \in \mathcal{S}^{\otimes}_{(\mathcal{L},\mathcal{H},\Phi)}$, integers $m, n \in \mathbb{N}$, and reals $\epsilon_1, \ldots, \epsilon_{m+n} > 0$ such that:

1. for any $\eta \in \mathcal{L}$ there exists $m$ linear forms $a^{(1)}_\eta, \ldots, a^{(m)}_\eta$ and $n$ (symmetric) non-negative bilinear forms $b^{(1)}_\eta, \ldots, b^{(n)}_\eta$ on $\mathcal{C}^*_\eta \times \mathcal{P}^*_\eta$;
2. for any $\eta \preccurlyeq \eta' \in \mathcal{L}$:

$$\forall k \in \{1, \ldots, m\},\ a^{(k)}_\eta = a^{(k)}_{\eta'} \circ (Q^*_{\eta'\to\eta} \times P^*_{\eta'\to\eta})$$
$$\&\quad \forall l \in \{1, \ldots, n\},\ b^{(l)}_\eta = b^{(l)}_{\eta'} \circ \big((Q^*_{\eta'\to\eta} \times P^*_{\eta'\to\eta}) \times (Q^*_{\eta'\to\eta} \times P^*_{\eta'\to\eta})\big)\,;$$

3. for any $\eta \in \mathcal{L}$ and any $(s, t) \in \mathcal{C}^*_\eta \times \mathcal{P}^*_\eta$:

$$\Big(\forall k \in \{1, \ldots, m\},\ \big|a^{(k)}_\eta(s, t)\big|^2 < \epsilon_k \quad \& \quad \forall l \in \{1, \ldots, n\},\ b^{(l)}_\eta(s, t;\, s, t) < \epsilon_{m+l}\Big)$$
$$\Rightarrow\ \big|W_{\rho_\eta}(s, t) - 1\big| < \frac{1}{2}\,.$$

Then, $\hat{\mathcal{S}}^{\otimes}_{(\mathcal{L},\mathcal{H},\Phi)} \neq \varnothing$.

**Proof** For any $\eta \in \mathcal{L}$ we define:

$$\forall (s, t), (s', t') \in \mathcal{C}^*_\eta \times \mathcal{P}^*_\eta,$$
$$B_\eta(s, t;\, s', t') := \frac{1}{4\pi\, \min(\epsilon_1, \ldots, \epsilon_{m+n})} \left(\sum_{k=1}^{m} a^{(k)}_\eta(s, t)\, a^{(k)}_\eta(s', t') + \sum_{l=1}^{n} b^{(l)}_\eta(s, t;\, s', t')\right),$$

as well as:

$$\forall s, s' \in \mathcal{C}^*_\eta,\ V_\eta(s, s') := B_\eta(s, 0;\, s', 0) \quad \& \quad \forall t, t' \in \mathcal{P}^*_\eta,\ U_\eta(t, t') := B_\eta(0, t;\, 0, t')\,.$$

$V_\eta$, resp. $U_\eta$, is a (symmetric) non-negative bilinear form on $\mathcal{C}^*_\eta$, resp. $\mathcal{P}^*_\eta$, and for any $\eta \preccurlyeq \eta' \in \mathcal{L}$, we have:

$$V_\eta = V_{\eta'} \circ (Q^*_{\eta'\to\eta} \times Q^*_{\eta'\to\eta}) \quad \& \quad U_\eta = U_{\eta'} \circ (P^*_{\eta'\to\eta} \times P^*_{\eta'\to\eta})\,.$$

Let $\eta \in \mathcal{L}$. Reasoning by contradiction, we suppose that there exists $(s', t') \in \mathcal{C}^*_\eta \times \mathcal{P}^*_\eta$ such that:

$$V_\eta(s', s') + U_\eta(t', t') < t'\big(\Xi^{-1}_\eta(s')\big)\,.$$

In particular, this implies $\epsilon := t'\big(\Xi_\eta^{-1}(s')\big) > 0$. We define $s := \sqrt{\frac{2\pi}{\epsilon}}s'$ and $t := \sqrt{\frac{2\pi}{\epsilon}}t'$, so that:

$$V_\eta(s,s) + U_\eta(t,t) < 2\pi \quad \& \quad t\big(\Xi_\eta^{-1}(s)\big) = 2\pi\,.$$

Then, we have:

$$B_\eta(s,0;\,s,0) = V_\eta(s,s) < 4\pi \quad \& \quad B_\eta(0,t;\,0,t) = U_\eta(t,t) < 4\pi\,,$$

as well as:

$$B_\eta(-s,t;\,-s,t) \leqslant B_\eta(-s,t;\,-s,t) + B_\eta(s,t;\,s,t) = 2\,\big(V_\eta(s,s) + U_\eta(t,t)\big) < 4\pi\,.$$

Thus, we get:

$$\big|W_{\rho_\eta}(s,0) - 1\big| < \frac{1}{2}\,, \quad \big|W_{\rho_\eta}(0,t) - 1\big| < \frac{1}{2} \quad \& \quad \big|W_{\rho_\eta}(-s,t) - 1\big| < \frac{1}{2}\,. \tag{2.15.1}$$

Now, the positivity condition eq. (2.2.1) applied to the points $(0,0)$, $(s,0)$ and $(0,t)$ can be rewritten as:

$$\begin{aligned}&\forall z_o\,,\, z_1\,,\, z_2 \in \mathbb{C}, \\ &\qquad |z_o|^2 + |z_1|^2 + |z_2|^2 + 2\,\mathrm{Re}\,\big(\overline{z_o}\,z_1\,W_{\rho_\eta}(s,0) + \overline{z_o}\,z_2\,W_{\rho_\eta}(0,t) - \overline{z_1}z_2\,W_{\rho_\eta}(-s,t)\big) \geqslant 0\end{aligned} \tag{2.15.2}$$

where we have used $t\big(\Xi_\eta^{-1}(s)\big) = 2\pi$ and $W_{\rho_\eta}(0,0) = 1$ (for $\rho \in \mathcal{S}^{\otimes}_{(\mathcal{L},\mathcal{H},\Phi)}$, hence $\mathrm{Tr}_{\mathcal{H}_\eta}\,\rho_\eta = 1$). Applying eq. (2.15.2) with $z_o = 1$ and $z_1 = z_2 = -1$ yields:

$$\begin{aligned}0 &\leqslant 3 - \mathrm{Re}\,\big(2\,W_{\rho_\eta}(s,0) - 2\,W_{\rho_\eta}(0,t) - 2\,W_{\rho_\eta}(-s,t)\big) \\ &\leqslant 2\,\big|1 - W_{\rho_\eta}(s,0)\big| + 2\,\big|1 - W_{\rho_\eta}(0,t)\big| - 1 - 2\,\mathrm{Re}\,\big(W_{\rho_\eta}(-s,t)\big) \\ &\quad < \mathrm{Re}\,\big(1 - W_{\rho_\eta}(-s,t)\big) - \mathrm{Re}\,\big(W_{\rho_\eta}(-s,t)\big)\,.\end{aligned}$$

Thus, $-\mathrm{Re}\big(1 - W_{\rho_\eta}(-s,t)\big) < \mathrm{Re}\big(W_{\rho_\eta}(-s,t)\big) < \mathrm{Re}\big(1 - W_{\rho_\eta}(-s,t)\big)$, and since we also have $\mathrm{Im}\big(W_{\rho_\eta}(-s,t)\big) = -\mathrm{Im}\big(1 - W_{\rho_\eta}(-s,t)\big)$, this implies $\big|W_{\rho_\eta}(-s,t)\big| < \big|1 - W_{\rho_\eta}(-s,t)\big| < \frac{1}{2} < 1$.

Next, we apply eq. (2.15.2) with:

$$\begin{aligned}&z_o = 1 - \big|W_{\rho_\eta}(-s,t)\big|^2, \\ &\qquad z_1 = -\overline{W_{\rho_\eta}(s,0)} - \overline{W_{\rho_\eta}(0,t)}\,W_{\rho_\eta}(-s,t) \quad \& \quad z_2 = -\overline{W_{\rho_\eta}(0,t)} - \overline{W_{\rho_\eta}(s,0)}\,\overline{W_{\rho_\eta}(-s,t)}\end{aligned}$$

(these values arise by optimizing on $z_1$, $z_2$, keeping $z_o$ fixed). After simplifications, we get:

$$0 \leqslant z_o^2 - z_o\,\Big[\big|W_{\rho_\eta}(s,0)\big|^2 + \big|W_{\rho_\eta}(0,t)\big|^2\Big] - 2\,z_o\,\mathrm{Re}\,\big(W_{\rho_\eta}(s,0)\,\overline{W_{\rho_\eta}(0,t)}\,W_{\rho_\eta}(-s,t)\big)\,.$$

Since $\big|W_{\rho_\eta}(-s,t)\big| < 1$, $z_o > 0$, so this leads to:

$$\big|W_{\rho_\eta}(-s,t)\big|^2 + \big|W_{\rho_\eta}(s,0)\big|^2 + \big|W_{\rho_\eta}(0,t)\big|^2 + 2\,\mathrm{Re}\,\big(W_{\rho_\eta}(s,0)\,\overline{W_{\rho_\eta}(0,t)}\,W_{\rho_\eta}(-s,t)\big) \leqslant 1\,.$$

Finally, we rewrite this inequality in terms of $w_o := W_{\rho_\eta}(-s,t) - 1$, $w_1 := W_{\rho_\eta}(s,0) - 1$ and $w_2 := W_{\rho_\eta}(0,t) - 1$:

$$\begin{aligned}0 &\geqslant 4 + 4\mathrm{Re}\,(w_o + w_1 + \overline{w_2}) + |w_o + w_1 + \overline{w_2}|^2 + 2\,\mathrm{Re}\,(w_o w_1 \overline{w_2}) \\ &\geqslant 4 - 4\,|w_o + w_1 + \overline{w_2}| + |w_o + w_1 + \overline{w_2}|^2 - 2\,|w_o w_1 \overline{w_2}|\,.\end{aligned}$$

However, this contradicts eq. (2.15.1) which requires $|w_o + w_1 + \overline{w_2}| < 3/2$ and $|w_o w_1 \overline{w_2}| < 1/8$,

since $(x \mapsto x^2 - 4\,x)\left\langle\, [0,\, 3/2[\, \right\rangle = ]-15/4,\, 0]$.

Thus, we have proven that:

$$\forall (s,t) \in \mathcal{C}^*_\eta \times \mathcal{P}^*_\eta,\ V_\eta(s,s) + U_\eta(t,t) - t\left(\Xi^{-1}_\eta(s)\right) \geqslant 0\,.$$

In particular, for any $s \neq 0 \in \mathcal{C}^*_\eta$, there exists $x \in \mathcal{C}_\eta$ such that $s(x) \neq 0$, so defining $t := \Xi^*_\eta\left(\frac{x}{s(x)}\right) \in \mathcal{P}^*_\eta$, we have:

$$\forall \lambda \in \mathbb{R},\ \lambda^2\, V_\eta(s,s) + U_\eta(t,t) - \lambda \geqslant 0\,,$$

hence the symmetric bilinear form $V_\eta$ is strictly positive. Similarly, $U_\eta$ is also strictly positive. Therefore, $\left(V_\eta\,,\, U_\eta\right)_{\eta \in \mathcal{L}} \in \mathcal{V}^{\downarrow}_{(\mathcal{L},(\mathcal{C},\mathcal{P}),(Q,P))}$, so by prop. 2.9, $\hat{\mathcal{S}}^{\otimes}_{(\mathcal{L},\mathcal{H},\Phi)} \neq \varnothing$. □

The previous result can also be seen as excluding a different notion of state confinement, focusing on *quantiles* rather than *moments* (in other words, working with the cumulative distribution function rather than with the characteristic or moment-generating one).

**Proposition 2.16** We consider the same objects a in prop. 2.4. Let $\mathcal{C}^*_{\text{cyl}}$, resp. $\mathcal{P}^*_{\text{cyl}}$, be the inductive limit (*without any completion*) of the inductive system $\left(\left(\mathcal{C}^*_\eta\right)_{\eta \in \mathcal{L}},\, \left(Q^*_{\eta' \to \eta}\right)_{\eta \preccurlyeq \eta'}\right)$, resp. $\left(\left(\mathcal{P}^*_\eta\right)_{\eta \in \mathcal{L}},\, \left(P^*_{\eta' \to \eta}\right)_{\eta \preccurlyeq \eta'}\right)$. For any $\eta \in \mathcal{L}$, let $Q^*_{\text{cyl} \to \eta}$, resp. $P^*_{\text{cyl} \to \eta}$, be the canonical injection of $\mathcal{C}^*_\eta$ in $\mathcal{C}^*_{\text{cyl}}$, resp. of $\mathcal{P}^*_\eta$ in $\mathcal{P}^*_{\text{cyl}}$.

We suppose that there exists a state $\rho \in \mathcal{S}^{\otimes}_{(\mathcal{L},\mathcal{H},\Phi)}$, a real $p > 3/4$ and a non-negative (symmetric) bilinear form $B_{\text{cyl}}$ on $\mathcal{C}^*_{\text{cyl}} \times \mathcal{P}^*_{\text{cyl}}$ such that, for any $\eta \in \mathcal{L}$ and any $(s,t) \in \mathcal{C}^*_\eta \times \mathcal{P}^*_\eta$:

$$\mathrm{Tr}_{\mathcal{H}_\eta}\left[\rho_\eta\, \Pi_\eta(s,t)\right] \geqslant p\,,$$

where $\Pi_\eta(s,t)$, is the spectral projector of $\mathrm{X}_\eta(s,t)$ on $\left[-A,\, A\right]$ with:

$$A = \sqrt{B_{\text{cyl}}\left(Q^*_{\text{cyl}\to\eta}(s), P^*_{\text{cyl}\to\eta}(t);\ Q^*_{\text{cyl}\to\eta}(s), P^*_{\text{cyl}\to\eta}(t)\right)}\,.$$

Then, $\hat{\mathcal{S}}^{\otimes}_{(\mathcal{L},\mathcal{H},\Phi)} \neq \varnothing$.

**Proof** For any $\eta \in \mathcal{L}$, we define:

$$\forall (s,t),\, (s',t') \in \mathcal{L},\ b^{(1)}_\eta(s,t;\, s',t') := B_{\text{cyl}}\left(Q^*_{\text{cyl}\to\eta}(s), P^*_{\text{cyl}\to\eta}(t);\ Q^*_{\text{cyl}\to\eta}(s'), P^*_{\text{cyl}\to\eta}(t')\right).$$

By construction, $\left(b^{(1)}_\eta\right)_{\eta \in \mathcal{L}}$ fulfills prop. 2.15.1 and 2.15.2.

Let $\epsilon_1 := \left(2\,p - \frac{3}{2}\right)^2 > 0$. For any $\eta \in \mathcal{L}$ and any $(s,t) \in \mathcal{C}^*_\eta \times \mathcal{P}^*_\eta$ such that $b^{(1)}_\eta(s,t;\, s,t) < \epsilon_1$, we have:

$$\left\|\Pi_\eta(s,t)\ \left(\mathrm{T}_\eta(s,t) - \mathrm{id}_{\mathcal{H}_\eta}\right)\right\| \leqslant \sup_{x \in \left[-\sqrt{b^{(1)}_\eta(s,t;\,s,t)},\ \sqrt{b^{(1)}_\eta(s,t;\,s,t)}\right]} \left|e^{i\,x} - 1\right|$$

$$= 2\sin\left(\frac{\sqrt{b_\eta^{(1)}(s,t;\,s,t)}}{2}\right) < 2p - \tfrac{3}{2}\,,$$

as well as:

$$\left\|\mathrm{T}_\eta(s,t) - \mathrm{id}_{\mathcal{H}_\eta}\right\| \leqslant \left\|\mathrm{T}_\eta(s,t)\right\| + \left\|\mathrm{id}_{\mathcal{H}_\eta}\right\| = 2\,.$$

Thus, we get:

$$\begin{aligned}\left|W_{\rho_\eta}(s,t) - 1\right| &= \left|Tr_{\mathcal{H}_\eta}\,\rho_\eta\,\left(\mathrm{T}_\eta(s,t) - \mathrm{id}_{\mathcal{H}_\eta}\right)\right| \\ &< 2p - \tfrac{3}{2} + 2\,Tr_{\mathcal{H}_\eta}\,\rho_\eta\,\left(\mathrm{id}_{\mathcal{H}_\eta} - \Pi_\eta(s,t)\right) < \frac{1}{2}\,.\end{aligned}$$

Hence, prop. 2.15 implies $\hat{\mathcal{S}}^{\otimes}_{(\mathcal{L},\mathcal{H},\Phi)} \neq \varnothing$. □

# 3 Outlook

The intention to obtain better *semi-classical states* for Loop Quantum Gravity was an important motivation for the state space extension considered in [13]. However, as just shown, it turns out that going beyond the Ashtekar-Lewandowski sector is only one part of the solution, and that further *obstructions* to the existence of such states on the holonomy-flux algebra need to be addressed as well.

Note that, while the argument cannot be easily generalized beyond the $G = \mathbb{R}$ case (because the systematic study of narrow states in subsection 2.1 makes heavy use of the linearity, viz. the discussion preceding prop. 2.9), we can nevertheless take it as a hint on what will be required to make the construction of semi-classical states easier – or even possible – in the *general case*. On the one hand, the negative result of subsection 2.2 casts serious doubts on the possibility to design admissible semi-classical states on the full holonomy-flux algebra, even in the compact group case: although diverging variances are of course excluded in this case (at least for the configuration variables), we expect that, if states very peaked around the group identity *could* be designed, the group structure should *not* matter much to them (heuristically, the portion of the configuration space seen by such states would be nearly linear). On the other hand, the approach we will develop in [16] to go around the obstruction should simplify the construction of projective states to the point that keeping $G$ *arbitrary* would become effortless.

The exclusion of large classes of states on the real-valued holonomy-flux algebra, discussed at the end of subsection 2.2, also raises an additional question, namely whether the projective state space could possibly be *empty* in this case. While projective systems built on *countable* label sets always yield non-empty quantum state spaces (because projective states can then be constructed recursively, see [16, subsection 2.1]), the situation is less clear on *uncountable* labels sets (see [28] for an instructive example of a seemingly healthy, yet empty projective limit). In the case of the holonomy-flux algebra on a compact group $G$, we have proved in [13, theorem 3.20] that all states on the corresponding AL Hilbert space belong to the projective state space, which is thus guaranteed to be non-empty, but we do not have such a proof if the group is *non-compact*, because the AL Hilbert space does not exist there.

To analyze this question more closely in the $G = \mathbb{R}$ case, one can observe that any state on the algebra $\overline{\mathcal{A}}^{\otimes}_{(\mathcal{L},\mathcal{H},\Phi)}$ (aka. positive linear form of norm 1, see [7, part III, def. 2.2.8] and [15, prop. 2.4]) yields a $\mathbb{C}$-valued function on the inductive limit $\mathcal{C}^*_{\mathrm{cyl}} \times \mathcal{P}^*_{\mathrm{cyl}}$, which satisfies suitable *positivity requirements* (in accordance with def. 2.2.2). Since the space of states on a $C^*$-algebra is never empty [25, theorem I.9.18], we are assured that there do exist such functions of positive type on $\mathcal{C}^*_{\mathrm{cyl}} \times \mathcal{P}^*_{\mathrm{cyl}}$. However, a function of positive type comes from a projective state in $\mathcal{S}^{\otimes}_{(\mathcal{L},\mathcal{H},\Phi)}$ if and only if it is *continuous* with respect to the inductive limit topology on $\mathcal{C}^*_{\mathrm{cyl}} \times \mathcal{P}^*_{\mathrm{cyl}}$ (this reflects the characterization of normal states over a $W^*$-algebra, see [25, corollary III.3.11]). What we have established above is that, in the case of the projective structure constructed as in [13, subsection 3.1] for the holonomy-flux algebra with $G = \mathbb{R}$:

- there do not exist any function of positive type *twice-differentiable* at 0 with respect to the *inductive limit topology*;
- given some scalar product on $\mathcal{C}^*_{\mathrm{cyl}} \times \mathcal{P}^*_{\mathrm{cyl}}$, there do not exist any function of positive type *continuous* at 0 with respect to the *corresponding (strong) topology*.

This still leaves a fairly large window for projective quantum states to be found, however, it would be reassuring to have an actual proof of their existence, and, even better, *constructive techniques* to obtain them.

On the other hand, the first of these two points demonstrates that the failure to find admissible semi-classical states would *not* be solved by looking for an even larger state space: if there would exist such states on the algebra they would be twice-differentiable, hence continuous, at 0, and they would belong to the projective state space (thanks to the positivity condition, continuity at zero implies continuity everywhere). In other words, a state on the algebra that would have finite variances would *automatically* be regular enough to be representable as a projective family of density matrices (aka. projective quantum state). As announced in section 1, the problem we uncovered here has its roots in the holonomy-flux algebra *itself*. Therefore, the resolution we will propose in [16] is meant to act directly on the structure of this algebra. At the same time, it has the potential to bypass the concerns just raised about the projective state space being possibly empty, since it should provide us with a systematic procedure to construct arbitrary projective quantum states, no matter whether the gauge group is compact or not [16, subsection 2.1].

## Acknowledgements

This work has been financially supported by the Université François Rabelais, Tours, France (via a 3-years doctoral stipend from the French Ministry of Education – Contrat Doctoral Normalien), and by the Friedrich-Alexander-Universität Erlangen-Nürnberg, Germany (via the Bavarian Equal Opportunities Sponsorship – Förderung von Frauen in Forschung und Lehre (FFL) – Promoting Equal Opportunities for Women in Research and Teaching).

This research project has also been supported by funds to Emerging Field Project "Quantum Geometry" from the FAU Erlangen-Nürnberg within its Emerging Fields Initiative.

# A Appendix: Wigner characteristic functions

In this appendix, we give a brief overview of the Wigner-Weyl transform [29], that allows to represent quantum density matrices as functions on the phase space, mimicking the probability distributions of classical statistical physics. The presentation below follows very closely [11], merely adapting of the notations and definitions to the ones we will be using in section 2.

Note that, in view of applying this tool to projective systems of quantum state spaces (subsection 2.1), we will be doing only 'half' of the Wigner transform: going from the density matrix to the Wigner characteristic function. The second half would be to perform a Fourier transform yielding the Wigner quasi-probability distribution, in analogy to the reconstruction of a classical probability distribution from its characteristic function. However, Wigner characteristic functions are more convenient when working with partial traces of the underlying density matrices, exactly like classical characteristic functions are convenient for computing marginal probabilities.

## A.1 Weakly continuous representations of the Weyl algebra

We begin by recalling basic facts regarding the Weyl algebra and weakly continuous representations thereof. These representation-theoretic considerations will come into play to prove that the Wigner transform is surjective onto a suitably defined space of functions.

**Definition A.1** Let $\mathcal{C}$, $\mathcal{P}$ be two finite-dimensional real vector spaces and let $\Xi : \mathcal{P} \to \mathcal{C}^*$ be an invertible linear map (with $\mathcal{C}^*$ the dual of $\mathcal{C}$). We equip $\mathcal{C} \times \mathcal{P}$ with the symplectic form $\Omega$ given by:

$$\forall u, u' \in \mathcal{C},\ \forall v, v' \in \mathcal{P},\ \Omega(u, v;\, u', v') := \Xi(v')(u) - \Xi(v)(u').$$

The map $M := \mathrm{id}_{\mathcal{C}} \times \Xi$ is then a symplectomorphism $\mathcal{C} \times \mathcal{P} \to T^*(\mathcal{C})$ ($T^*(\mathcal{C})$ being equipped with its canonical symplectic structure, see eg. [14, eq. (2.16.1)]).

**Proposition A.2** Let $\mathcal{I}$ be the complex vector space of functions $\mathcal{C}^* \times \mathcal{P}^* \to \mathbb{C}$ with support on finitely many points, ie:

$$\forall \iota \in \mathcal{I},\quad \#\left\{(s, t) \in \mathcal{C}^* \times \mathcal{P}^* \mid \iota(s, t) \neq 0\right\} < \infty.$$

In particular, for any $(s, t) \in \mathcal{C}^* \times \mathcal{P}^*$, we define $\delta_{(s,t)} \in \mathcal{I}$ by:

$$\forall (s', t') \in \mathcal{C}^* \times \mathcal{P}^*,\ \delta_{(s,t)}(s', t') := \begin{cases} 1 \text{ if } (s', t') = (s, t) \\ 0 \text{ else} \end{cases}.$$

For $\iota_1$, $\iota_2 \in \mathcal{I}$, we define their product $\iota_1 \star \iota_2 \in \mathcal{I}$ by:

$$\forall (s, t) \in \mathcal{C}^* \times \mathcal{P}^*,\ (\iota_1 \star \iota_2)(s, t) := \sum_{\substack{(s_1,t_1),(s_2,t_2) \in \mathcal{C}^* \times \mathcal{P}^* \\ s = s_1 + s_2,\ t = t_1 + t_2}} e^{i\xi(s_1,t_1;s_2,t_2)}\, \iota_1(s_1, t_1)\, \iota_2(s_2, t_2),$$

where, for any $(s_1, t_1), (s_2, t_2) \in \mathcal{C}^* \times \mathcal{P}^*$:

$$\xi(s_1, t_1; s_2, t_2) := \frac{t_1\big(\Xi^{-1}(s_2)\big) - t_2\big(\Xi^{-1}(s_1)\big)}{2}\,.$$

We also define, for any $\iota \in \mathcal{I}$, its conjugate $\iota^*$ by:

$$\forall (s, t) \in \mathcal{C}^* \times \mathcal{P}^*,\ \iota^*(s, t) := \overline{\iota(-s, -t)}\,,$$

where $\overline{\,\cdot\,}$ stands for complex conjugation.

$\mathcal{I}, \star, {}^*$ is a $^*$-algebra [7, section III.2.2], with unit $\delta_{(0,0)}$.

**Proof** Let $\iota_1$, $\iota_2 \in \mathcal{I}$. Since $\iota_1$ and $\iota_2$ are supported on finitely many points, the sum defining $\iota_1 \star \iota_2$ is finite and $\iota_1 \star \iota_2$ is again supported on finitely many points, ie. it is in $\mathcal{I}$. The operation $\star$ is bilinear, and, for any $\iota_1$, $\iota_2$, $\iota_3 \in \mathcal{I}$, we have:

$$\begin{aligned}
\forall (s, t) \in \mathcal{C}^* \times \mathcal{P}^*,\ & \big(\iota_1 \star (\iota_2 \star \iota_3)\big)(s,\ t) = \\
&= \sum_{\substack{s = s_1 + s'_1 \\ t = t_1 + t'_1}} e^{i\,\xi(s_1, t_1; s'_1, t'_1)}\, \iota_1(s_1\,,\ t_1) \sum_{\substack{s'_1 = s_2 + s_3 \\ t'_1 = t_2 + t_3}} e^{i\,\xi(s_2, t_2; s_3, t_3)}\, \iota_2(s_2\,,\ t_2)\, \iota_3(s_3\,,\ t_3) \\
&= \sum_{\substack{s = s_1 + s_2 + s_3 \\ t = t_1 + t_2 + t_3}} e^{i\,\xi(s_1, t_1; s_2, t_2) + i\,\xi(s_1, t_1; s_3, t_3) + i\,\xi(s_2, t_2; s_3, t_3)}\, \iota_1(s_1\,,\ t_1)\, \iota_2(s_2\,,\ t_2)\, \iota_3(s_3\,,\ t_3) \\
&= \big((\iota_1 \star \iota_2) \star \iota_3\big)(s,\ t)\,,
\end{aligned}$$

so it is associative as well. We can check that $\delta_{(0,0)}$ is a unit for $\mathcal{I}, \star$. Finally, the operation $(\,\cdot\,)^*$ is an antilinear involution by construction, and, for any $\iota_1$, $\iota_2 \in \mathcal{I}$, we have:

$$\forall (s, t) \in \mathcal{C}^* \times \mathcal{P}^*,\ (\iota_1 \star \iota_2)^*(s,\ t) = \sum_{\substack{s = s_1 + s_2 \\ t = t_1 + t_2}} e^{-i\,\xi(s_1, t_1; s_2, t_2)}\, \iota_1^*(s_1\,,\ t_1)\, \iota_2^*(s_2\,,\ t_2) = (\iota_2^* \star \iota_1^*)(s,\ t)\,.$$

□

Note that the elementary observables $\mathrm{X}_\kappa(s, t)$ in a representation $\kappa$ are labeled by elements of the dual space $\mathcal{C}^* \times \mathcal{P}^*$ of the phase space $\mathcal{M} \approx \mathcal{C} \times \mathcal{P}$, because they are actually labeled by the linear observables of which they are the quantization (in the spirit of [15, def. A.3], up to a difference in sign convention). Therefore, the unitary operators $\mathrm{T}_\kappa(s, t)$, which are obtained by exponentiating these observables, are labeled by elements of the dual as well. This observation will be important for the development of subsection 2.1.

**Definition A.3** Let $\mathcal{H}_\kappa$ be a complex Hilbert space. A weakly continuous, unitary representation of $\mathcal{C}^* \times \mathcal{P}^*$ on $\mathcal{H}_\kappa$ is a map $\mathrm{T}_\kappa : \mathcal{C}^* \times \mathcal{P}^* \to \mathcal{A}_\kappa$ (where $\mathcal{A}_\kappa$ is the algebra of bounded linear operators over $\mathcal{H}_\kappa$), satisfying the following properties:

1. $\forall (s, t) \in \mathcal{C}^* \times \mathcal{P}^*$, $\mathrm{T}_\kappa(s, t)$ is a unitary operator on $\mathcal{H}_\kappa$;
2. $\forall (s_1, t_1), (s_2, t_2) \in \mathcal{C}^* \times \mathcal{P}^*$, $\mathrm{T}_\kappa(s_1, t_1)\, \mathrm{T}_\kappa(s_2, t_2) = e^{i\xi(s_1, t_1; s_2, t_2)}\, \mathrm{T}_\kappa(s_1 + s_2, t_1 + t_2)$;
3. $\mathrm{T}_\kappa(0, 0) = \mathrm{id}_{\mathcal{H}_\kappa}$;
4. $\forall \varphi, \varphi' \in \mathcal{H}_\kappa$, the map $(s, t) \mapsto \langle \varphi' \mid \mathrm{T}_\kappa(s, t)\, \varphi \rangle$ is a continuous function $\mathcal{C}^* \times \mathcal{P}^* \to \mathbb{C}$.

This provides a representation of the $^*$-algebra $\mathcal{I}$ on $\mathcal{H}_\kappa$:

$$\forall \iota \in \mathcal{I},\ \mathrm{T}_\kappa\left\{\iota\right\} := \sum_{(s,t)\in\mathcal{C}^*\times\mathcal{P}^*} \iota(s,t)\, \mathrm{T}_\kappa(s,t)$$

(using in particular that points A.3.1 to A.3.3 imply $\forall (s,t) \in \mathcal{C}^* \times \mathcal{P}^*$, $\mathrm{T}_\kappa(-s,-t) = \mathrm{T}_\kappa(s,t)^{-1} = \mathrm{T}_\kappa(s,t)^\dagger$ ).

**Proposition A.4** Let $\mathcal{H}_\kappa$, $\mathrm{T}_\kappa$ be as in def. A.3. Then, there exists, for any $(s,t) \in \mathcal{C}^* \times \mathcal{P}^*$ a densely defined (possibly unbounded), self-adjoint operator $\mathrm{X}_\kappa(s,t)$ on $\mathcal{H}_\kappa$ such that:

$$\forall \tau \in \mathbb{R},\ \mathrm{T}_\kappa(\tau s, \tau t) = \exp(i\,\tau\, \mathrm{X}_\kappa(s,t))$$

(the exponential being defined via spectral resolution [23, theorem VIII.5]).

Moreover, there exists a dense vector subspace $\mathcal{D}_\kappa \subset \mathcal{H}_\kappa$ such that:

1. for each $(s,t) \in \mathcal{C}^* \times \mathcal{P}^*$, $\mathcal{D}_\kappa \subset \mathrm{Dom}\big(\mathrm{X}_\kappa(s,t)\big)$ (where $\mathrm{Dom}\big(\mathrm{X}_\kappa(s,t)\big)$ denotes the dense domain of $\mathrm{X}_\kappa(s,t)$) and $\mathrm{X}_\kappa(s,t)|_{\mathcal{D}_\kappa}$ is essentially self-adjoint;
2. $\forall (s,t) \in \mathcal{C}^* \times \mathcal{P}^*$, $\mathrm{X}_\kappa(s,t)\left\langle \mathcal{D}_\kappa \right\rangle \subset \mathcal{D}_\kappa$ ;
3. $\forall (s,t), (s',t') \in \mathcal{C}^* \times \mathcal{P}^*$, $\forall \tau, \tau' \in \mathbb{R}$, $\mathrm{X}_\kappa(\tau s + \tau' s',\ \tau t + \tau' t')|_{\mathcal{D}_\kappa} = \tau\ \mathrm{X}_\kappa(s,t)|_{\mathcal{D}_\kappa} + \tau'\ \mathrm{X}_\kappa(s',t')|_{\mathcal{D}_\kappa}$.

**Proof** Let $(s,t) \in \mathcal{C}^* \times \mathcal{P}^*$. For any $\tau \in \mathbb{R}$, $\mathrm{T}_\kappa(\tau s, \tau t)$ is unitary (def. A.3.1), and $\mathrm{T}_\kappa(0,0) = \mathrm{id}_{\mathcal{H}_\kappa}$ (def. A.3.3). Moreover, using def. A.3.2, we have, for any $\tau_1, \tau_2 \in \mathbb{R}$:

$$\mathrm{T}_\kappa(\tau_1 s, \tau_1 t)\, \mathrm{T}_\kappa(\tau_2 s, \tau_2 t) = \mathrm{T}_\kappa\big((\tau_1 + \tau_2)s,\ (\tau_1 + \tau_2)t\big)\,, \tag{A.4.1}$$

since $\xi(\tau_1 s, \tau_1 t;\ \tau_2 s, \tau_2 t) = \tau_1\, \tau_2\, \xi(s,t;\ s,t) = 0$.

Let $\varphi \in \mathcal{H}_\kappa$, $(s',t') \in \mathcal{C}^* \times \mathcal{P}^*$ and $\epsilon > 0$. Let $\varphi' := \mathrm{T}_\kappa(s',t')\,\varphi$. From def. A.3.4, there exists an open neighborhood $U$ of $(s',t')$ in $\mathcal{C}^* \times \mathcal{P}^*$, such that:

$$\forall (s'',t'') \in U,\ \left| \langle \varphi' \mid \mathrm{T}_\kappa(s'',t'')\,\varphi\rangle - \|\varphi'\|^2 \right| < \frac{\epsilon}{2}\,.$$

Hence, thanks to the unitarity of $\mathrm{T}_\kappa(s'',t'')$ for any $(s'',t'') \in \mathcal{C}^* \times \mathcal{P}^*$, we get:

$$\forall (s'',t'') \in U,\ \left\|\mathrm{T}_\kappa(s'',t'')\,\varphi - \mathrm{T}_\kappa(s',t')\,\varphi\right\|^2 \leqslant 2\left| \langle \varphi' \mid \mathrm{T}_\kappa(s'',t'')\,\varphi\rangle - \|\varphi'\|^2 \right| < \epsilon\,. \tag{A.4.2}$$

Let $\varphi \in \mathcal{H}_\kappa$, $\tau_o \in \mathbb{R}$ and $\epsilon > 0$. Applying the previous point with $(s',t') = (\tau_o s, \tau_o t)$, there exists $\epsilon' > 0$ such that:

$$\forall \tau \in \left]\tau_{o-\epsilon'}, \tau_{o+\epsilon'}\right[\,,\ \left\|\mathrm{T}_\kappa(\tau s, \tau t)\,\varphi - \mathrm{T}_\kappa(\tau_o s, \tau_o t)\,\varphi\right\|^2 < \epsilon\,.$$

Thus, $\tau \mapsto \mathrm{T}_\kappa(\tau s, \tau t)$ is a strongly continuous one-parameter unitary group, so, by Stone's theorem [23, theorem VIII.8], there exists a densely defined, self-adjoint operator $\mathrm{X}_\kappa(s,t)$ on $\mathcal{H}_\kappa$ such that, for any $\tau \in \mathbb{R}$, $\mathrm{T}_\kappa(\tau s, \tau t) = \exp\big(i\,\tau\, \mathrm{X}_\kappa(s,t)\big)$.

Let $\widetilde{\mu}$ be a Lebesgue measure on the finite-dimensional real vector space $\mathcal{C}^* \times \mathcal{P}^*$. For any $f \in C^\infty_o(\mathcal{C}^* \times \mathcal{P}^*,\ \mathbb{C})$ (where $C^\infty_o(\mathcal{C}^* \times \mathcal{P}^*,\ \mathbb{C})$ denotes the space of smooth, complex valued, compactly supported functions on $\mathcal{C}^* \times \mathcal{P}^*$), and any $\varphi \in \mathcal{H}_\kappa$, we define $\varphi\left\{f\right\} \in \mathcal{H}_\kappa$ as:

$$\varphi\left\{f\right\} := \int_{\mathcal{C}^*\times\mathcal{P}^*} d\widetilde{\mu}(s',t')\, f(s',t')\, \mathrm{T}_\kappa(s',t')\,\varphi\,.$$

As shown above $(s', t') \mapsto \mathrm{T}_\kappa(s', t')\,\varphi$ is a continuous map $\mathcal{C}^*\times\mathcal{P}^* \to \mathcal{H}_\kappa$, so $(s', t') \mapsto f(s', t')\,\mathrm{T}_\kappa(s', t')\,\varphi$ is a continuous, compactly supported map $\mathcal{C}^* \times \mathcal{P}^* \to \mathcal{H}_\kappa$, hence it is integrable on $\mathcal{C}^* \times \mathcal{P}^*$, $\widetilde{\mu}$ [10, section III.1]. Therefore, $\varphi\,\{f\}$ is well-defined as an element of $\mathcal{H}_\kappa$.

We define the vector subspace $\mathcal{D}_\kappa \subset \mathcal{H}_\kappa$ as:

$$\mathcal{D}_\kappa := \mathrm{Vect}\left\{\varphi\,\{f\} \mid f \in C_o^\infty(\mathcal{C}^* \times \mathcal{P}^*, \mathbb{C}),\ \varphi \in \mathcal{H}_\kappa\right\}.$$

Let $\varphi \in \mathcal{H}_\kappa$ and $\epsilon > 0$. From def. A.3.3 and eq. (A.4.2), there exists an open neighborhood $U$ of $(0, 0)$ in $\mathcal{C}^* \times \mathcal{P}^*$ such that:

$$\forall (s', t') \in U,\ \left\|\mathrm{T}_\kappa(s', t')\,\varphi - \varphi\right\|^2 < \epsilon\,.$$

Let $f \in C_o^\infty(\mathcal{C}^* \times \mathcal{P}^*, \mathbb{C})$ be a bump function with:

$$f \geqslant 0,\quad \mathrm{supp}(f) \subset U \quad \& \quad \int_{\mathcal{C}^*\times\mathcal{P}^*} d\widetilde{\mu}\ \ f = 1\,,$$

where $\mathrm{supp}(f)$ denotes the support of $f$. Then, we have:

$$\|\varphi - \varphi\,\{f\}\| \leqslant \int_{\mathcal{C}^*\times\mathcal{P}^*} d\widetilde{\mu}(s', t')\, f(s', t')\ \|\varphi - \mathrm{T}_\kappa(s', t')\,\varphi\| < \epsilon\,.$$

Thus, $\mathcal{D}_\kappa$ is a dense subspace of $\mathcal{H}_\kappa$.

Let $\varphi \in \mathcal{H}_\kappa$, $f \in C_o^\infty(\mathcal{C}^* \times \mathcal{P}^*, \mathbb{C})$ and $(s, t) \in \mathcal{C}^* \times \mathcal{P}^*$. We have, for any $\tau \in \mathbb{R}$:

$$\begin{aligned}\mathrm{T}_\kappa(\tau s, \tau t)\,\varphi\,\{f\} &= \int_{\mathcal{C}^*\times\mathcal{P}^*} d\widetilde{\mu}(s', t')\, f(s', t')\,\mathrm{T}_\kappa(\tau s, \tau t)\,\mathrm{T}_\kappa(s', t')\,\varphi \\ &= \int_{\mathcal{C}^*\times\mathcal{P}^*} d\widetilde{\mu}(s'', t'')\, f(s'' - \tau s, t'' - \tau t)\, e^{i\tau\,\xi(s,t;s'',t'')}\,\mathrm{T}_\kappa(s'', t'')\,\varphi\,,\end{aligned}$$

where we have used the translation invariance of the Lebesgue measure and the fact that $\forall (s'', t'') \in \mathcal{C}^* \times \mathcal{P}^*$, $\xi(\tau s, \tau t; s'' - \tau s, t'' - \tau t) = \tau\,\xi(s, t; s'', t'')$. Defining $f^{(\tau)}_{(s,t)} \in C_o^\infty(\mathcal{C}^* \times \mathcal{P}^*, \mathbb{C})$ by:

$$\forall (s', t') \in \mathcal{C}^* \times \mathcal{P}^*,\ f^{(\tau)}_{(s,t)}(s', t') := e^{i\tau\,\xi(s,t;s',t')}\, f(s' - \tau s, t' - \tau t)\,,$$

we thus get $\mathrm{T}_\kappa(\tau s, \tau t)\,\varphi\,\{f\} = \varphi\left\{f^{(\tau)}_{(s,t)}\right\}$.

We now define, for any $\tau \in \mathbb{R}$, $f^{1,(\tau)}_{(s,t)}, f^{2,(\tau)}_{(s,t)} \in C_o^\infty(\mathcal{C}^* \times \mathcal{P}^*, \mathbb{C})$ by:

$$\forall (s', t') \in \mathcal{C}^* \times \mathcal{P}^*,\ f^{1,(\tau)}_{(s,t)}(s', t') := \left.\frac{d}{d\tau'} f^{(\tau')}_{(s,t)}(s', t')\right|_{\tau'=\tau} \quad \& \quad f^{2,(\tau)}_{(s,t)}(s', t') := \left.\frac{d}{d\tau'} f^{1,(\tau')}_{(s,t)}(s', t')\right|_{\tau'=\tau}.$$

Then, we have:

$$\forall \tau \in \mathbb{R},\ \forall (s', t') \in \mathcal{C}^* \times \mathcal{P}^*,$$
$$g^{(\tau)}_{(s,t)}(s', t') := \frac{f^{(\tau)}_{(s,t)}(s', t') - f(s', t')}{\tau} - f^{1,(0)}_{(s,t)}(s', t') = \frac{1}{\tau}\int_0^\tau d\tau' \int_0^{\tau'} d\tau''\ \ f^{2,(\tau'')}_{(s,t)}(s', t')\,.$$

Let $K := \{(s' + \tau s, t' + \tau t) \mid (s', t') \in \mathrm{supp}(f),\ \tau \in [-1, 1]\}$. $K$ is compact as the continuous image of the compact $\mathrm{supp}(f) \times [-1, 1]$ and we have:

$$\forall \tau \in [-1, 1],\ \forall (s', t') \notin K,\ g^{(\tau)}_{(s,t)}(s', t') = 0\,.$$

Moreover, the map $s', t'; \tau'' \mapsto f^{2,(\tau'')}_{(s,t)}(s', t')$ is continuous on $\mathcal{C}^* \times \mathcal{P}^* \times \mathbb{R}$, hence bounded on $K \times [-1, 1]$, so there exists $M > 0$ such that:

$$\forall \tau \in [-1, 1]\,,\ \forall (s', t') \in \mathcal{C}^* \times \mathcal{P}^*,\ \left| g^{(\tau)}_{(s,t)}(s', t') \right| \leqslant M\, \mathbf{1}_K(s', t')\,,$$

where $\mathbf{1}_K$ denotes the indicator function of $K$. So, the dominated convergence theorem yields:

$$\lim_{\tau \to 0} \int_{\mathcal{C}^* \times \mathcal{P}^*} d\widetilde{\mu}(s', t')\ \left\| g^{(\tau)}_{(s,t)}(s', t')\, \mathrm{T}_\kappa(s', t')\, \varphi \right\| = 0\,,$$

and therefore:

$$\lim_{\tau \to 0} \left\| \frac{\mathrm{T}_\kappa(\tau s, \tau t)\, \varphi\,\{f\} - \varphi\,\{f\}}{\tau} - \varphi\,\{f_{(s,t)}\} \right\| = 0\,,$$

where:

$$\forall (s', t') \in \mathcal{C}^* \times \mathcal{P}^*,\ f_{(s,t)}(s', t') := f^{1,(0)}_{(s,t)}(s', t') = i\, \xi(s, t; s', t')\, f(s', t') - \left[T_{(s',t')} f\right](s, t)$$

(with $T_{(s',t')} f$ the differential of $f$ at $(s', t')$).

By definition of $\mathrm{X}_\kappa(s, t)$, this implies [23, theorem VIII.7]:

$$\varphi\,\{f\} \in \mathrm{Dom}\big(\mathrm{X}_\kappa(s, t)\big) \quad \& \quad \mathrm{X}_\kappa(s, t)\, \varphi\,\{f\} = -i\, \varphi\,\{f_{(s,t)}\}\,.$$

Hence, $\mathcal{D}_\kappa \subset \mathrm{Dom}\big(\mathrm{X}_\kappa(s, t)\big)$, and, since $f_{(s,t)} \in C^\infty_o(\mathcal{C}^* \times \mathcal{P}^*, \mathbb{C})$ for any $f \in C^\infty_o(\mathcal{C}^* \times \mathcal{P}^*, \mathbb{C})$, $\mathrm{X}_\kappa(s, t) \langle \mathcal{D}_\kappa \rangle \subset \mathcal{D}_\kappa$. Next, $\mathrm{X}_\kappa(s, t)|_{\mathcal{D}_\kappa}$ is symmetric (as the restriction of a self-adjoint operator), and we have, for $\varphi' \in \mathcal{D}_\kappa$:

$$\mathrm{X}_\kappa(s, t)|_{\mathcal{D}_\kappa}\ \varphi' = -i\, \lim_{\tau \to 0} \frac{\mathrm{T}_\kappa(\tau s, \tau t)\, \varphi' - \varphi'}{\tau}\,,$$

so:

$$\begin{aligned}
&\varphi \in \mathrm{Ker}\left( \left[ \mathrm{X}_\kappa(s, t)|_{\mathcal{D}_\kappa} \right]^\dagger \pm i \right) \Leftrightarrow \\
&\qquad \Leftrightarrow\ \forall \varphi' \in \mathcal{D}_\kappa\,,\ -i\, \lim_{\tau \to 0} \frac{\langle \varphi \mid \mathrm{T}_\kappa(\tau s, \tau t)\, \varphi' \rangle - \langle \varphi \mid \varphi' \rangle}{\tau} \mp i\, \langle \varphi \mid \varphi' \rangle = 0 \\
&\qquad \Leftrightarrow\ \forall \varphi' \in \mathcal{D}_\kappa\,,\ \frac{d}{d\tau} \langle \varphi \mid \mathrm{T}_\kappa(\tau s, \tau t)\, \varphi' \rangle \Big|_{\tau = 0} = \mp \langle \varphi \mid \varphi' \rangle \\
&\qquad \Leftrightarrow\ \forall \varphi' \in \mathcal{D}_\kappa\,,\ \forall \tau \in \mathbb{R}\,,\ \frac{d}{d\tau'} \langle \varphi \mid \mathrm{T}_\kappa(\tau' s, \tau' t)\, \varphi' \rangle \Big|_{\tau' = \tau} = \mp \langle \varphi \mid \mathrm{T}_\kappa(\tau s, \tau t)\, \varphi' \rangle
\end{aligned}$$

(using $\mathrm{T}_\kappa(\tau s, \tau t) \langle \mathcal{D}_\kappa \rangle \subset \mathcal{D}_\kappa$ and eq. (A.4.1))

$$\begin{aligned}
&\qquad \Leftrightarrow\ \forall \varphi' \in \mathcal{D}_\kappa\,,\ \forall \tau \in \mathbb{R}\,,\ \langle \varphi \mid \mathrm{T}_\kappa(\tau s, \tau t)\, \varphi' \rangle = e^{\mp \tau}\, \langle \varphi \mid \varphi' \rangle \\
&\qquad \Leftrightarrow\ \forall \varphi' \in \mathcal{D}_\kappa\,,\ \langle \varphi \mid \varphi' \rangle = 0
\end{aligned}$$

(since, for each $\varphi' \in \mathcal{D}_\kappa$, $\tau \mapsto \langle \varphi \mid \mathrm{T}_\kappa(\tau s, \tau t)\, \varphi' \rangle$ is bounded, $\mathrm{T}_\kappa(\tau s, \tau t)$ being unitary for any $\tau \in \mathbb{R}$).

Thus, $\mathrm{Ker}\left( \left[ \mathrm{X}_\kappa(s, t)|_{\mathcal{D}_\kappa} \right]^\dagger \pm i \right) = \mathcal{D}_\kappa^\perp = \{0\}$ (for $\mathcal{D}_\kappa$ is dense in $\mathcal{H}_\kappa$), and therefore $\mathrm{X}_\kappa(s, t)|_{\mathcal{D}_\kappa}$ is essentially self-adjoint [23, theorem VIII.3].

Finally, for any $f \in C^\infty_o(\mathcal{C}^* \times \mathcal{P}^*, \mathbb{C})$, $(s, t), (s', t') \in \mathcal{C}^* \times \mathcal{P}^*$ and $\tau, \tau' \in \mathbb{R}$, we have:

$$f_{(\tau s+\tau' s',\tau t+\tau' t')} = \tau\, f_{(s,t)} + \tau'\, f_{(s',t')}\,,$$

which proves point A.4.3. □

Of course, the usual Schrödinger representation is such a weakly continuous representation of the Weyl algebra, and, in fact, the Stone-von Neumann theorem [27], which we will recall below (prop. A.10, following the proof given in [11, theorem 15a]), tells us that it is the only one (more precisely, it is the only *irreducible* one: an arbitrary weakly continuous representation thus decomposes as a direct sum of independent copies of the Schrödinger representation).

**Proposition A.5** Let $\mathcal{H}_o := L_2(\mathcal{C},\, d\mu)$ with $\mu$ a Lebesgue measure on the finite-dimensional real vector space $\mathcal{C}$. For any $(s, t) \in \mathcal{C}^* \times \mathcal{P}^*$, we define a unitary operator $T_o(s, t)$ by:

$$\forall \varphi \in \mathcal{H}_o,\ \forall x \in \mathcal{C},\ [T_o(s, t)\, \varphi](x) := e^{i\, s(x) + \frac{i}{2}\, t\left(\Xi^{-1}(s)\right)}\, \varphi\left(x + \Xi^{*,-1}(t)\right),$$

where $\Xi^* : \mathcal{C} \to \mathcal{P}^*$ is the dual of the map $\Xi$ and $\Xi^{*,-1} = (\Xi^*)^{-1} = \left(\Xi^{-1}\right)^*$.

$T_o$ is a weakly continuous, unitary representation of $\mathcal{C}^* \times \mathcal{P}^*$.

Let $\mathcal{D}^{\circ}_o$ be the space of smooth, compactly supported, complex-valued functions on $\mathcal{C}$. $\mathcal{D}^{\circ}_o$ is a dense vector subspace in $\mathcal{H}_o$, and, for any $(s, t) \in \mathcal{C}^* \times \mathcal{P}^*$, the linear operator $X^{\circ}_o(s, t)$ defined on $\mathcal{D}^{\circ}_o$ by:

$$\forall \varphi \in \mathcal{D}^{\circ}_o,\ \forall x \in \mathcal{C},\ \left[X^{\circ}_o(s, t)\, \varphi\right](x) := s(x)\, \varphi(x) - i\, [T_x\varphi]\left(\Xi^{*,-1}(t)\right),$$

is essentially self-adjoint (with $T_x\varphi$ the differential of $\varphi$ at $x$). Moreover, we have:

$$\forall (s, t) \in \mathcal{C}^* \times \mathcal{P}^*,\ \mathcal{D}^{\circ}_o \subset \mathrm{Dom}\left(X_o(s, t)\right) \quad \& \quad X^{\circ}_o(s, t) = X_o(s, t)\big|_{\mathcal{D}^{\circ}_o},$$

where, for any $(s, t) \in \mathcal{C}^* \times \mathcal{P}^*$, $X_o(s, t)$ is the self-adjoint generator introduced in prop. A.4.

**Proof** Let $\varphi \in \mathcal{H}_o$ and $(s, t) \in \mathcal{C}^* \times \mathcal{P}^*$. Let $\varphi'$ be given by:

$$\forall x \in \mathcal{C},\ \varphi'(x) := e^{i\, s(x) + \frac{i}{2}\, t\left(\Xi^{-1}(s)\right)}\, \varphi\left(x + \Xi^{*,-1}(t)\right).$$

Using the translation invariance of the Lebesgue measure, we have:

$$\int_{\mathcal{C}} d\mu(x)\ |\varphi'(x)|^2 = \int_{\mathcal{C}} d\mu(x')\ \left|\varphi\left(x'\right)\right|^2 = \|\varphi\|^2,$$

hence $T_o(s, t)$ is well-defined as a linear operator $\mathcal{H}_o \to \mathcal{H}_o$ and is isometric.

Moreover, we have $T_o(0, 0) = \mathrm{id}_{\mathcal{H}_o}$ and, for any $(s_1, t_1), (s_2, t_2) \in \mathcal{C}^* \times \mathcal{P}^*$:

$\forall \varphi \in \mathcal{H}_o,\ \forall x \in \mathcal{C},$

$$\begin{aligned} [T_o(s_1, t_1)\, T_o(s_2, t_2)\, \varphi](x) &= e^{i\, s_1(x) + \frac{i}{2}\, t_1\left(\Xi^{-1}(s_1)\right)}\, [T_o(s_2, t_2)\, \varphi]\left(x + \Xi^{*,-1}(t_1)\right) \\ &= e^{i\, (s_1+s_2)(x) + \frac{i}{2}\left[(t_1+t_2)\left(\Xi^{-1}(s_1+s_2)\right) - t_2\left(\Xi^{-1}(s_1)\right) + t_1\left(\Xi^{-1}(s_2)\right)\right]}\, \varphi\left(x + \Xi^{*,-1}(t_1 + t_2)\right) \\ &= e^{i\, \xi(s_1, t_1; s_2, t_2)}\, [T_o(s_1 + s_2, t_1 + t_2)\, \varphi](x), \end{aligned}$$

where we have used $s_2\left(\Xi^{*,-1}(t_1)\right) = t_1\left(\Xi^{-1}(s_2)\right)$. In particular, we thus have, for any $(s, t) \in \mathcal{C}^* \times \mathcal{P}^*$, $T_o(s, t)\, T_o(-s, -t) = T_o(-s, -t)\, T_o(s, t) = \mathrm{id}_{\mathcal{H}_o}$, so $T_o(s, t)$ is invertible, and, being isometric, it is a unitary operator on $\mathcal{H}_o$.

Let $\varphi \in \mathcal{D}^{\circ}_o$, $(s,t) \in \mathcal{C}^* \times \mathcal{P}^*$ and $\epsilon > 0$. In particular, $\varphi$ is bounded and has compact support, so:

$$M := \int_{\mathcal{C}} d\mu(x)\ |\varphi(x)| < \infty\,.$$

Moreover, $\varphi$ is also absolutely continuous, so there exists an open neighborhood $U_1$ of 0 in $\mathcal{C}$ such that:

$$\forall x \in \mathcal{C},\ \forall x' \in U_1,\ |\varphi(x + x') - \varphi(x)| < \frac{\epsilon}{4M+1}\,.$$

The map $\Xi^{*,-1}$ being linear on the finite-dimensional vector space $\mathcal{P}^*$, it is continuous, therefore $U_2 := \Xi^* \langle U_1 \rangle$ is an open neighborhood of 0 in $\mathcal{P}^*$.

Next, the map $z : \mathcal{C} \times \mathcal{C}^* \times \mathcal{P}^* \to \mathbb{C}$, defined by:

$$\forall x \in \mathcal{C},\ \forall (s',t') \in \mathcal{C}^* \times \mathcal{P}^*,\ z(x;\, s',t') := e^{-i\xi(s,t;\,s',t') + i\,s'(x) + \frac{i}{2}\,t'\left(\Xi^{-1}(s')\right)}\,,$$

is continuous, so, for any $x \in \mathcal{C}$, there exists an open neighborhood $V_x$ of $x$ in $\mathcal{C}$ and an open neighborhood $U_3^{(x)}$ of $(0,0)$ in $\mathcal{C}^* \times \mathcal{P}^*$ such that:

$$\forall x \in V_x,\ \forall (s',t') \in U_3^{(x)},\ |z(x;\, s',t') - 1| < \frac{\epsilon}{4\left\|\varphi\right\|^2 + 1}\,.$$

Since the support $\operatorname{supp}(\varphi)$ of $\varphi$ is compact, there exist $n \in \mathbb{N}$ and $x_1, \dots, x_n \in \mathcal{C}$ such that $\operatorname{supp}(\varphi) \subset \bigcup_{i=1}^n V_{x_i}$. Thus, defining $U_3 := \bigcap_{i=1}^n U_3^{(x_i)}$, $U_3$ is an open neighborhood of $(0,0)$ in $\mathcal{C}^* \times \mathcal{P}^*$, and:

$$\forall x \in \operatorname{supp}(\varphi),\ \forall (s',t') \in U_3,\ |z(x;\, s',t') - 1| < \frac{\epsilon}{4\left\|\varphi\right\|^2 + 1}\,.$$

Then, $U_4 := (\mathcal{C}^* \times U_2) \cap U_3$ is an open neighborhood of $(0,0)$ in $\mathcal{C}^* \times \mathcal{P}^*$, and, for any $(s',t') \in U_4$:

$$\left\|\mathrm{T}_o(s+s', t+t')\,\varphi - \mathrm{T}_o(s,t)\,\varphi\right\|^2$$

$$= 2\left\|\varphi\right\|^2 - 2\,\mathrm{Re}\left\langle \varphi \,\middle|\, e^{-i\xi(s,t;\,s',t')}\,\mathrm{T}_o(s',t')\,\varphi \right\rangle$$

(where we have used $\mathrm{T}_o(s+s', t+t') = e^{-i\xi(s,t;\,s',t')}\,\mathrm{T}_o(s,t)\,\mathrm{T}_o(s',t')$ and the fact that both $\mathrm{T}_o(s,t)$ and $\mathrm{T}_o(s',t')$ are unitary)

$$\leqslant 2\left|\left\langle \varphi \,\middle|\, e^{-i\xi(s,t;\,s',t')}\,\mathrm{T}_o(s',t')\,\varphi \right\rangle - \left\|\varphi\right\|^2\right|$$

$$\leqslant 2\int_{\mathcal{C}} d\mu(x)\ |\varphi(x)|\ \left|z(x;\, s',t')\,\varphi\left(x + \Xi^{*,-1}(t')\right) - \varphi(x)\right|$$

$$\leqslant 2\int_{\mathcal{C}} d\mu(x)\ |\varphi(x)|\ \left|\varphi\left(x + \Xi^{*,-1}(t')\right) - \varphi(x)\right| + 2\int_{\operatorname{supp}(\varphi)} d\mu(x)\ |\varphi(x)|^2\ |z(x;\, s',t') - 1|$$

(for $\forall x \in \mathcal{C}$, $|z(x;\, s',t')| = 1$)

$$< \epsilon\,.$$

Hence, for any $\varphi \in \mathcal{D}^{\circ}_o$, the map $(s,t) \mapsto \mathrm{T}_o(s,t)\,\varphi$ is continuous $\mathcal{C}^* \times \mathcal{P}^* \to \mathcal{H}_o$.

Now, the smooth, compactly supported functions are dense in $\mathcal{H}_o$, so for any $\varphi, \varphi' \in \mathcal{H}_o$, and any $\epsilon > 0$, there exist $\widetilde{\varphi} \in \mathcal{D}^{\circ}_o$ such that:

$$\|\varphi - \widetilde{\varphi}\| < \frac{\epsilon}{3\,\|\varphi'\| + 1}\,.$$

Next, for any $(s,t) \in \mathcal{C}^* x \mathcal{P}^*$, there exists an open neighborhood $U_5$ of $(s,t)$ in $\mathcal{C}^* x \mathcal{P}^*$ such that:

$$\forall s'', t'' \in U_5,\ \|\mathrm{T}_o(s'',t'')\,\widetilde{\varphi} - \mathrm{T}_o(s,t)\,\widetilde{\varphi}\| < \frac{\epsilon}{3\,\|\varphi'\| + 1}\,.$$

Thus, for any $s'', t'' \in U_5$,

$$|\langle \varphi' \mid \mathrm{T}_o(s'',t'')\,\varphi\rangle - \langle \varphi' \mid \mathrm{T}_o(s,t)\,\varphi\rangle| \leqslant \|\varphi'\| \,\left[2\,\|\varphi - \widetilde{\varphi}\| + \|\mathrm{T}_o(s'',t'')\,\widetilde{\varphi} - \mathrm{T}_o(s,t)\,\widetilde{\varphi}\|\,\right] < \epsilon$$

(since $\mathrm{T}_o(s,t)$ and $\mathrm{T}_o(s'',t'')$ are both unitary), which proves def. A.3.4.

Let $\varphi \in \mathcal{D}^{\circ}_o$ and $(s,t) \in \mathcal{C}^* \times \mathcal{P}^*$. For any $\tau \in \mathbb{R}$, we have $[\mathrm{T}_o(\tau s, \tau t)\,\varphi] = \varphi^{(\tau)}_{(s,t)} \in \mathcal{D}^{\circ}_o$, where:

$$\forall x \in \mathcal{C},\ \varphi^{(\tau)}_{(s,t)}(x) := e^{i\,\tau\,s(x) + \frac{i}{2}\,\tau^2\,t\left(\Xi^{-1}(s)\right)}\,\varphi\left(x + \tau\,\Xi^{*,-1}(t)\right),$$

and $\mathrm{X}^{\circ}_o(s,t)\,\varphi = -i\,\varphi_{(s,t)} \in \mathcal{D}^{\circ}_o$, where:

$$\forall x \in \mathcal{C},\ \varphi_{(s,t)}(x) := \left.\frac{d}{d\tau}\varphi^{(\tau)}_{(s,t)}(x)\right|_{\tau=0}.$$

In particular, $\mathrm{X}^{\circ}_o(s,t)\,\varphi$ belongs to $\mathcal{H}_o$, thus $\mathrm{X}^{\circ}_o(s,t)$ is well-defined as a linear operator on $\mathcal{D}^{\circ}_o$.

Moreover, like in the proof of prop. A.4, the dominated convergence theorem yields:

$$\lim_{\tau\to 0}\int_{\mathcal{C}} d\mu(x)\,\left|\frac{\varphi^{(\tau)}_{(s,t)}(x) - \varphi(x)}{\tau} - \varphi_{(s,t)}(x)\right|^2 = 0\,,$$

and therefore:

$$\lim_{\tau\to 0}\left\|\frac{\mathrm{T}_o(\tau s, \tau t)\,\varphi - \varphi}{\tau} - i\,\mathrm{X}^{\circ}_o(s,t)\,\varphi\right\| = 0\,.$$

Hence, we have [23, theorem VIII.7]:

$$\varphi \in \mathrm{Dom}\,(\mathrm{X}_o(s,t)) \quad \& \quad \mathrm{X}^{\circ}_o(s,t)\,\varphi = \mathrm{X}_o(s,t)\,\varphi\,,$$

so $\mathcal{D}^{\circ}_o \subset \mathrm{Dom}\,(\mathrm{X}_o(s,t))$ and $\mathrm{X}^{\circ}_o(s,t) = \mathrm{X}_o(s,t)|_{\mathcal{D}^{\circ}_o}$.

Finally, using: □

$$\forall \varphi \in \mathcal{D}^{\circ}_o,\ \mathrm{X}^{\circ}_o(s,t)\,\varphi = -i\,\lim_{\tau\to 0}\frac{\mathrm{T}_o(\tau s, \tau t)\,\varphi - \varphi}{\tau}\,,$$

together with $\forall \tau \in \mathbb{R},\ \mathrm{T}_o(\tau s, \tau t)\left\langle \mathcal{D}^{\circ}_o \right\rangle \subset \mathcal{D}^{\circ}_o$ and the density of $\mathcal{D}^{\circ}_o$ in $\mathcal{H}_o$, we get, via the same argument as in the proof of prop. A.4, that $\mathrm{X}^{\circ}_o$ is essentially self-adjoint.

## A.2 The Wigner-Weyl transform

We now come to the Wigner transform itself. The Wigner characteristic function associated to a quantum state maps to a point $(s,t)$ the expectation value of $\mathrm{T}_\kappa(s,t)$ in this state. In particular, it is defined as a function on the *dual* space $\mathcal{C}^* \times \mathcal{P}^*$, since, as underlined above, the operators $\mathrm{T}_\kappa(s,t)$ are labeled by elements of the dual. Note that if we would complete the Wigner transform, by Fourier-transforming this characteristic function, we would pass to the dual again and obtain a quasi-probability defined on the *phase space*, in correct analogy with the probability distributions considered in classical statistical physics.

Insisting on a strict distinction between $\mathcal{M}$ and its dual, may seems an unnecessary care in the case of a linear space, since we could simply choose some identification between the two. However, it allows us to keep track of the nature of the objects we are considering, and in particular of the natural direction of their transformations under morphisms. Probability measures are naturally push-forwarded along a projection, while functions on $\mathcal{M}$ should be pull-backed, but functions on the dual, being pull-backed by the dual map, effectively flow in the same direction as probability measures (viz. def. 2.2): thus, defining characteristic functions as a functions on the dual is consistent with the fact that they encode the same information as probability measures.

We begin by characterizing the image of the (first half of) the Wigner transform. The positivity condition eq. (A.6.1) reflects the requirement for density matrices to be non-negative operators and it manifests the truly quantum nature of the Wigner quasi-probability: a similar equation for a classical characteristic function, reflecting the positivity of its associated probability distribution, would not have the twisting phase $e^{i\xi(s,t;s',t')}$. In other words, eq. (A.6.1) captures the non-commutation of the position and momentum variables (viz. prop. A.2), and it will play a central role in the derivation of the negative result in subsection 2.2. Finally, the continuity of the Wigner characteristic function expresses the fact that it comes from a density matrix on a *weakly continuous* representation (this is closely related to the characterization of normal states on a $W^*$-algebra, see [25, corollary III.3.11]).

**Proposition A.6** A continuous function $W : \mathcal{C}^* \times \mathcal{P}^* \to \mathbb{C}$ is said of positive type if, for any $\iota \in \mathcal{I}$:

$$\sum_{(s,t),(s',t') \in \mathcal{C}^* \times \mathcal{P}^*} \overline{\iota(s',t')}\, \iota(s,t)\, e^{i\xi(s,t;s',t')}\, W(s-s',t-t') \geqslant 0. \tag{A.6.1}$$

In particular, this implies:

$$\forall (s,t) \in \mathcal{C}^* \times \mathcal{P}^*,\ W(-s,-t) = \overline{W(s,t)} \quad \& \quad |W(s,t)| \leqslant W(0,0)\,.$$

We denote by $\mathcal{W}$ the space of all continuous functions of positive type.

**Proof** Let $(s,t) \in \mathcal{C}^* \times \mathcal{P}^*$ and let $\lambda \in \mathbb{C}$. Applying eq. (A.6.1) to $\delta_{(0,0)} + \lambda\, \delta_{(s,t)} \in \mathcal{I}$ yields:

$$\left(1 + |\lambda|^2\right) W(0,0) + \lambda\, W(s,t) + \overline{\lambda}\, W(-s,-t) \geqslant 0\,,$$

where we have used $\xi(0,0;0,0) = \xi(s,t;0,0) = \xi(0,0;s,t) = \xi(s,t;s,t) = 0$. In particular, $W(0,0)$ is real positive (as follows from setting $\lambda = 0$). Thus, $\lambda\, W(s,t) + \overline{\lambda}\, W(-s,-t)$ is real for any $\lambda \in \mathbb{C}$, so $\overline{W(s,t)} = W(-s,-t)$. Now, $\left(1 + |\lambda|^2\right) W(0,0) + 2\,\mathrm{Re}\left[\lambda\, W(s,t)\right]$ is positive for any $\lambda \in \mathbb{C}$, hence $|W(s,t)| \leqslant W(0,0)$. □

**Proposition A.7** Let $\mathcal{H}_\kappa$, $\mathrm{T}_\kappa$ be as in def. A.3 and let $\rho$ be a traceclass, (self-adjoint) positive semi-definite operator on $\mathcal{H}_\kappa$. The map $W_\rho : \mathcal{C}^* \times \mathcal{P}^* \to \mathbb{C}$, defined by:

$$\forall (s,t) \in \mathcal{C}^* \times \mathcal{P}^*,\ W_\rho(s,t) := \mathrm{Tr}_{\mathcal{H}_\kappa}\left[\rho\, \mathrm{T}_\kappa(s,t)\right],$$

is a continuous function of positive type.

**Proof** For any $(s,t) \in \mathcal{C}^* \times \mathcal{P}^*$, $\mathrm{T}_\kappa(s,t)$ is a bounded operator, hence $\rho\, \mathrm{T}_\kappa(s,t)$ is traceclass, therefore $W_\rho$ is well-defined. From the spectral theorem, there exist $N \in \mathbb{N} \cup \{\infty\}$, an orthonormal family $(\varphi_k)_{k \leqslant N}$ in $\mathcal{H}_\kappa$ and a family of non-negative reals $(p_k)_{k \leqslant N}$ such that:

$$\rho = \sum_{k \leqslant N} p_k \, |\varphi_k\,\rangle\langle\,\varphi_k| \quad \& \quad \sum_{k \leqslant N} p_k = \mathrm{Tr}_{\mathcal{H}_\kappa} \rho.$$

Let $(s,t) \in \mathcal{C}^* \times \mathcal{P}^*$ and $\epsilon > 0$. Let $K \in \mathbb{N}$ with $K \leqslant N$ such that:

$$\sum_{K < k \leqslant N} p_k < \frac{\epsilon}{3}.$$

From def. A.3.4, the map $(s,t) \mapsto \langle \varphi_k \mid \mathrm{T}_\kappa(s,t)\, \varphi_k \rangle$ is continuous for every $k \leqslant K$. Hence, for each $k \leqslant K$, there exists an open neighborhood $U_k$ of $(s,t)$ in $\mathcal{C}^* \times \mathcal{P}^*$ such that:

$$\forall (s',t') \in U_k\,,\ \big| \langle \varphi_k \mid \mathrm{T}_\kappa(s',t')\, \varphi_k \rangle - \langle \varphi_k \mid \mathrm{T}_\kappa(s,t)\, \varphi_k \rangle \big| < \frac{\epsilon}{3\,\mathrm{Tr}_{\mathcal{H}_\kappa} \rho + 1}.$$

Since $K$ is finite, $U := \bigcap_{k \leqslant K} U_k$ is an open neighborhood of $(s,t)$ in $\mathcal{C}^* \times \mathcal{P}^*$ and, for any $(s',t') \in U$, we have:

$$\begin{aligned}
|W_\rho(s',t') - W_\rho(s,t)| &\leqslant \sum_{k \leqslant N} p_k \big| \langle \varphi_k \mid \mathrm{T}_\kappa(s',t')\, \varphi_k \rangle - \langle \varphi_k \mid \mathrm{T}_\kappa(s,t)\, \varphi_k \rangle \big| \\
&\leqslant 2 \sum_{K < k \leqslant N} p_k + \sum_{k \leqslant K} p_k \big| \langle \varphi_k \mid \mathrm{T}_\kappa(s',t')\, \varphi_k \rangle - \langle \varphi_k \mid \mathrm{T}_\kappa(s,t)\, \varphi_k \rangle \big| \\
&< 2\,\frac{\epsilon}{3} + \sum_{k \leqslant K} p_k \, \frac{\epsilon}{3\,\mathrm{Tr}_{\mathcal{H}_\kappa} \rho + 1} \leqslant \epsilon\,,
\end{aligned}$$

where we have used that $\mathrm{T}_\kappa(s,t)$ and $\mathrm{T}_\kappa(s',t')$ are unitary operators and each $\varphi_k$ for $k \leqslant N$ is a normalized vector, hence $|\langle \varphi_k \mid \mathrm{T}_\kappa(s,t)\, \varphi_k \rangle| \leqslant 1$ and $|\langle \varphi_k \mid \mathrm{T}_\kappa(s',t')\, \varphi_k \rangle| \leqslant 1$. Thus, $W_\rho$ is a continuous function $\mathcal{C}^* \times \mathcal{P}^* \to \mathbb{C}$.

Let $\iota \in \mathcal{I}$. We have:

$$\sum_{(s,t),(s',t') \in \mathcal{C}^* \times \mathcal{P}^*} \overline{\iota(s',t')}\, \iota(s,t)\, e^{i \xi(s,t;s',t')}\, W_\rho(s-s', t-t') \;=\; \sum_{(s'',t'') \in \mathcal{C}^* \times \mathcal{P}^*} [\iota^* \star \iota](s'',t'')\, W_\rho(s'',t'')$$

$$= \mathrm{Tr}_{\mathcal{H}_\kappa} [\rho\, \mathrm{T}_\kappa \{\iota^* \star \iota\}]$$

(where $\mathrm{T}_\kappa \{\iota'\}$ has been defined for $\iota' \in \mathcal{I}$ in def. A.3, as a *finite* linear combination of unitary operators belonging to $\mathrm{T}_\kappa \langle \mathcal{C}^* \times \mathcal{P}^* \rangle$)

$$= \sum_{k \leqslant N} p_k \left\langle \varphi_k \,\middle|\, \mathrm{T}_\kappa \{\iota\}^\dagger\, \mathrm{T}_\kappa \{\iota\}\, \varphi_k \right\rangle \geqslant 0.$$

Therefore, $W_\rho$ fulfills eq. (A.6.1). □

Like its classical analogue, the Wigner characteristic function is related to the moment-generating function (precisely, the latter, when it exists, is the restriction to the complex axis of the analytical expansion of the former). In particular, a quantum state exhibits finite variances for the elementary observables $X_\kappa(s,t)$ if its characteristic function is twice-differentiable at 0, and the covariance matrix can then be recovered from the corresponding Hessian.

**Proposition A.8** Let $\mathcal{H}_\kappa$, $T_\kappa$, $\rho$ and $W_\rho$ be as in prop. A.7. We moreover assume that there exist a linear form $W_\rho^{(1)}$ and a symmetric bilinear form $W_\rho^{(2)}$ on $\mathcal{C}^* \times \mathcal{P}^*$ such that:

$$\forall (s,t) \in \mathcal{C}^* \times \mathcal{P}^*,\ W_\rho(\tau s, \tau t) = W_\rho(0,0) + i\,\tau\, W_\rho^{(1)}(s,t) - \frac{\tau^2}{2} W_\rho^{(2)}(s,t;s,t) + o(\tau^2)\,. \qquad \text{(A.8.1)}$$

Then, for any $(s,t)$, $(s',t') \in \mathcal{C}^* \times \mathcal{P}^*$, the operators $\rho\, X_\kappa(s,t)$ (densely defined on $\mathrm{Dom}\big(X_\kappa(s,t)\big)$) and $\frac{X_\kappa(s,t)\,\rho\,X_\kappa(s',t') + X_\kappa(s',t')\,\rho\,X_\kappa(s,t)}{2}$ (which can be defined at least as a sesquilinear form on the dense subset $\mathcal{D}_\kappa \subset \mathcal{H}_\kappa$) admit a traceclass extension on $\mathcal{H}_\kappa$, with:

$$\mathrm{Tr}_{\mathcal{H}_\kappa} \rho\, X_\kappa(s,t) = W_\rho^{(1)}(s,t) \quad \& \quad \mathrm{Tr}_{\mathcal{H}_\kappa} \frac{X_\kappa(s,t)\,\rho\,X_\kappa(s',t') + X_\kappa(s',t')\,\rho\,X_\kappa(s,t)}{2} = W_\rho^{(2)}(s,t;s',t')\,.$$

**Proof** Let $(s,t) \in \mathcal{C}^* \times \mathcal{P}^*$. For any $\tau \neq 0$, we define:

$$Y_\kappa^{(\tau)}(s,t) := \frac{2\,\mathrm{id}_{\mathcal{H}_\kappa} - T_\kappa(\tau s, \tau t) - T_\kappa(\tau s, \tau t)^\dagger}{\tau^2}\,.$$

Since $T_\kappa(\tau s, \tau t)$ is a unitary operator, $Y_\kappa^{(\tau)}(s,t)$ is a bounded, (self-adjoint) positive semi-definite operator on $\mathcal{H}_\kappa$. Using $T_\kappa(0,0) = \mathrm{id}_{\mathcal{H}_\kappa}$ and $T_\kappa(\tau s, \tau t)^\dagger = T_\kappa(-\tau s, -\tau t)$, we get:

$$\mathrm{Tr}_{\mathcal{H}_\kappa} \rho\, Y_\kappa^{(\tau)}(s,t) = \frac{2\,W_\rho(0,0) - W_\rho(\tau s, \tau t) - W_\rho(-\tau s, -\tau t)}{\tau^2}\,.$$

Hence, eq. (A.8.1) implies:

$$\lim_{\tau \to 0} \mathrm{Tr}_{\mathcal{H}_\kappa} \rho\, Y_\kappa^{(\tau)}(s,t) = W_\rho^{(2)}(s,t;s,t)\,.$$

Let $\tau_1 > 0$ such that:

$$\forall \tau \in \,]0, \tau_1[\;,\ \mathrm{Tr}_{\mathcal{H}_\kappa} \rho\, Y_\kappa^{(\tau)}(s,t) \leqslant W_\rho^{(2)}(s,t;s,t) + 1 =: A\,.$$

Applying the spectral theorem to the self-adjoint operator $X_\kappa(s,t)$, we denote by $d\Pi_\kappa(s,t)$ its projection-valued measure, with:

$$X_\kappa(s,t) = \int_{-\infty}^{+\infty} \sigma\, d\Pi_\kappa(s,t)[\sigma]\,.$$

Then, for any $\tau \neq 0$, we have, using $T_\kappa(\tau s, \tau t) = \exp\left(i\,\tau\, X_\kappa(s,t)\right)$:

$$Y_\kappa^{(\tau)}(s,t) := \frac{2}{\tau^2} \int_{-\infty}^{+\infty} [1 - \cos(\tau\sigma)]\, d\Pi_\kappa(s,t)[\sigma]\,.$$

There exists $\epsilon > 0$ such that:

$$\forall x \in \,]-\epsilon, \epsilon[\;,\ 1 - \cos(x) > \frac{x^2}{4}\,,$$

hence, for any $\tau \neq 0$, the operator:

$$\mathrm{Y}_\kappa^{(\tau)}(s,t) - \frac{2}{\tau^2}\int_{-\epsilon/\tau}^{\epsilon/\tau} \frac{(\tau\sigma)^2}{4}\, d\Pi_\kappa(s,t)[\sigma],$$

is a bounded, (self-adjoint) positive semi-definite operator on $\mathcal{H}_\kappa$. Defining, for any $B \leqslant B' \in \mathbb{R}$, the spectral projector $\Pi_\kappa^{[B,B']}(s,t) := \int_B^{B'} d\Pi_\kappa(s,t)[\sigma]$, we thus have:

$$\forall B > \frac{\epsilon}{\tau_1}\,,\ \mathrm{Tr}_{\mathcal{H}_\kappa}\, \rho\, \Pi_\kappa^{[-B,B]}(s,t)\, \mathrm{X}_\kappa(s,t)^2\, \Pi_\kappa^{[-B,B]}(s,t) \leqslant 2\,A\,. \tag{A.8.2}$$

Now, from the spectral theorem, there exist $N \in \mathbb{N} \cup \{\infty\}$, an orthonormal family $(\varphi_k)_{k \leqslant N}$ in $\mathcal{H}_\kappa$ and a family of strictly positive reals $(p_k)_{k\leqslant N}$ such that:

$$\rho = \sum_{k\leqslant N} p_k\, |\varphi_k\,\rangle\langle\,\varphi_k| \quad \& \quad \sum_{k\leqslant N} p_k = \mathrm{Tr}_{\mathcal{H}_\kappa}\, \rho\,.$$

For each $k \leqslant N$ and each $B \geqslant 0$, we have:

$$\left\langle \varphi_k \,\middle|\, \Pi_\kappa^{[-B,B]}(s,t)\, \mathrm{X}_\kappa(s,t)^2\, \Pi_\kappa^{[-B,B]}(s,t)\, \varphi_k \right\rangle = \left\| \mathrm{X}_\kappa(s,t)\, \Pi_\kappa^{[-B,B]}(s,t)\, \varphi_k \right\|^2 \geqslant 0,$$

so, for any $B > \epsilon/\tau_1$ and any $k \leqslant N$,

$$\left\| \mathrm{X}_\kappa(s,t)\, \Pi_\kappa^{[-B,B]}(s,t)\, \varphi_k \right\|^2 \leqslant \frac{2\,A}{p_k}\,.$$

Therefore, $\varphi_k \in \mathrm{Dom}\big(\mathrm{X}_\kappa(s,t)\big)$. Moreover, we have [23, theorem VIII.7]:

$$\lim_{\tau\to 0} \frac{\mathrm{T}_\kappa(\tau s, \tau t)\, \varphi_k - \varphi_k}{\tau} = i\, \mathrm{X}_\kappa(s,t)\, \varphi_k\,,$$

hence, we get:

$$\lim_{\tau\to 0} \left\langle \varphi_k \,\middle|\, \mathrm{Y}_\kappa^{(\tau)}(s,t)\, \varphi_k \right\rangle = \lim_{\tau\to 0} \left\| \frac{\mathrm{T}_\kappa(\tau s, \tau t)\, \varphi_k - \varphi_k}{\tau} \right\|^2 = \|\mathrm{X}_\kappa(s,t)\, \varphi_k\|^2\,.$$

Fatou's lemma then yields:

$$\sum_{k\leqslant N} p_k\, \|\mathrm{X}_\kappa(s,t)\, \varphi_k\|^2 \leqslant \liminf_{\tau\to 0} \sum_{k\leqslant N} p_k\, \left\langle \varphi_k \,\middle|\, \mathrm{Y}_\kappa^{(\tau)}(s,t)\, \varphi_k \right\rangle = W_\rho^{(2)}(s,t;\, s,t)\,.$$

Next, this provides the bound:

$$\sum_{k\leqslant N} p_k\, \|\varphi_k\|\, \|\mathrm{X}_\kappa(s,t)\, \varphi_k\| \leqslant \sqrt{W_\rho^{(2)}(s,t;\, s,t)\, \mathrm{Tr}_{\mathcal{H}_\kappa}\, \rho}\,,$$

where we have used $\|\varphi_k\| = 1$ and the Cauchy-Schwarz inequality (with $\sum_{k\leqslant N} p_k = \mathrm{Tr}_{\mathcal{H}_\kappa}\, \rho$). Hence, we can define a bounded operator $\{\rho\, \mathrm{X}_\kappa(s,t)\}$ on $\mathcal{H}_\kappa$ by:

$$\{\rho\, \mathrm{X}_\kappa(s,t)\} := \sum_{k\leqslant N} p_k\, |\varphi_k\,\rangle\langle\, \mathrm{X}_\kappa(s,t)\, \varphi_k|\,,$$

and this operator coincides with $\rho\, \mathrm{X}_\kappa(s,t)$ on $\mathrm{Dom}\big(\mathrm{X}_\kappa(s,t)\big)$, since $\mathrm{X}_\kappa(s,t)$ is self-adjoint. In addition, we have:

$$\big\|\, \{\rho\, \mathrm{X}_\kappa(s,t)\}\, \big\|_1 \leqslant \sum_{k\leqslant N} p_k\, \big\|\, |\varphi_k\,\rangle\langle\, \mathrm{X}_\kappa(s,t)\, \varphi_k|\, \big\|_1 = \sum_{k\leqslant N} p_k\, \|\varphi_k\|\, \|\mathrm{X}_\kappa(s,t)\, \varphi_k\| < \infty\,,$$

where $\|\cdot\|_1$ denotes the trace norm [23, theorem VI.20], so $\{\rho\, X_\kappa(s,t)\}$ is traceclass. Its trace is given by:

$$\mathrm{Tr}_{\mathcal{H}_\kappa}\ \{\rho\, X_\kappa(s,t)\} = \sum_{k\leqslant N} p_k\ \langle X_\kappa(s,t)\,\varphi_k \mid \varphi_k\rangle = \sum_{k\leqslant N} p_k\ \langle \varphi_k \mid X_\kappa(s,t)\,\varphi_k\rangle\,.$$

Now, from:

$$\forall x\in\mathbb{R},\ \left|e^{ix}-1\right|\leqslant |x|\,,$$

together with the spectral decomposition of $X_\kappa(s,t)$, we get, for any $\tau\neq 0$:

$$\left\|\frac{T_\kappa(\tau s,\tau t)\,\varphi_k-\varphi_k}{\tau}\right\|\leqslant \|X_\kappa(s,t)\,\varphi_k\|\,.$$

Thus, the dominated convergence theorem yields:

$$\begin{aligned} i\sum_{k\leqslant N} p_k\ \langle\varphi_k \mid X_\kappa(s,t)\,\varphi_k\rangle &= \sum_{k\leqslant N} p_k\ \lim_{\tau\to 0}\left\langle \varphi_k\ \middle|\ \frac{T_\kappa(\tau s,\tau t)\,\varphi_k-\varphi_k}{\tau}\right\rangle \\ &= \lim_{\tau\to 0}\sum_{k\leqslant N} p_k\ \left\langle \varphi_k\ \middle|\ \frac{T_\kappa(\tau s,\tau t)\,\varphi_k-\varphi_k}{\tau}\right\rangle, \end{aligned}$$

which can be rewritten, using eq. (A.8.1), as:

$$\mathrm{Tr}_{\mathcal{H}_\kappa}\ \{\rho\, X_\kappa(s,t)\} = W^{(1)}_\rho(s,t)\,.$$

(note that $W_\rho(0,0)=\mathrm{Tr}_{\mathcal{H}_\kappa}\rho$ for $T_\kappa(0,0)=\mathrm{id}_{\mathcal{H}_\kappa}$).

Similarly, the inequality:

$$\sum_{k\leqslant N} p_k\ \|X_\kappa(s,t)\,\varphi_k\|^2\leqslant W^{(2)}_\rho(s,t;\ s,t)<\infty\,,$$

ensures that we can define a traceclass operator $\{X_\kappa(s,t)\,\rho\, X_\kappa(s,t)\}$ on $\mathcal{H}_\kappa$ by:

$$\{X_\kappa(s,t)\,\rho\, X_\kappa(s,t)\} := \sum_{k\leqslant N} p_k\ |X_\kappa(s,t)\,\varphi_k\ \rangle\langle\ X_\kappa(s,t)\,\varphi_k|\,,$$

and we have:

$$\forall\varphi',\varphi''\in\mathrm{Dom}\big(X_\kappa(s,t)\big),\ \langle\varphi''\mid\{X_\kappa(s,t)\,\rho\, X_\kappa(s,t)\}\ \varphi'\rangle = \langle X_\kappa(s,t)\,\varphi''\mid\rho\, X_\kappa(s,t)\,\varphi'\rangle\,.$$

Moreover, using again the dominated convergence theorem, with:

$$\forall\tau\neq 0,\ \left\langle\varphi_k\ \middle|\ Y^{(\tau)}_\kappa(s,t)\,\varphi_k\right\rangle = \left\|\frac{T_\kappa(\tau s,\tau t)\,\varphi_k-\varphi_k}{\tau}\right\|^2\leqslant\|X_\kappa(s,t)\,\varphi_k\|^2\,,$$

we get:

$$\mathrm{Tr}_{\mathcal{H}_\kappa}\ \{X_\kappa(s,t)\,\rho\, X_\kappa(s,t)\} = \lim_{\tau\to 0}\mathrm{Tr}_{\mathcal{H}_\kappa}\,\rho\, Y^{(\tau)}_\kappa(s,t) = W^{(2)}_\rho(s,t;\ s,t)\,.$$

Finally, for any $(s,t)$, $(s',t')\in\mathcal{C}^*\times\mathcal{P}^*$, we have, thanks to prop. A.4.3:

$$\begin{aligned} &\forall\varphi,\varphi'\in\mathcal{D}_\kappa\,,\ \langle X_\kappa(s,t)\,\varphi''\mid\rho\, X_\kappa(s',t')\,\varphi'\rangle + \langle X_\kappa(s',t')\,\varphi''\mid\rho\, X_\kappa(s,t)\,\varphi'\rangle = \\ &\qquad = \langle\varphi''\mid\{X_\kappa(s+s',t+t')\,\rho\, X_\kappa(s+s',t+t')\}\ \varphi'\rangle + \\ &\qquad\quad - \langle\varphi''\mid\{X_\kappa(s,t)\,\rho\, X_\kappa(s,t)\}\ \varphi'\rangle - \langle\varphi''\mid\{X_\kappa(s',t')\,\rho\, X_\kappa(s',t')\}\ \varphi'\rangle\,. \end{aligned}$$

Hence, $\frac{\{X_\kappa(s+s',t+t')\,\rho\, X_\kappa(s+s',t+t')\} - \{X_\kappa(s,t)\,\rho\, X_\kappa(s,t)\} - \{X_\kappa(s',t')\,\rho\, X_\kappa(s',t')\}}{2}$ provides a traceclass extension of the sesquilinear form $\frac{X_\kappa(s,t)\,\rho\, X_\kappa(s',t') + X_\kappa(s',t')\,\rho\, X_\kappa(s,t)}{2}$ on $\mathcal{D}_\kappa$ and its trace is given by:

$$\mathrm{Tr}_{\mathcal{H}_\kappa} \frac{\{X_\kappa(s+s',t+t')\,\rho\, X_\kappa(s+s',t+t')\} - \{X_\kappa(s,t)\,\rho\, X_\kappa(s,t)\} - \{X_\kappa(s',t')\,\rho\, X_\kappa(s',t')\}}{2} = W^{(2)}_\rho(s,t;\, s',t')\,.$$

□

As a first step toward recovering the density matrix from its Wigner characteristic function, we observe that the positivity condition eq. (A.6.1) is exactly what we need to turn a function on $\mathcal{C}^* \times \mathcal{P}^*$ into a state (aka. a normalized positive linear functional, see [7, part III, def. 2.2.8]) on the Weyl algebra. Then, from any state we can reconstruct a representation of the algebra, via the GNS construction [5], and the continuity of the characteristic function ensures that this representation will be weakly continuous.

**Proposition A.9** Let $W \in \mathcal{W}$. Then, there exist a weakly continuous, unitary representation $\mathcal{H}_W$, $\mathrm{T}_W$ of $\mathcal{C}^* \times \mathcal{P}^*$ and a vector $\zeta_W \in \mathcal{H}_W$ such that:

1. $\mathcal{H}_W = \overline{\mathrm{Vect}\,\{\mathrm{T}_W(s,t)\,\zeta_W \mid (s,t) \in \mathcal{C}^* \times \mathcal{P}^*\}}$ (ie. $\zeta_W$ is a cyclic vector for $\mathcal{H}_W$, $\mathrm{T}_W$);
2. $\forall (s,t) \in \mathcal{C}^* \times \mathcal{P}^*$, $W(s,t) = \langle \zeta_W \mid \mathrm{T}_W(s,t)\,\zeta_W \rangle_{\mathcal{H}_W}$.

**Proof** Let $\iota_1, \iota_2 \in \mathcal{I}$. We define:

$$\langle \iota_1 \mid \iota_2 \rangle_W := \sum_{(s,t) \in \mathcal{C}^* \times \mathcal{P}^*} (\iota_1^* \star \iota_2)(s,t)\, W(s,t)\,.$$

This is well defined, for the sum has only finitely many non-zero contributions ($\iota_1^* \star \iota_2 \in \mathcal{I}$), and eq. (A.6.1) ensures that $\langle \cdot \mid \cdot \rangle_W$ provides an Hermitian, positive semi-definite sesquilinear form on the complex vector space $\mathcal{I}$ (the hermiticity comes from $(\iota_1^* \star \iota_2)^* = \iota_2^* \star \iota_1$ together with $\forall (s,t) \in \mathcal{C}^* \times \mathcal{P}^*$, $W(-s,-t) = \overline{W(s,t)}$, as was proven in prop. A.6).

Let $\mathcal{N}_W := \{\upsilon \in \mathcal{I} \mid \langle \upsilon \mid \upsilon \rangle_W = 0\}$. For any $\iota \in \mathcal{I}$ and any $\upsilon \in \mathcal{N}_W$, we have:

$$\forall \lambda \in \mathbb{C},\ \langle \iota \mid \iota \rangle + 2\,\mathrm{Re}\big[\lambda\, \langle \iota \mid \upsilon \rangle\big] = \langle \iota + \lambda\,\upsilon \mid \iota + \lambda\,\upsilon \rangle_W \geqslant 0\,,$$

therefore $\langle \iota \mid \upsilon \rangle = 0$. Hence, for any $\iota_1, \iota_2 \in \mathcal{I}$ and any $\upsilon_1, \upsilon_2 \in \mathcal{N}_W$:

$$\langle \iota_1 + \upsilon_1 \mid \iota_2 + \upsilon_2 \rangle_W = \langle \iota_1 \mid \iota_2 \rangle_W\,,$$

so $\langle \cdot \mid \cdot \rangle_W$ induces an inner product on the quotient $\mathcal{I}/\mathcal{N}_W$. We denote by $\mathcal{H}_W$ the completion of $\mathcal{I}/\mathcal{N}_W$ with respect to the corresponding norm.

Let $(s,t) \in \mathcal{C}^* \times \mathcal{P}^*$. For any $\iota_1, \iota_2 \in \mathcal{I}$, we have:

$$\langle \delta_{(s,t)} \star \iota_1 \mid \delta_{(s,t)} \star \iota_2 \rangle_W = \langle \iota_1 \mid \iota_2 \rangle_W$$

(for $(\delta_{(s,t)} \star \iota_1)^* \star (\delta_{(s,t)} \star \iota_2) = \iota_1^* \star (\delta_{(-s,-t)} \star \delta_{(s,t)}) \star \iota_2 = \iota_1^* \star \iota_2$), so in particular $[\delta_{(s,t)} \star \cdot\,]\langle \mathcal{N}_W \rangle \subset \mathcal{N}_W$, and $\delta_{(s,t)} \star \cdot$ induces a unitary operator $\mathrm{T}_W(s,t)$ on $\mathcal{H}_W$. Then, def. A.3.2 and A.3.3 come from:

$$\forall (s_1,t_1),\,(s_2,t_2) \in \mathcal{C}^* \times \mathcal{P}^*,\ \delta_{(s_1,t_1)} \star \delta_{(s_2,t_2)} = e^{i\xi(s_1,t_1;s_2,t_2)}\, \delta_{(s_1+s_2,\,t_1+t_2)}\,,$$

and from the fact that $\delta_{(0,0)}$ is the unit of the $^*$-algebra $\mathcal{I}$.

Let $\iota_1$, $\iota_2 \in \mathcal{I}$. For any $(s,t) \in \mathcal{C}^* \times \mathcal{P}^*$, we have:

$$\langle \iota_1 \mid \delta_{(s,t)} \star \iota_2 \rangle_W = \sum_{\substack{(s_1,t_1),(s_2,t_2) \\ \in \mathcal{C}^* \times \mathcal{P}^*}} e^{i\xi(s_2,t_2;s_1,t_1)+i\xi(s,t;s_1+s_2,t_1+t_2)} \, \overline{\iota_1(s_1,t_1)} \, \iota_2(s_2,t_2) \, W(s+s_2-s_1, t+t_2-t_1) \,.$$

Thus, the map $(s,t) \mapsto \langle \iota_1 \mid \delta_{(s,t)} \star \iota_2 \rangle_W$ is a finite linear combination of translations of $W$, and those are continuous by definition of $\mathcal{W}$. So, for any $\widetilde{\iota_1}, \widetilde{\iota_2} \in \mathcal{I}/\mathcal{N}_W$, the map $(s,t) \mapsto \langle \widetilde{\iota_1} \mid \mathrm{T}_W(s,t)\, \widetilde{\iota_2} \rangle_{\mathcal{H}_W}$ is continuous. Now, $\mathcal{I}/\mathcal{N}_W$ being dense in $\mathcal{H}_W$, there exist, for any $\varphi_1, \varphi_2 \in \mathcal{H}_W$ and any $\epsilon > 0$, $\widetilde{\iota_1}, \widetilde{\iota_2} \in \mathcal{I}/\mathcal{N}_W$ such that:

$$\|\varphi_1 - \widetilde{\iota_1}\|_{\mathcal{H}_W} < \frac{\epsilon}{6\, \|\varphi_2\|_{\mathcal{H}_W} + \sqrt{6\epsilon}} \quad \& \quad \|\varphi_2 - \widetilde{\iota_2}\|_{\mathcal{H}_W} < \frac{\epsilon}{6\, \|\varphi_1\|_{\mathcal{H}_W} + \sqrt{6\epsilon}} \,.$$

Next, for any $(s,t) \in \mathcal{C}^* \times \mathcal{P}^*$, there exists an open neighborhood $U$ of $(s,t)$ in $\mathcal{C}^* \times \mathcal{P}^*$ such that:

$$\forall (s',t') \in U,\ \left| \langle \widetilde{\iota_1} \mid \mathrm{T}_W(s',t')\, \widetilde{\iota_2} \rangle_{\mathcal{H}_W} - \langle \widetilde{\iota_1} \mid \mathrm{T}_W(s,t)\, \widetilde{\iota_2} \rangle_{\mathcal{H}_W} \right| < \frac{\epsilon}{3} \,.$$

Hence, for any $(s',t') \in U$, we have:

$$\left| \langle \varphi_1 \mid \mathrm{T}_W(s',t')\, \varphi_2 \rangle - \langle \varphi_1 \mid \mathrm{T}_W(s,t)\, \varphi_2 \rangle \right| < \epsilon \,,$$

where we have used that $\mathrm{T}_W(s,t)$ and $\mathrm{T}_W(s',t')$ are unitary. This proves def. A.3.4.

Finally, we define $\zeta_W$ to be the equivalence class of $\delta_{(0,0)}$ in $\mathcal{I}/\mathcal{N}_W$. For any $(s,t) \in \mathcal{C}^* \times \mathcal{P}^*$, we have $\delta_{(s,t)} \star \delta_{(0,0)} = \delta_{(s,t)}$, so $\mathrm{T}_W(s,t)\, \zeta_W$ is the equivalence class of $\delta_{(s,t)}$ in $\mathcal{I}/\mathcal{N}_W$. But since $\mathcal{I} = \mathrm{Vect}\left\{ \delta_{(s,t)} \mid (s,t) \in \mathcal{C}^* \times \mathcal{P}^* \right\}$, we have:

$$\mathcal{I}/\mathcal{N}_W = \mathrm{Vect}\left\{ \mathrm{T}_W(s,t)\, \zeta_W \mid (s,t) \in \mathcal{C}^* \times \mathcal{P}^* \right\},$$

which proves point A.9.1. Moreover, we also have:

$$\langle \zeta_W \mid \mathrm{T}_W(s,t)\, \zeta_W \rangle_{\mathcal{H}_W} = \langle \delta_{(0,0)} \mid \delta_{(s,t)} \rangle_W = W(s,t) \,,$$

which proves point A.9.2. □

Finally, the invertibility of Wigner transform can be seen as a consequence of the Stone-von Neumann theorem [27] (which we implicitly prove below, following closely [11, theorem 15a]). Indeed, the Schrödinger representation being the unique irreducible, weakly continuous representation of the Weyl algebra, the Hilbert space $\mathcal{H}_W$ constructed from $W$ above can be written as a direct sum of independent copies of $\mathcal{H}_o$. Projecting the vector $\zeta_W$ representing $W$ in $\mathcal{H}_W$ on each of these copies, we obtain a collection of vectors in $\mathcal{H}_o$, and we can reconstruct a density matrix on $\mathcal{H}_o$ as the statistical superposition of the corresponding pure states (this is consistent with the general result that a state is pure if and only if its GNS representation is irreducible, see [7, part III, theorem 2.2.17]).

**Proposition A.10** Let $W \in \mathcal{W}$. Then, there exists a *unique* traceclass, (self-adjoint) positive semi-definite operator $\rho$ on $\mathcal{H}_o$ such that $W = W_\rho$ (with $W_\rho$ defined as in prop. A.7).

**Lemma A.11** Let $( \cdot \mid \cdot )$ be a real inner product on $\mathcal{C}$. Using the resulting identification of $\mathcal{C}^*$ with $\mathcal{C}$ and identifying $\mathcal{P}^*$ with $\mathcal{C}$ through the dual map $\Xi^*$, we are provided with a corresponding real inner product on $\mathcal{C}^* \times \mathcal{P}^*$, which we will denote by $( \cdot, \cdot \mid \cdot, \cdot )$. Let $\widetilde{\mu}$ be the Lebesgue measure on $\mathcal{C}^* \times \mathcal{P}^*$ normalized with respect to this Euclidean structure. We define a map $\Psi$ on $\mathcal{C}^* \times \mathcal{P}^*$

through:

$$\forall (s,t) \in \mathcal{C}^* \times \mathcal{P}^*,\ \Psi(s,t) := \frac{1}{\alpha}\, \exp\left(-\frac{(s,t \mid s,t)}{4}\right),$$

where $\alpha := \int_{\mathcal{C}^*\times\mathcal{P}^*} d\widetilde{\mu}(s,t)\, \exp\left(-\frac{(s,t \mid s,t)}{2}\right)$.

Let $W$ be a bounded, continuous function on $\mathcal{C}^* \times \mathcal{P}^*$ such that:

$$\forall (s_o,t_o),\ (s_1,t_1) \in \mathcal{C}^* \times \mathcal{P}^*,\ \int_{\mathcal{C}^*\times\mathcal{P}^*} d\widetilde{\mu}(s,t)\, \Psi(s,t)\, e^{i\,\xi(s,t;s_1,t_1)}\, W(s_o+s,\ t_o+t) = 0\,. \tag{A.11.1}$$

Then, $W \equiv 0$.

**Proof** Let $(e_i)_{i\in\{1,\dots,p\}}$ be an orthonormal basis of $\mathcal{C}$, $(\,\cdot\,,\,\cdot\,)$, let $(f_i)_{i\in\{1,\dots,p\}}$ be the corresponding dual basis in $\mathcal{C}^*$ and let $(g_i)_{i\in\{1,\dots,p\}}$ be the basis in $\mathcal{P}^*$ given by:

$$\forall i \in \{1,\dots,p\}\,,\ g_i := \Xi^*(e_i)\,.$$

For any $(s,t) \in \mathcal{C}^* \times \mathcal{P}^*$, we will write $s =: s^i f_i$ and $t =: t^i g_i$ (with implicit summation). In particular, we then have:

$$\forall (s,t),\ (s',t') \in \mathcal{C}^* \times \mathcal{P}^*,\ \xi(s,t;s',t') = \frac{t^i\, s'^{\,i} - s^i\, t'^{\,i}}{2} \quad \& \quad \big(s,t \mid s',t'\big) = s^i\, s'^{\,i} + t^i\, t'^{\,i}\,,$$

as well as $d\widetilde{\mu}(s,t) = ds^1 \dots ds^p\, dt^1 \dots dt^p$ . Therefore, $\alpha = (2\pi)^p$.

Now, for any $\beta > 0$ and any $(s_o,t_o) \in \mathcal{C}^* \times \mathcal{P}^*$, we have:

$$\int_{\mathcal{C}^*\times\mathcal{P}^*} d\widetilde{\mu}(s_1,t_1) \int_{\mathcal{C}^*\times\mathcal{P}^*} d\widetilde{\mu}(s,t)\ \left|\Psi(\beta s_1,\beta t_1)\, \Psi(s,t)\, e^{i\xi(s,t;s_1,t_1)}\, W\left(s_o+s,\ t_o+t\right)\right|$$
$$\leqslant \left[\frac{2}{\beta}\right]^{2p} \|W\|_\infty < \infty\,,$$

where $\|W\|_\infty$ is the sup norm of the bounded function $W$. Thus, Fubini's theorem, together with eq. (A.11.1) yields:

$$0 = \int_{\mathcal{C}^*\times\mathcal{P}^*} d\widetilde{\mu}(s,t)\, \Psi(s,t)\, W\left(s_o+s,\ t_o+t\right) \int_{\mathcal{C}^*\times\mathcal{P}^*} d\widetilde{\mu}(s_1,t_1) \Psi(\beta s_1,\beta t_1)\, e^{i\xi(s,t;s_1,t_1)}\,.$$

Performing the Gaussian integration, we get:

$$0 = \int_{\mathcal{C}^*\times\mathcal{P}^*} d\widetilde{\mu}(s,t)\ \left[\frac{\gamma^2-1}{\pi}\right]^p \exp\left(-\frac{\gamma^2}{4}\ (s,t \mid s,t)\right)\ W\left(s_o+s,\ t_o+t\right),$$

where $\gamma := \sqrt{1 + {}^1\!/\!_{\beta^2}}$.

Let $\epsilon > 0$. There exists $A > 0$ such that:

$$\int_A^\infty du\, u^{2p-1}\, e^{-\frac{u^2}{4}} < \frac{\epsilon}{4\,\|W\|_\infty + 1} \int_0^\infty du\, u^{2p-1}\, e^{-\frac{u^2}{4}}\,.$$

$W$ being continuous, there also exists $\delta \in \,]0,\, A[$ such that:

$$\forall (s,t) \in \mathcal{C}^* \times \mathcal{P}^* \big/\ (s,t \mid s,t) \leqslant \delta^2,\ \left|W(s_o+s,\ t_o+t) - W(s_o,\ t_o)\right| < \frac{\epsilon}{2}\,.$$

Hence, applying the previous equality with $\beta = \dfrac{\delta}{\sqrt{A^2 - \delta^2}}$, we have:

$$
\begin{aligned}
&|W(s_o,\, t_o)| \int_{\mathcal{C}^* \times \mathcal{P}^*} d\widetilde{\mu}(s,t)\, e^{-\frac{\gamma^2}{4}(s,t\,|\,s,t)} = \\
&\quad = \left| \int_{\mathcal{C}^* \times \mathcal{P}^*} d\widetilde{\mu}(s,t)\, e^{-\frac{\gamma^2}{4}(s,t\,|\,s,t)} \left[ W(s_o + s,\, t_o + t) - W(s_o,\, t_o) \right] \right| \\
&\qquad < \frac{\epsilon}{2} \int_{(s,t\,|\,s,t) \leqslant \delta^2} d\widetilde{\mu}(s,t)\, e^{-\frac{\gamma^2}{4}(s,t\,|\,s,t)} + 2\, \|W\|_\infty \int_{(s,t\,|\,s,t) \geqslant \delta^2} d\widetilde{\mu}(s,t)\, e^{-\frac{\gamma^2}{4}(s,t\,|\,s,t)} .
\end{aligned}
$$

Changing to spherical coordinates and dropping an overall constant, this can be rewritten as:

$$
\begin{aligned}
|W(s_o,\, t_o)| \int_0^\infty du\, u^{2p-1}\, e^{-\frac{u^2}{4}} &< \frac{\epsilon}{2} \int_0^{\gamma\delta} du\, u^{2p-1}\, e^{-\frac{u^2}{4}} + 2\, \|W\|_\infty \int_{\gamma\delta}^\infty du\, u^{2p-1}\, e^{-\frac{u^2}{4}} \\
&< \epsilon \int_0^\infty du\, u^{2p-1}\, e^{-\frac{u^2}{4}} .
\end{aligned}
$$

where we have used that $\gamma\delta = A$. Thus, we have, for any $(s_o, t_o) \in \mathcal{C}^* \times \mathcal{P}^*$ and any $\epsilon > 0$, $|W(s_o, t_o)| < \epsilon$, and therefore $W \equiv 0$. □

**Proof of prop. A.10** *Existence.* Let $W \in \mathcal{W}$ and let $\mathcal{H}_W$, $\mathrm{T}_W$ be the weakly continuous, unitary representation introduced in prop. A.9, with cyclic vector $\zeta_W$. For any $\varphi, \varphi' \in \mathcal{H}_W$, the map $(s,t) \mapsto \Psi(s,t)\, \langle \varphi' \mid \mathrm{T}_W(s,t)\, \varphi \rangle_{\mathcal{H}_W}$ is continuous, hence measurable, and we have:

$$
\int_{\mathcal{C}^* \times \mathcal{P}^*} d\widetilde{\mu}(s,t)\, \left| \Psi(s,t)\, \langle \varphi' \mid \mathrm{T}_W(s,t)\, \varphi \rangle_{\mathcal{H}_W} \right| \leqslant 2^p\, \|\varphi'\|_{\mathcal{H}_W}\, \|\varphi\|_{\mathcal{H}_W} < \infty .
$$

Hence, there exists a bounded operator $\mathrm{T}_W\{\Psi\}$ on $\mathcal{H}_W$ such that, for any $\varphi, \varphi' \in \mathcal{H}_W$:

$$
\langle \varphi' \mid \mathrm{T}_W\{\Psi\}\, \varphi \rangle_{\mathcal{H}_W} := \int_{\mathcal{C}^* \times \mathcal{P}^*} d\widetilde{\mu}(s,t)\, \Psi(s,t)\, \langle \varphi' \mid \mathrm{T}_W(s,t)\, \varphi \rangle_{\mathcal{H}_W} .
$$

In particular, $\mathrm{T}_W\{\Psi\}$ is symmetric (since, for any $(s,t) \in \mathcal{C}^* \times \mathcal{P}^*$, $\overline{\Psi(s,t)} = \Psi(s,t) = \Psi(-s,-t)$ and $\mathrm{T}_W(s,t)^\dagger = \mathrm{T}_W(-s,-t)$). Being bounded, it is therefore self-adjoint.

Now, for any $(s_1, t_1) \in \mathcal{C}^* \times \mathcal{P}^*$ and any $\varphi, \varphi' \in \mathcal{H}_W$, we have:

$$
\begin{aligned}
\left\langle \varphi' \,\middle|\, \left[ \mathrm{T}_W\{\Psi\}\, \mathrm{T}_W(s_1,t_1) \right] \varphi \right\rangle_{\mathcal{H}_W} &= \left\langle \varphi' \,\middle|\, \mathrm{T}_W\{\Psi\}\, \left[ \mathrm{T}_W(s_1,t_1)\, \varphi \right] \right\rangle_{\mathcal{H}_W} \\
&= \int_{\mathcal{C}^* \times \mathcal{P}^*} d\widetilde{\mu}(s,t)\, \Psi(s,t)\, \langle \varphi' \mid \mathrm{T}_W(s,t)\, \mathrm{T}_W(s_1,t_1)\, \varphi \rangle_{\mathcal{H}_W} \\
&= \int_{\mathcal{C}^* \times \mathcal{P}^*} d\widetilde{\mu}(s,t)\, \Psi(s,t)\, e^{i\xi(s,t;s_1,t_1)}\, \langle \varphi' \mid \mathrm{T}_W(s+s_1, t+t_1)\, \varphi \rangle_{\mathcal{H}_W} \\
&= \int_{\mathcal{C}^* \times \mathcal{P}^*} d\widetilde{\mu}(s,t)\, \Psi(s-s_1, t-t_1)\, e^{i\xi(s,t;s_1,t_1)}\, \langle \varphi' \mid \mathrm{T}_W(s,t)\, \varphi \rangle_{\mathcal{H}_W} ,
\end{aligned}
$$

where we have used def. A.3.2 and $\forall (s,t) \in \mathcal{C}^* \times \mathcal{P}^*$, $\xi(s-s_1, t-t_1; s_1, t_1) = \xi(s,t;s_1,t_1)$. Since $\mathrm{T}_W\{\Psi\}^\dagger = \mathrm{T}_W\{\Psi\}$ and $\mathrm{T}_W(s_1,t_1)^\dagger = \mathrm{T}_W(-s_1,-t_1)$, we also have:

$$
\left\langle \varphi' \,\middle|\, \left[ \mathrm{T}_W(s_1,t_1)\, \mathrm{T}_W\{\Psi\} \right] \varphi \right\rangle_{\mathcal{H}_W} = \int_{\mathcal{C}^* \times \mathcal{P}^*} d\widetilde{\mu}(s,t)\, \Psi(s-s_1, t-t_1)\, e^{-i\xi(s,t;s_1,t_1)}\, \langle \varphi' \mid \mathrm{T}_W(s,t)\, \varphi \rangle_{\mathcal{H}_W} .
$$

Applying these two formulas successively, we get, for any $(s_1, t_1) \in \mathcal{C}^* \times \mathcal{P}^*$ and any $\varphi, \varphi' \in \mathcal{H}_W$ :

$$\begin{aligned}
&\left\langle \varphi' \,\middle|\, \left[\mathrm{T}_W\left\{\Psi\right\}\, \mathrm{T}_W(s_1,t_1)\, \mathrm{T}_W\left\{\Psi\right\}\right] \varphi \right\rangle = \\
&\quad = \left\langle \varphi' \,\middle|\, \mathrm{T}_W\left\{\Psi\right\}\, \mathrm{T}_W(s_1,t_1) \left[\mathrm{T}_W\left\{\Psi\right\}\, \varphi\right]\right\rangle \\
&\quad = \int_{\mathcal{C}^*\times\mathcal{P}^*} d\widetilde{\mu}(s,t)\, \Psi(s-s_1, t-t_1)\, e^{i\xi(s,t;s_1,t_1)}\, \left\langle \varphi' \mid \mathrm{T}_W(s,t)\, \mathrm{T}_W\left\{\Psi\right\}\, \varphi\right\rangle_{\mathcal{H}_W} \\
&\quad = \int_{\mathcal{C}^*\times\mathcal{P}^*} d\widetilde{\mu}(s,t) \int_{\mathcal{C}^*\times\mathcal{P}^*} d\widetilde{\mu}(s',t')\, \Psi(s-s_1, t-t_1)\, \Psi(s'-s, t'-t)\, e^{i\xi(s,t;s_1+s',t_1+t')}\, \left\langle \varphi' \mid \mathrm{T}_W(s',t')\, \varphi\right\rangle_{\mathcal{H}_W}.
\end{aligned}$$

Invoking Fubini's theorem and performing the Gaussian integration, like in the proof of lemma A.11, we obtain:

$$\begin{aligned}
&\left\langle \varphi' \,\middle|\, \left[\mathrm{T}_W\left\{\Psi\right\}\, \mathrm{T}_W(s_1,t_1)\, \mathrm{T}_W\left\{\Psi\right\}\right] \varphi \right\rangle = \\
&\qquad = \int_{\mathcal{C}^*\times\mathcal{P}^*} d\widetilde{\mu}(s',t')\, \exp\left(-\frac{(s_1,t_1 \mid s_1,t_1)}{4}\right)\, \Psi(s',t')\, \left\langle \varphi' \mid \mathrm{T}_W(s',t')\, \varphi\right\rangle_{\mathcal{H}_W} \\
&\qquad = \exp\left(-\frac{(s_1,t_1 \mid s_1,t_1)}{4}\right) \left\langle \varphi' \mid \mathrm{T}_W\left\{\Psi\right\}\, \varphi\right\rangle_{\mathcal{H}_W}.
\end{aligned}$$

Since this holds for any $\varphi, \varphi' \in \mathcal{H}_W$ , we have:

$$\forall (s,t) \in \mathcal{C}^* \times \mathcal{P}^*,\ \mathrm{T}_W\left\{\Psi\right\}\, \mathrm{T}_W(s,t)\, \mathrm{T}_W\left\{\Psi\right\} = \exp\left(-\frac{(s,t \mid s,t)}{4}\right) \mathrm{T}_W\left\{\Psi\right\}. \tag{A.10.1}$$

In particular, applying with $(s,t) = (0,0)$, and using that $\mathrm{T}_W(0,0) = \mathrm{id}_{\mathcal{H}_W}$ , this implies:

$$\mathrm{T}_W\left\{\Psi\right\}\, \mathrm{T}_W\left\{\Psi\right\} = \mathrm{T}_W\left\{\Psi\right\}.$$

$\mathrm{T}_W\left\{\Psi\right\}$ being idempotent and self-adjoint, it is an orthogonal projection.

Next, we define $\psi_o \in \mathcal{H}_o$ by:

$$\forall x \in \mathcal{C},\ \psi_o(x) = \frac{1}{\pi^{p/4}} \exp\left(-\frac{(x \mid x)}{2}\right)$$

(choosing the Lebesgue measure $\mu$, entering the definition of $\mathcal{H}_o$ in prop. A.5, to be normalized with respect to the scalar product $(\cdot \mid \cdot)$ on $\mathcal{C}$; this normalization can be chosen without loss of generality since the Hilbert spaces obtained using different normalizations of $\mu$ are unitarily identified in a natural way through a corresponding renormalization of the wave-functions). We can check that, for any $(s,t) \in \mathcal{C}^* \times \mathcal{P}^*$:

$$\left\langle \psi_o \mid \mathrm{T}_o(s,t)\, \psi_o\right\rangle_{\mathcal{H}_o} = \exp\left(-\frac{(s,t \mid s,t)}{4}\right). \tag{A.10.2}$$

Let $\left(\psi_W^{(a)}\right)_a$ be an orthonormal basis of the image of $\mathrm{T}_W\left\{\Psi\right\}$. Then, for any $a, b$ and any $(s,t)$, $(s',t') \in \mathcal{C}^* \times \mathcal{P}^*$, we have, using the hermiticity and idempotence of $\mathrm{T}_W\left\{\Psi\right\}$ together with the properties of the representations $\mathrm{T}_W$ and $\mathrm{T}_o$ :

$$\begin{aligned}
&\left\langle \mathrm{T}_W(s',t')\, \psi_W^{(b)} \,\middle|\, \mathrm{T}_W(s,t)\, \psi_W^{(a)} \right\rangle_{\mathcal{H}_W} = \\
&\qquad = \left\langle \mathrm{T}_W\left\{\Psi\right\}\, \psi_W^{(b)} \,\middle|\, \mathrm{T}_W(-s',-t')\, \mathrm{T}_W(s,t)\, \mathrm{T}_W\left\{\Psi\right\}\, \psi_W^{(a)} \right\rangle_{\mathcal{H}_W}
\end{aligned}$$

$$= e^{i\xi(s,t;s',t')} \left\langle \psi_W^{(b)} \,\middle|\, \mathrm{T}_W\{\Psi\}\, \mathrm{T}_W(s-s',t-t')\, \mathrm{T}_W\{\Psi\}\, \psi_W^{(a)} \right\rangle_{\mathcal{H}_W}$$

$$= e^{i\xi(s,t;s',t')} \left\langle \psi_o \,\middle|\, \mathrm{T}_o(s-s',t-t')\, \psi_o \right\rangle_{\mathcal{H}_o} \left\langle \psi_W^{(b)} \,\middle|\, \mathrm{T}_W\{\Psi\}\, \psi_W^{(a)} \right\rangle_{\mathcal{H}_W}$$

(combining eqs. (A.10.1) and (A.10.2))

$$= \delta_{ab} \left\langle \mathrm{T}_o(s',t')\, \psi_o \,\middle|\, \mathrm{T}_o(s,t)\, \psi_o \right\rangle_{\mathcal{H}_o}.$$

For any $a$, we define $\mathcal{H}_W^{(a)} := \overline{\mathrm{Vect}\left\{\mathrm{T}_W(s,t)\,\psi_W^{(a)} \,\middle|\, (s,t) \in \mathcal{C}^* \times \mathcal{P}^*\right\}}$. Each $\mathcal{H}_W^{(a)}$ is thus stable under $\mathrm{T}_W\langle \mathcal{C}^* \times \mathcal{P}^*\rangle$. The previous computation shows that the $\mathcal{H}_W^{(a)}$ are mutually orthogonal and that there exists, for each $a$, an isometric injection $I^{(a)} : \mathcal{H}_W^{(a)} \to \mathcal{H}_o$ such that:

$$\forall \iota \in \mathcal{I},\ I^{(a)} \left[\mathrm{T}_W\{\iota\}\ \psi_W^{(a)}\right] = \mathrm{T}_o\{\iota\}\ \psi_o\,.$$

Then, using that:

$$\forall \iota_1, \iota_2 \in \mathcal{I},\ \mathrm{T}_W\{\iota_1\}\ \mathrm{T}_W\{\iota_2\} = \mathrm{T}_W\{\iota_1 \star \iota_2\} \quad \& \quad \mathrm{T}_o\{\iota_1\}\ \mathrm{T}_o\{\iota_2\} = \mathrm{T}_o\{\iota_1 \star \iota_2\}\,,$$

together with the density of $\left\{\mathrm{T}_W\{\iota\}\ \psi_W^{(a)} \,\middle|\, \iota \in \mathcal{I}\right\} = \mathrm{Vect}\left\{\mathrm{T}_W(s,t)\,\psi_W^{(a)} \,\middle|\, (s,t) \in \mathcal{C}^* \times \mathcal{P}^*\right\}$ in $\mathcal{H}_W^{(a)}$, we get:

$$\forall \iota \in \mathcal{I},\ I^{(a)}\,\mathrm{T}_W\{\iota\} = \mathrm{T}_o\{\iota\}\ I^{(a)}\,.$$

Now, we define:

$$\mathcal{H}_W^{\mathrm{rest}} := \left(\bigoplus_a \mathcal{H}_W^{(a)}\right)^{\perp},$$

so that:

$$\mathcal{H}_W = \mathcal{H}_W^{\mathrm{rest}} \oplus \overline{\bigoplus_a \mathcal{H}_W^{(a)}}$$

(as can be seen as a consequence of Riesz lemma). Since each $\mathcal{H}_W^{(a)}$ is stable under $\mathrm{T}_W\langle \mathcal{C}^* \times \mathcal{P}^*\rangle$, so is $\mathcal{H}_W^{\mathrm{rest}}$. Next, we write:

$$\zeta_W = \zeta_W^{\mathrm{rest}} + \left(\sum_a \zeta_W^{(a)}\right),$$

with $\zeta_W^{\mathrm{rest}} \in \mathcal{H}_W^{\mathrm{rest}}$ and $\forall a,\ \zeta_W^{(a)} \in \mathcal{H}_W^{(a)}$. In particular, we have:

$$\|\zeta_W\|^2 = \left\|\zeta_W^{\mathrm{rest}}\right\|^2 + \sum_a \left\|\zeta_W^{(a)}\right\|^2,$$

so there can be only countably many non-zero $\zeta_W^{(a)}$ (actually, the cyclicity of $\zeta_W$ together with the stability of each $\mathcal{H}_W^{(a)}$ requires all $\zeta_W^{(a)}$ to be non-zero, so this even implies that $a$ takes value in a countable set). Using prop. A.9.2, we have:

$$\forall (s,t) \in \mathcal{C}^* \times \mathcal{P}^*,\ W(s,t) = \left\langle \zeta_W \,\middle|\, \mathrm{T}_W(s,t)\,\zeta_W \right\rangle_{\mathcal{H}_W}$$

$$= \left\langle \zeta_W^{\text{rest}} \;\middle|\; \mathrm{T}_W(s,t)\, \zeta_W^{\text{rest}} \right\rangle_{\mathcal{H}_W} + \sum_a \left\langle \zeta_W^{(a)} \;\middle|\; \mathrm{T}_W(s,t)\, \zeta_W^{(a)} \right\rangle_{\mathcal{H}_W}.$$

Thus, defining $\zeta_o^{(a)} := I^{(a)}\, \zeta_W^{(a)} \in \mathcal{H}_o$ for each $a$, the isometric and intertwining properties of $I^{(a)}$ yield:

$$\forall (s,t) \in \mathcal{C}^* \times \mathcal{P}^*,\ W(s,t) = \left\langle \zeta_W^{\text{rest}} \;\middle|\; \mathrm{T}_W(s,t)\, \zeta_W^{\text{rest}} \right\rangle_{\mathcal{H}_W} + \sum_a \left\langle \zeta_o^{(a)} \;\middle|\; \mathrm{T}_o(s,t)\, \zeta_o^{(a)} \right\rangle_{\mathcal{H}_o}.$$

The bound:

$$\sum_a \left\| \zeta_o^{(a)} \right\|_{\mathcal{H}_o}^2 = \sum_a \left\| \zeta_W^{(a)} \right\|_{\mathcal{H}_W}^2 \leqslant \left\| \zeta_W \right\|^2 < \infty,$$

allows us to define a traceclass, (self-adjoint) positive semi-definite operator $\rho$ on $\mathcal{H}_o$ by:

$$\rho := \sum_a \left| \zeta_o^{(a)} \right\rangle \left\langle \zeta_o^{(a)} \right|.$$

Then, we get:

$$\forall (s,t) \in \mathcal{C}^* \times \mathcal{P}^*,\ W(s,t) = W^{\text{rest}}(s,t) + \mathrm{Tr}_{\mathcal{H}_o}\left[\rho\, \mathrm{T}_o(s,t)\right] = W^{\text{rest}}(s,t) + W_\rho(s,t),$$

where, for any $(s,t) \in \mathcal{C}^* \times \mathcal{P}^*$, $W^{\text{rest}}(s,t) := \left\langle \zeta_W^{\text{rest}} \;\middle|\; \mathrm{T}_W(s,t)\, \zeta_W^{\text{rest}} \right\rangle_{\mathcal{H}_W}$.

$\mathrm{T}_W$ being a weakly continuous, unitary representation of $\mathcal{C}^* \times \mathcal{P}^*$ on $\mathcal{H}_W$, the function $W^{\text{rest}}$ is continuous $\mathcal{C}^* \times \mathcal{P}^* \to \mathbb{C}$ and it is bounded by $\left\| \zeta_W^{\text{rest}} \right\|_{\mathcal{H}_W}^2$. Now, for any $(s',t') \in \mathcal{C}^* \times \mathcal{P}^*$, $\mathrm{T}_W(s',t')\, \zeta_W^{\text{rest}} \in \mathcal{H}_W^{\text{rest}}$, so it is in particular orthogonal to each $\psi_W^{(a)}$, as $\psi_W^{(a)} \in \mathcal{H}_W^{(a)}$. But since $\left(\psi_W^{(a)}\right)_a$ is an orthonormal basis of the image of the orthogonal projection $\mathrm{T}_W\left\{\Psi\right\}$, this implies:

$$\forall (s',t') \in \mathcal{C}^* \times \mathcal{P}^*,\ \mathrm{T}_W\left\{\Psi\right\}\, \mathrm{T}_W(s',t')\, \zeta_W^{\text{rest}} = 0,$$

and therefore:

$$\forall (s',t'),\, (s'',t'') \in \mathcal{C}^* \times \mathcal{P}^*,\ \left\langle \mathrm{T}_W(s'',t'')\, \zeta_W^{\text{rest}} \;\middle|\; \mathrm{T}_W\left\{\Psi\right\}\, \mathrm{T}_W(s',t')\, \zeta_W^{\text{rest}} \right\rangle_{\mathcal{H}_W} = 0.$$

On the other hand, we have, for any $\varphi', \varphi'' \in \mathcal{H}_W$:

$$\begin{aligned}
&\forall (s',t'),\, (s'',t'') \in \mathcal{C}^* \times \mathcal{P}^*,\ \langle \mathrm{T}_W(s'',t'')\, \varphi'' \mid \mathrm{T}_W\left\{\Psi\right\}\, \mathrm{T}_W(s',t')\, \varphi' \rangle_{\mathcal{H}_W} = \\
&\quad = \int_{\mathcal{C}^* \times \mathcal{P}^*} d\widetilde{\mu}(s,t)\, \Psi(s,t)\, \langle \mathrm{T}_W(s'',t'')\, \varphi'' \mid \mathrm{T}_W(s,t)\, \mathrm{T}_W(s',t')\, \varphi' \rangle_{\mathcal{H}_W} \\
&\quad = \int_{\mathcal{C}^* \times \mathcal{P}^*} d\widetilde{\mu}(s,t)\, \Psi(s,t)\, e^{-i\xi(s'',t'';s,t)}\, e^{i\xi(s-s'',t-t'';s',t')}\, \langle \varphi'' \mid \mathrm{T}_W(s+s'-s'', t+t'-t'')\, \varphi' \rangle_{\mathcal{H}_W} \\
&\quad = e^{-i\xi(s'',t'';s',t')} \int_{\mathcal{C}^* \times \mathcal{P}^*} d\widetilde{\mu}(s,t)\, \Psi(s,t)\, e^{i\xi(s,t;s'+s'',t'+t'')}\, \langle \varphi'' \mid \mathrm{T}_W(s+s'-s'', t+t'-t'')\, \varphi' \rangle_{\mathcal{H}_W}.
\end{aligned}$$

Thus, we have:

$$\forall (s_o,t_o),\, (s_1,t_1) \in \mathcal{C}^* \times \mathcal{P}^*,\ \int_{\mathcal{C}^* \times \mathcal{P}^*} d\widetilde{\mu}(s,t)\, \Psi(s,t)\, e^{i\xi(s,t;s_1,t_1)}\, W^{\text{rest}}(s+s_o, t+t_o) = 0,$$

so, from lemma A.11, $W^{\text{rest}} \equiv 0$, hence $W = W_\rho$.

*Uniqueness.* Let $\rho_1$ , $\rho_2$ be two traceclass, (self-adjoint) positive semi-definite operators on $\mathcal{H}_o$ such that $W_{\rho_1} = W_{\rho_2}$ . Like above, we define a bounded linear operator $\mathrm{T}_o\{\Psi\}$ on $\mathcal{H}_o$ satisfying, for any $\varphi_1, \varphi_2 \in \mathcal{H}_o$:

$$\langle \varphi_1 \mid \mathrm{T}_o\{\Psi\}\ \varphi_2\rangle_{\mathcal{H}_o} := \int_{\mathcal{C}^*\times\mathcal{P}^*} d\widetilde{\mu}(s,t)\,\Psi(s,t)\ \langle \varphi_1 \mid \mathrm{T}_o(s,t)\,\varphi_2\rangle_{\mathcal{H}_o}\,.$$

Using the bases introduced in the proof of lemma A.11 together with the expression for $\mathrm{T}_o(s,t)$ from prop. A.5, this can be rewritten as:

$$\begin{aligned}
&\langle \varphi_1 \mid \mathrm{T}_o\{\Psi\}\ \varphi_2\rangle_{\mathcal{H}_o} = \\
&= \frac{1}{(2\pi)^p} \int ds^1 \dots ds^p\, dt^1 \dots dt^p \int dx_1^1 \dots dx_1^p\ \overline{\varphi_1(x_1^i\, e_i)}\,\varphi_2\big((x_1^i + t^i)\, e_i\big) \times \\
&\qquad \times \exp\left(-\frac{s^i\, s^i + t^i\, t^i}{4}\right)\ \exp\left(i\, s^i\, x_1^i + \frac{i}{2} t^i\, s^i\right) \\
&= \frac{1}{(2\pi)^p} \int dx_1^1 \dots dx_1^p\, dt^1 \dots dt^p\ \overline{\varphi_1(x_1^i\, e_i)}\,\varphi_2\big((x_1^i + t^i)\, e_i\big)\ \exp\left(-\frac{t^i\, t^i}{4}\right) \times \\
&\qquad \times \int ds^1 \dots ds^p\ \ \exp\left(-\frac{s^i\, s^i}{4} + i\, s^i\, x_1^i + \frac{i}{2} t^i\, s^i\right)
\end{aligned}$$

(using Fubini's theorem as the integral is absolutely convergent)

$$\begin{aligned}
&= \frac{1}{\sqrt{\pi}^p} \int dx_1^1 \dots dx_1^p\, dt^1 \dots dt^p\ \overline{\varphi_1(x_1^i\, e_i)}\,\varphi_2\big((x_1^i + t^i)\, e_i\big)\ \exp\left(-\frac{t^i\, t^i}{4}\right) \times \\
&\qquad \times \exp\left(-\frac{(2x_1^i + t^i)\,(2x_1^i + t^i)}{4}\right) \\
&= \frac{1}{\sqrt{\pi}^p} \int dx_1^1 \dots dx_1^p\, dx_2^1 \dots dx_2^p\ \overline{\varphi_1(x_1^i\, e_i)}\,\varphi_2(x_2^i\, e_i)\ \exp\left(-\frac{x_1^i\, x_1^i + x_2^i\, x_2^i}{2}\right)
\end{aligned}$$

(with the change of variables $x_2^i := x_1^i + t^i$)

$$= \langle \varphi_1 \mid \psi_o\rangle_{\mathcal{H}_o}\ \langle \psi_o \mid \varphi_2\rangle_{\mathcal{H}_o}\,.$$

Since this holds for any $\varphi_1, \varphi_2 \in \mathcal{H}_o$ , we have $\mathrm{T}_o\{\Psi\} = |\psi_o\,\rangle\langle\,\psi_o|$ .

Now, from the spectral theorem, there exist $N_1 \in \mathbb{N} \cup \{\infty\}$, an orthonormal family $\left(\varphi_k^{(1)}\right)_{k<N_1}$ in $\mathcal{H}_o$ and a family of strictly positive reals $\left(p_k^{(1)}\right)_{k<N_1}$ such that:

$$\rho_1 = \sum_{k<N_1} p_k^{(1)} \left|\varphi_k^{(1)}\right\rangle \left\langle \varphi_k^{(1)}\right| \quad \& \quad \sum_{k<N_1} p_k^{(1)} = \mathrm{Tr}_{\mathcal{H}_o}\, \rho_1\,.$$

Thus, we get, for any $(s_1, t_1)$, $(s_2, t_2) \in \mathcal{C}^* \times \mathcal{P}^*$:

$$\begin{aligned}
&\langle \mathrm{T}_o(s_1, t_1)\,\psi_o \mid \rho_1\, \mathrm{T}_o(s_2, t_2)\,\psi_o\rangle_{\mathcal{H}_o} = \\
&= \sum_{k<N_1} p_k^{(1)} \left\langle \mathrm{T}_o(s_1, t_1)\,\psi_o \,\middle|\, \varphi_k^{(1)}\right\rangle \left\langle \varphi_k^{(1)} \,\middle|\, \mathrm{T}_o(s_2, t_2)\,\psi_o\right\rangle_{\mathcal{H}_o} \\
&= \sum_{k<N_1} p_k^{(1)} \left\langle \varphi_k^{(1)} \,\middle|\, \mathrm{T}_o(s_2, t_2)\,\mathrm{T}_o\{\Psi\}\ \mathrm{T}_o(-s_1, -t_1)\,\varphi_k^{(1)}\right\rangle_{\mathcal{H}_o}
\end{aligned}$$

$$= \sum_{k<N_1} p_k^{(1)} \int_{\mathcal{C}^*\times\mathcal{P}^*} d\widetilde{\mu}(s,t)\, \Psi(s,t)\, e^{-i\xi(s,t;\,s_1+s_2,t_1+t_2)+i\xi(s_1,t_1;\,s_2,t_2)} \left\langle \varphi_k^{(1)} \,\middle|\, \mathrm{T}_o(s+s_{2-s_1}, t+t_{2-t_1})\, \varphi_k^{(1)} \right\rangle_{\mathcal{H}_o}$$

$$= \int_{\mathcal{C}^*\times\mathcal{P}^*} d\widetilde{\mu}(s,t)\, \Psi(s,t)\, e^{-i\xi(s,t;\,s_1+s_2,t_1+t_2)+i\xi(s_1,t_1;\,s_2,t_2)} \sum_{k<N_1} p_k^{(1)} \left\langle \varphi_k^{(1)} \,\middle|\, \mathrm{T}_o(s+s_{2-s_1}, t+t_{2-t_1})\, \varphi_k^{(1)} \right\rangle_{\mathcal{H}_o}$$

(using Fubini's theorem with the discrete measure on $\{k < N\}$)

$$= \int_{\mathcal{C}^*\times\mathcal{P}^*} d\widetilde{\mu}(s,t)\, \Psi(s,t)\, e^{-i\xi(s,t;\,s_1+s_2,t_1+t_2)+i\xi(s_1,t_1;\,s_2,t_2)}\, \mathrm{Tr}_{\mathcal{H}_o} \left[\rho_1\, \mathrm{T}_o(s+s_{2-s_1}, t+t_{2-t_1})\right]$$

$$= e^{i\xi(s_1,t_1;\,s_2,t_2)} \int_{\mathcal{C}^*\times\mathcal{P}^*} d\widetilde{\mu}(s,t)\, \Psi(s,t)\, e^{-i\xi(s,t;\,s_1+s_2,t_1+t_2)}\, W_{\rho_1}(s+s_{2-s_1}, t+t_{2-t_1})\,.$$

As the same holds for $\rho_2$, $W_{\rho_1} = W_{\rho_2}$ implies:

$$\forall \varphi, \varphi' \in \mathcal{H}_o^{(0)},\ \langle \varphi' \mid \rho_1\, \varphi\rangle_{\mathcal{H}_o} = \langle \varphi' \mid \rho_2\, \varphi\rangle_{\mathcal{H}_o}\,,$$

where $\mathcal{H}_o^{(0)} := \overline{\mathrm{Vect}\ \{\mathrm{T}_o(s,t)\,\psi_o \mid (s,t) \in \mathcal{C}^* \times \mathcal{P}^*\}}\,.$

Finally, let $\varphi \in \left[\mathcal{H}_o^{(0)}\right]^\perp$. Like above, we then have, for any $(s',t')$, $(s'',t'') \in \mathcal{C}^* \times \mathcal{P}^*$:

$$0 = \langle \mathrm{T}_o(s'',t'')\,\psi_o \mid \varphi\rangle_{\mathcal{H}_o} \langle \varphi \mid \mathrm{T}_o(s',t')\,\psi_o\rangle_{\mathcal{H}_o}$$

$$= e^{-i\xi(s',t';\,s'',t'')} \int_{\mathcal{C}^*\times\mathcal{P}^*} d\widetilde{\mu}(s,t)\, \Psi(s,t)\, e^{-i\xi(s,t;\,s'+s'',t'+t'')} \langle \varphi \mid \mathrm{T}_o(s+s'-s'', t+t'-t'')\,\varphi\rangle_{\mathcal{H}_o}\,,$$

so applying lemma A.11 to the bounded, continuous function $(s,t) \mapsto \langle \varphi \mid \mathrm{T}_o(s,t)\,\varphi\rangle_{\mathcal{H}_o}$ yields:

$$\forall (s,t) \in \mathcal{C}^* \times \mathcal{P}^*,\ \langle \varphi \mid \mathrm{T}_o(s,t)\,\varphi\rangle_{\mathcal{H}_o} = 0,$$

and in particular $\|\varphi\|^2 = \langle \varphi \mid \mathrm{T}_o(0,0)\,\varphi\rangle_{\mathcal{H}_o} = 0$, so $\varphi = 0$. Hence, $\mathcal{H}_o^{(0)} = \mathcal{H}_o$ and, therefore, $\rho_1 = \rho_2$. □

# B References